\documentclass[preprint,3p,review,authoryear]{elsarticle}

\usepackage{graphicx}
\usepackage{subfigure}
\usepackage{float}
\usepackage{amsmath}
\usepackage{color}

\usepackage{amssymb}

 \usepackage{lineno}

\journal{Ocean Modelling}

\begin{document}
\begin{frontmatter}



\cortext[cor1]{Corresponding author.}

\title{Approximate deconvolution large eddy simulation\\ of a barotropic ocean circulation model}


\author{Omer San\corref{cor1}\fnref{address1}}
\ead{omersan@vt.edu}
\author{Anne E. Staples\fnref{address1}}
\author{Zhu Wang\fnref{address2}}
\author{Traian Iliescu\fnref{address2}}

\address[address1]{Department of Engineering Science and Mechanics, Virginia Tech, Blacksburg, VA, USA}

\address[address2]{Department of Mathematics, Virginia Tech, Blacksburg, VA, USA}

\begin{abstract}
This paper puts forth a new large eddy simulation closure modeling strategy for two-dimensional
turbulent geophysical flows. This closure modeling approach utilizes {\em approximate deconvolution}, which is based
solely on mathematical approximations and does not employ additional phenomenological arguments to the model.
The new approximate deconvolution model is tested in the numerical simulation of the wind-driven
circulation in a shallow ocean basin, a standard prototype of more realistic ocean dynamics.
The model employs the barotropic vorticity equation driven by a symmetric double-gyre wind
forcing, which yields a four-gyre circulation in the time mean. The approximate deconvolution model yields the correct four-gyre circulation structure predicted by a direct numerical simulation, on a coarser mesh but at a fraction of the
computational cost. This first step in the numerical assessment of the new model shows that approximate deconvolution
could represent a viable tool for under-resolved computations in the large eddy
simulation of more realistic turbulent geophysical flows.
\\
\end{abstract}


\end{frontmatter}


\section{Introduction}
\label{sec:intro}

As a first approximation, the mean near-surface currents in the oceans are driven by the
mean effects of winds.
Wind-driven flows of mid-latitude ocean basins have been studied by modelers using
idealized single and double-gyre wind forcing.
This type of double-gyre circulation characterizes all mid-latitude ocean basins, including those in the North
and South Atlantic, as well as the North and South Pacific.
The {\em barotropic vorticity equation (BVE)} represents one of the most commonly used
mathematical models for this type of geostrophic flows with various dissipative and forcing
terms \citep{majda2006non}.
For more details on the physical mechanism and formulations of the BVE, the reader is
referred to \cite{holland1980example,munk1982observing,griffa1989wind,cummins1992inertial,
greatbatch2000four,nadiga2001dispersive,fox2005reevaluating,cushman1994introduction}.

The main assumptions that go into the BVE model are hydrostatic balance, the $\beta$-plane approximation, geostrophic balance, and horizontal eddy viscosity parametrization.
Despite the fact that the BVE models are a simplified version of the full-fledged equations of
geophysical flows, their numerical simulation is still computationally challenging when
long-time integration is required, as is the case in climate modeling.
Thus, to further reduce the computational cost of the BVE, {\em large eddy simulation
(LES)} appears to be a natural approach. In LES only the large spatial structures are approximated, whereas the small scales are modeled. This allows for much coarser spatial meshes and thus a computational cost  that is significantly lower than that of a {\em direct numerical simulation (DNS)}. To achieve the same order of physical accuracy as DNS, however, LES needs to correctly treat the {\em closure problem} \citep{meneveau2000scale,sagaut2006large,berselli2006mathematics}:
the effect of the small scales on the large ones needs to be modeled.
This closure problem represents a significant challenge for quasi-two-dimensional turbulent
flows, such as those in the ocean and atmosphere \citep{mcwilliams1989statistical,danilov2000quasi,
smith2002turbulent,lesieur2008turbulence,salmon1998lectures,majda2003introduction,
holton2004introduction,mcwilliams2006fundamentals,vallis2006atmospheric}.

Two-dimensional turbulence is a fundamental topic for understanding of geophysical flows and behaves in a profoundly different way from the three-dimensional turbulence due to different energy cascade behavior \citep{kraichnan1967inertial,batchelor1969computation,leith1971atmospheric}, which is described in the Kraichnan-Batchelor-Leith two-dimensional turbulence theory. In three-dimensional turbulence, energy is transferred forward, from large scales to smaller scales, via vortex stretching. In two dimensions that mechanism is absent, and under most forcing and dissipation conditions energy is transferred
from smaller scales to larger scales, largely because of the potential enstrophy, a quadratic invariant defined as the integral of the square of the potential vorticity.  The physical mechanism behind the enstrophy cascade is the stretching of small-scale vorticity gradients by the strain arising from larger-scale vortices \citep{PhysRevLett.91.214501}. This results in energy being trapped at large scales in forced-dissipative two-dimensional flows. Additionally, enstrophy is transferred to the smaller scales from larger scales, and is finally dissipated at large wavenumbers. \cite{danilov2000quasi} and \cite{tabeling2002two} reviewed both theoretical and experimental two-dimensional turbulence studies and provided extensive insights into the applicability of two-dimensional turbulence theory to geophysical flows. We also refer the reader to the energy-enstrophy conservation arguments provided by \cite{vallis2006atmospheric}.

Modeling the ocean and atmosphere inspired the first LES models, but most of the subsequent
development of LES has taken place in the engineering community \citep{sagaut2006large,berselli2006mathematics}. The majority of LES models have been developed for three-dimensional turbulent flows, such as those encountered in engineering applications. These LES models fundamentally rely on the concept of the forward energy cascade and so their extension to geophysical flows is beset with difficulties. The effective viscosity values in oceanic models are much greater than the molecular viscosity of seawater, hence a uniform eddy viscosity coefficient is generally used to parameterize the unresolved, subfilter-scale effects in most oceanic models \citep{mcwilliams2006fundamentals,vallis2006atmospheric}. LES models specifically developed for two-dimensional turbulent flows, such as those in the ocean and atmosphere, are relatively scarce \citep{fox2008can,awad2009large,ozgokmen2009large,
chen2010scale}, at least when compared to the plethora of LES models developed for three-dimensional turbulent
flows. \cite{holm2003modeling} combine the uniform eddy viscosity parametrization with the alpha regularization LES approach to capture the under-resolved flow where the grid length becomes greater than the specified Munk scale of the problem. In that work the structural alpha parameterization was tested on the BVE in an ocean basin with double-gyre wind forcing, which displays a four-gyre mean ocean circulation pattern. It was found that the alpha models provide a promising approach to LES closure modeling of the
barotropic ocean circulation by predicting the correct four-gyre circulation structure for under-resolved flows.

This paper puts forth a new LES closure modeling strategy for two-dimensional turbulent geophysical flows. The new closure modeling approach utilizes {\em approximate deconvolution (AD)}, which is particularly appealing for geophysical flows because of no additional phenomenological approximations to the BVE. The AD approach can achieve high accuracy by employing repeated filtering, which is computationally efficient and easy to implement. The AD method has been used successfully in LES of three-dimensional turbulent engineering flows \citep{stolz1999approximate,stolz2001approximate,stolz2001approximatec,stolz2004approximatei}.
We emphasize, however, that to the best of our knowledge, this is the {\em first time} that the AD
methodology is used in LES of large scale geophysical flows, such as the barotropic ocean circulation flow we consider in this paper. To assess the new AD closure modeling approach, we test it on the same two-dimensional barotropic flow problem as that employed in \cite{nadiga2001dispersive} and in \cite{holm2003modeling}.

The rest of the paper is organized as follows:
The BVE, the mathematical model used in this report, is presented in Section \ref{section_bve}.
Section \ref{section_ad} presents the AD methodology and introduces the new closure model.
The numerical methods used in our simulations are briefly discussed in Section \ref{section_methods}.
The results for the new AD model are presented in Section \ref{section_results}.
Finally, the conclusions are summarized in Section \ref{section_conclusions}.

\section{Barotropic Vorticity Equation}
\label{section_bve}

In this section, we present the BVE, the mathematical
model used in the numerical investigation of the new AD model.
The BVE is one of the most used mathematical models for geostrophic flows with various dissipative
and forcing terms \citep{majda2006non}.
Studies of wind-driven circulation using an idealized double-gyre wind forcing have played an important
role in understanding various aspects of ocean dynamics, including the role of mesoscale eddies and
their effect on mean circulation.
Following \cite{greatbatch2000four}, we briefly describe the BVE.
For more details on the physical mechanism and various formulations utilized, the reader is referred to
\cite{holland1980example,munk1982observing,griffa1989wind,cummins1992inertial,
greatbatch2000four,nadiga2001dispersive,fox2005reevaluating}.

The governing equations for two-dimensional incompressible barotropic flows can be written in
dimensionless form of the potential vorticity formulation in the beta plane as the BVE:
\begin{equation}
\frac{\partial q}{\partial t} + J = D + F .
\label{eq:ge}
\end{equation}
In Eq.~\eqref{eq:ge}, $q$ is the potential vorticity, defined as
\begin{equation}
q = \mbox{Ro} \, \omega + y ,
\label{eq:pv}
\end{equation}
where $\omega$ is the vorticity and $\mbox{Ro}$ is the Rossby number.

The nonlinear convection term in Eq.~\eqref{eq:ge}, called the Jacobian, is defined as
\begin{equation}
J = \frac{\partial \psi}{\partial y}\frac{\partial q}{\partial x} - \frac{\partial \psi}{\partial x}\frac{\partial q}{\partial y} ,
\label{eq:jac}
\end{equation}
where $\psi$ is the streamfunction. The kinematic relationship between the vorticity and the streamfunction yields the following Poisson equation:
\begin{equation}
\frac{\partial^2 \psi}{\partial x^2} + \frac{\partial^2 \psi}{\partial y^2} = -\omega .
\label{eq:ke}
\end{equation}

The viscous dissipation in Eq.~\eqref{eq:ge} has the form
\begin{equation}
D
= \left(\frac{\delta_{M}}{L}\right)^{3}
\left(\frac{\partial^2 \omega}{\partial x^2} + \frac{\partial^2 \omega}{\partial y^2}\right) ,
\label{eq:dis}
\end{equation}
where $\delta_{M}$ is the Munk scale and $L$ is the basin dimension.
The double-gyre wind forcing is given by
\begin{equation}
F = F_{0} \sin(\pi y) ,
\label{eq:forc}
\end{equation}
where $F_{0}=1$ due to the Sverdrup velocity scale used for nondimensionalization
\citep{greatbatch2000four}.
In this nondimensionalization, the velocity scale is
\begin{equation}
V = \frac{\pi \ \tau_0}{\rho H \beta L},
\label{eq:vscale}
\end{equation}
where $\tau_0$ is the maximum amplitude of double-gyre wind stress, $\rho$ is the mean density, $H$ is the mean depth of the basin, and $\beta$ is the gradient of the Coriolis parameter at the basin center ($y=0$). The dimensionless variables are then defined as
\begin{equation}
x = \frac{\tilde{x}}{L}, \ y = \frac{\tilde{y}}{L}, \ t = \frac{\tilde{t}}{L/V}, \ q = \frac{\tilde{q}}{\beta L}, \ \psi= \frac{\tilde{\psi}}{V L},
\label{eq:sverdrup}
\end{equation}
where the tilde denotes the corresponding dimensional variables. In dimensionless form, there are only two physical parameters, the Rhines scale and the Munk scale, which are related to the physical parameters in the following way:
\begin{equation}
\frac{\delta_{I}}{L} = \left(\frac{V}{\beta  L^2}\right)^{1/2}; \quad \quad \frac{\delta_{M}}{L} = \left(\frac{\nu}{\beta  L^3}\right)^{1/3},
\label{eq:relation}
\end{equation}
where $\nu$ is the uniform eddy viscosity coefficient.
The physical parameters in the BVE \eqref{eq:ge}, the Rhines scale
$\delta_{I}$
and the Munk scale $\delta_{M}$,
are related to the Reynolds and Rossby numbers through the following formulas:
\begin{equation}
\frac{\delta_{I}}{L}= (\mbox{Ro})^{1/2} ,
\label{eq:Rhines}
\end{equation}
\begin{equation}
\frac{\delta_{M}}{L}= (\mbox{Re}^{-1}\mbox{Ro})^{1/3} ,
\label{eq:Munk}
\end{equation}
where $\mbox{Re}$ is the Reynolds number based on the basin dimension, $L$.
We note that some authors use a boundary layer Reynolds number, which is written as
\begin{equation}
\mbox{Re}_B=\mbox{Re} \frac{\delta_{I}}{L} = \frac{\delta_{I}^{3}}{\delta_{M}^{3}} ,
\label{eq:Reb}
\end{equation}
where $\mbox{Re}_B \sim O(10)-O(10^3)$ for oceanic flows \citep{fox2005reevaluating}.
Finally, in order to completely specify the mathematical model, boundary and initial conditions need to be prescribed. In many theoretical studies of large scale ocean circulation, slip or no-slip boundary conditions are used.  Following these studies \citep[e.g.,][]{greatbatch2000four,nadiga2001dispersive,holm2003modeling,cummins1992inertial,
ozgokmen1998emergence,munk1950wind,bryan1963numerical}, we use slip boundary conditions
for the velocity, which translate into homogenous Dirichlet boundary conditions for the vorticity:
$\omega |_{\Omega} = 0$.
The impermeability boundary condition is imposed as $\psi |_{\Omega} = 0$.
For the initial condition, we start our computations from a quiescent state ($q=0$) and integrate Eq.~(\ref{eq:ge}) until a statistically steady state
is obtained in which the wind forcing, dissipation, and Jacobian balance each other.

\section{Approximate Deconvolution Model}
\label{section_ad}

The AD approach aims to obtain accurate and stable approximations of the original,
unfiltered flow variables when approximations of the filtered variables are available
\citep{stolz1999approximate,germano2009new}.
The AD methodology was developed in the image processing community, and has been
successfully adapted to the closure problem in turbulence modeling for engineering flows
\citep{stolz2001approximate,stolz2001approximatec,adams2002subgrid,stolz2004approximatei}.
We emphasize that this approach is purely mathematical, with no additional phenomenological arguments being
used.
This is particularly appealing for LES of geophysical flows in which different energy transfer characteristics are displayed than those in three-dimensional turbulent flows, for which successful phenomenological modeling has been done.
Next, we present the mathematical derivation of the new AD model for the BVE \eqref{eq:ge}.

To derive the equations for the filtered flow variables, the BVE \eqref{eq:ge} is first filtered with
a rapidly decaying spatial filter (to be specified later).
Thus, using a bar to denote the filtered quantities, the filtered BVE reads:
\begin{equation}
\frac{\partial \bar{q}}{\partial t} + J(\bar{q},\bar{\psi}) = \bar{D} + \bar{F} + S ,
\label{eq:gef}
\end{equation}
where $S$ is the subfilter-scale term, given by
\begin{equation}
S = -\overline{J(q,\psi)} + J(\bar{q},\bar{\psi}) .
\label{eq:ss1}
\end{equation}
It is precisely at this point in the LES model derivation that the celebrated closure problem
must be addressed.
In order to close the filtered BVE \eqref{eq:gef}, the subfilter-scale term $S$ in
Eq.~\eqref{eq:ss1} needs to be modeled in terms of the filtered flow variables, $\bar{q}$ and $\bar{\psi}$.

This paper proposes a new LES closure modeling approach for two-dimensional turbulent geophysical
flows. The AD approach can achieve high accuracy, is computationally efficient, and is easy to implement.
Although the AD methodology has already been successfully used in LES of three-dimensional turbulent
engineering flows
\citep{stolz1999approximate,stolz2001approximate,stolz2001approximatec,stolz2004approximatei},
this is the {\em first time} that it is used in LES of large scale geophysical flows, such as the barotropic ocean
circulation flow considered in this paper.

The goal in AD is to use repeated filtering in order to obtain approximations of the unfiltered flow
variables when approximations of the filtered flow variables are available.
These approximations of the unfiltered flow variables are then used in the subfilter-scale terms
to close the LES system.
To derive the new AD model, we start by denoting by $G$ the spatial filtering operator:
$Gq=\bar{q}$.
Since $G=I-(I-G)$, an inverse to $G$ can be written formally as the non-convergent Neumann series:
\begin{equation}
G^{-1} \sim \sum_{i=1}^{\infty}(I-G)^{i-1} .
\label{eq:2}
\end{equation}
Truncating the series gives the van Cittert approximate deconvolution operator, $Q_N$.
We truncate the series at $N$ and obtain $Q_N$ as an approximation of $G^{-1}$:
\begin{equation}
Q_N = \sum_{i=1}^{N}(I-G)^{i-1} ,
\label{eq:3}
\end{equation}
where $I$ is the identity operator.
The approximations $Q_N$ are not convergent as $N$ goes to infinity, but rather are asymptotic
as the filter radius, $\delta$, approaches zero \citep{berselli2006mathematics}.
An approximate deconvolution of $\bar{q}$ can now be obtained as follows:
\begin{equation}
q \approx q^*= Q_N \bar{q} .
\label{eq:4}
\end{equation}
For higher values of $N$ we get increasingly more accurate approximations of $q$:
\begin{eqnarray}
Q_1 &=& I \\
Q_2 &=& 2I -G \\
Q_3 &=& 3I-3G + G^2 \\
Q_4 &=& 4I-6G + 4G^2 -G^3 \\
Q_5 &=& 5I-10G + 10G^2 - 5G^3 + G^4 \\
\vdots \nonumber
\label{eq:5}
\end{eqnarray}
Following the same approach as that used in \cite{dunca2006stolz}, one can prove that these
models are highly accurate ($O(\delta^{2N+2})$ modeling consistency error) and stable.
We choose $N=5$ and find an AD approximation of the variable $q$ as
\begin{equation}
q \approx q^*=5 \bar{q} - 10 \bar{\bar{q}} + 10 \bar{\bar{\bar{q}}} -5 \bar{\bar{\bar{\bar{q}}}} + \bar{\bar{\bar{\bar{\bar{q}}}}}
\label{eq:6}
\end{equation}
and, similarly, an AD approximation of the variable $\psi$:
\begin{equation}
\psi \approx \psi^*=5 \bar{\psi} - 10 \bar{\bar{\psi}} + 10 \bar{\bar{\bar{\psi}}} -5 \bar{\bar{\bar{\bar{\psi}}}} + \bar{\bar{\bar{\bar{\bar{\psi}}}}}.
\label{eq:7}
\end{equation}
Using \eqref{eq:6} and \eqref{eq:7}, we can now approximate the nonlinear Jacobian:
\begin{equation}
\overline{J(q,\psi)} \approx \overline{J(q^*,\psi^*)} .
\label{eq:8}
\end{equation}
Finally, using the approximation Eq.~\eqref{eq:8} in Eq.~\eqref{eq:ss1}, closes the filtered BVE \eqref{eq:gef}
and yields the new {\em AD model}:
\begin{equation}
\frac{\partial \bar{q}}{\partial t} + J(\bar{q},\bar{\psi}) = \bar{D} + \bar{F} + S^* ,
\label{eq:gef_ad}
\end{equation}
where $S^*$ is the subfilter-scale term, given by
\begin{equation}
S^* = -\overline{J(q^*,\psi^*)} + J(\bar{q},\bar{\psi}) .
\label{eq:1_ad}
\end{equation}

To completely specify the new AD model \eqref{eq:gef_ad}-\eqref{eq:1_ad}, we need to choose
a computationally efficient filtering operator.
Following \cite{stolz1999approximate}, we use the following second-order accurate filtering
operator:
\begin{equation}
\alpha \bar{f}_{i-1}
+ \bar{f}_{i}
+ \alpha \bar{f}_{i+1}
= \left(\frac{1}{2} + \alpha\right)\left(f_{i}+\frac{ f_{i-1} + f_{i+1}}{2}\right),
\label{eq:9}
\end{equation}
where the subscript $i$ is the spatial index in the $x$-direction. This results in a tridiagonal system  of equations for each fixed value of $y$.
A generalization of Eq.~\eqref{eq:9}, the spectral-type high-order compact filter introduced in
\cite{visbal2002use}, is given by
\begin{equation}
\alpha \bar{f}_{i-1} + \bar{f}_{i} + \alpha \bar{f}_{i+1} = \sum_{n=0}^{M}\frac{a_n}{2}(f_{i+n}+f_{i-n}) ,
\label{eq:filter}
\end{equation}
where $f$ is the computed value, and $\bar{f}$ is the corresponding once filtered value.
This filter provides $2M$th-order accurate filtering on a $2M+1$ point stencil \citep{visbal2002use}.
The free parameter, $\alpha$, which is in the range $0 \leq |\alpha| \leq 0.5$, determines the filtering
properties, with high values of $\alpha$ yielding less dissipative results.
The coefficients for different order filters are given in Table~\ref{tab:filter}.
In our numerical tests, we found that for moderately fine meshes, the range of $0.25 \leq \alpha \leq 0.5$
is appropriate, but for coarser meshes, lower values (e.g., $\alpha=0.1$) should be used to
eliminate spurious oscillations. Most of our numerical simulations use the free parameter $\alpha=0.25$ \citep{stolz1999approximate}.
Since we used a second-order accurate space discretization in this study, the filtered simulations are
conducted either by using the second-order accurate filter given in Eq.~\eqref{eq:9} or the fourth-order accurate
compact filter given in Eq.~\eqref{eq:filter} with different parameter values for $\alpha$.
\begin{table}[h!]
\begin{tabular}{lcccc}
\hline \\[-0.3cm]
  & 2nd-order & 4th-order & 6th-order & 8th-order \\[0.1cm]
\hline \\[-0.3cm]
$a_0$ & $\frac{1}{2}+ \alpha $ & $\frac{5}{8} + \frac{3\alpha}{4}$ & $\frac{11}{16} + \frac{5\alpha}{8}$  & $\frac{93}{128} + \frac{70\alpha}{128}$ \\[0.2cm]
$a_1$ & $ \frac{1}{2}+\alpha $ & $\frac{1}{2} + \alpha $ & $\frac{15}{32} + \frac{17\alpha}{16}$ & $\frac{7}{16} + \frac{18\alpha}{16}$\\[0.2cm]
$a_2 $& 0& $-\frac{1}{8} + \frac{\alpha}{4}$ & $-\frac{3}{16} + \frac{3\alpha}{8}$ & $-\frac{7}{32} + \frac{14\alpha}{32}$ \\[0.2cm]
$a_3$ & 0& 0 & $\frac{1}{32} - \frac{\alpha}{16}$ & $\frac{1}{16} - \frac{\alpha}{8}$ \\[0.2cm]
$a_4$ & 0& 0 & 0 & $-\frac{1}{128} + \frac{\alpha}{64}$ \\[0.2cm]
\hline
\end{tabular}\vspace*{0.2cm}
\centering
\caption{Coefficients of the compact filters of various orders given by Eq.~\eqref{eq:filter}.}
\label{tab:filter}
\end{table}

\begin{figure}
\centering
\includegraphics[width=0.7\textwidth]{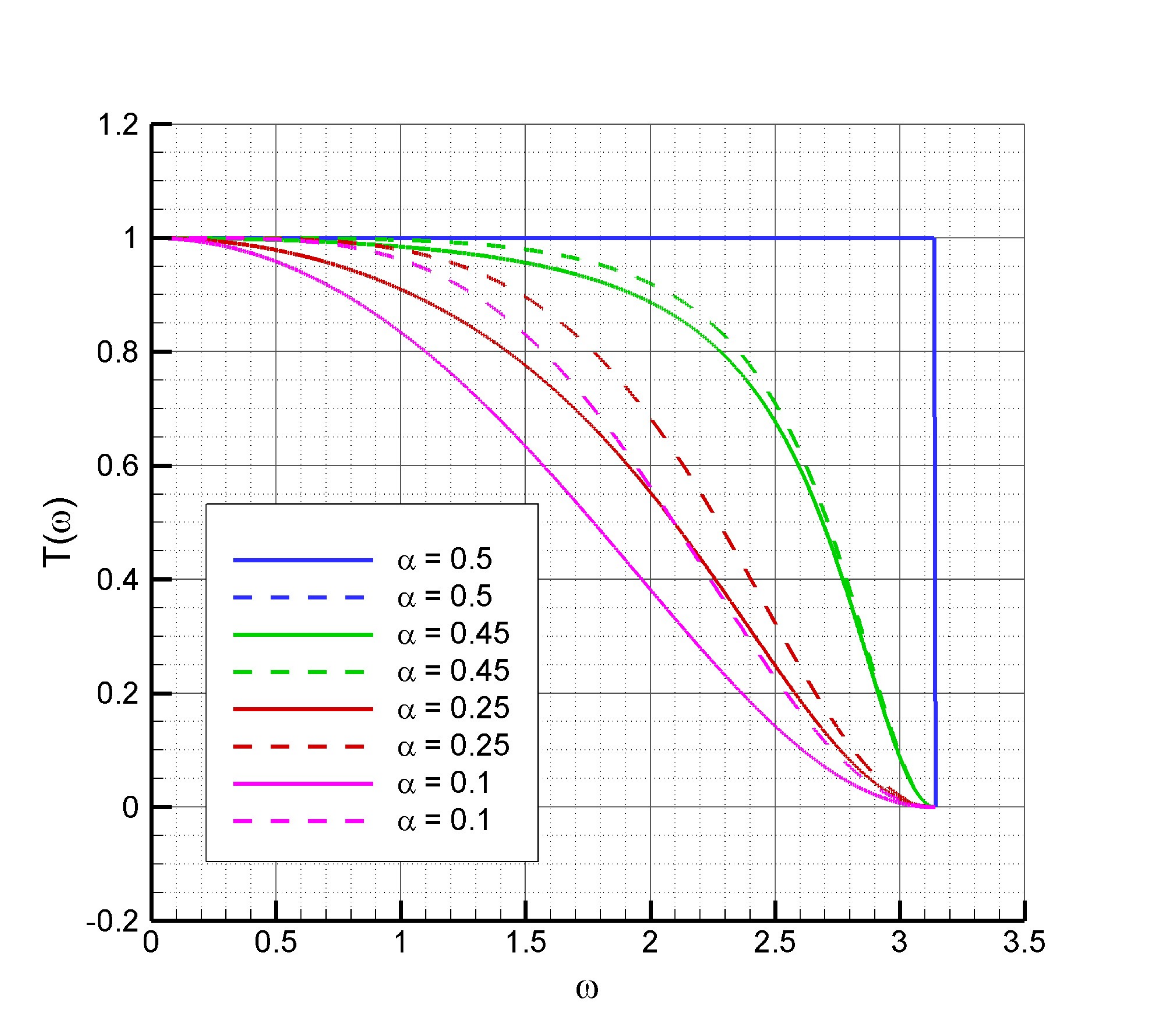}
\caption{
Transfer functions of the second-order (solid lines) and fourth-order (dashed lines) tridiagonal filters.
}
\label{fig:trfun}
\end{figure}

The existence and uniqueness of strong solutions of the BVE model using the AD closure was investigated by \cite{stanculescu2008existence}. According to this work, the transfer function of the filter used in the AD closure should be positive definite. The transfer function of the tridiagonal filter given in Eq.~\eqref{eq:9} can be written as
\begin{equation}
T(\omega) = (\frac{1}{2}+\alpha)\frac{1+\cos(\omega)}{1+2\alpha\cos(\omega)}.
\label{eq:tfun-1}
\end{equation}
The transfer function of the generalized filter given in \eqref{eq:filter} can be written as
\begin{equation}
T(\omega)=\frac{\sum_{n=0}^{M}a_n \cos(n \ \omega)}{1+2\alpha\cos(\omega)}.
\label{eq:tfun-2}
\end{equation}
The transfer functions corresponding to the second- and fourth-order filters are plotted in Fig.~(\ref{fig:trfun}) and are positive definite in the interval of $0 \leq |\alpha| \leq 0.5$, where $\alpha$ is a  free parameter of the filter. By substituting these transfer functions into the deconvolution operator, it can be easily seen that the transfer functions of the deconvolution operators also become positive definite. With these two conditions verified (the positive definiteness of the transfer functions of the filters and deconvolution operators), the theoretical results in \cite{stanculescu2008existence} provide solid mathematical support for our numerical investigations.

\section{Numerical Methods}
\label{section_methods}

In this section, we provide a brief description of the numerical methods employed in the investigation
of the new AD model.
For the time discretization, we employ an optimal third-order total variation diminishing Runge-Kutta (TVDRK3) scheme \citep{gottlieb1998total}. For clarity of notation, we rewrite the new AD model in the following form
\begin{equation}
\frac{d \bar{q}}{dt} = R ,
\label{eq:rk}
\end{equation}
where $R$ denotes the discrete spatial derivative operator, including the nonlinear
convective term, the linear diffusive term, the forcing term, and the subfilter-scale term. The TVDRK3 scheme then becomes
\begin{eqnarray}
\bar{q}^{(1)} &=& \bar{q}^{n} + \Delta t R^{(n)} \nonumber \\
\bar{q}^{(2)} &=& \frac{3}{4}  \bar{q}^{n} + \frac{1}{4} \bar{q}^{(1)} + \frac{1}{4}\Delta t R^{(1)} \\
\bar{q}^{n+1} &=& \frac{1}{3}  \bar{q}^{n} + \frac{2}{3} \bar{q}^{(2)} + \frac{2}{3}\Delta t R^{(2)}, \nonumber
\label{eq:TVDRK}
\end{eqnarray}
where $\Delta t$ is the adaptive time step size which can be computed at the end of each time step by:
\begin{equation}
\Delta t = c \frac{\mbox{min}(\Delta x,\Delta y)}{\mbox{max}\left(|\frac{\partial \bar{\psi}}{\partial x}|,|\frac{\partial \bar{\psi}}{\partial y}|\right)} ,
\label{eq:dt}
\end{equation}
where $c$ is known as Courant-–Friedrichs–-Lewy (CFL) number. We set $c=1.0$ for all the simulations reported here.

For the spatial discretization of linear operators, such as the viscous dissipation term and the Poisson equation, we use a second-order finite difference method.
For the nonlinear convection term, we utilize Arakawa's conservative scheme \citep{arakawa1966computational}.
Finally, a direct Poisson solver based on the fast sine transform is used to solve the
kinematic relationship between the vorticity and the streamfunction in Eq.~\eqref{eq:ke}. Specifically, we use a fast sine transform in one direction and a tridiagonal system
solver in the other direction \citep{moin2002fundamentals}.

For completeness, we summarize the solution algorithm for one time step.
We assume that the numerical approximation for time level $n$ is known, and
we seek the numerical approximation for time level $n+1$, after the time step $\Delta t$.
We write the algorithm for the new AD model \eqref{eq:gef_ad}-\eqref{eq:1_ad}, but we
emphasize that it is the same algorithm, with some obvious modifications, that is used for the high resolution well resolved simulation, which is referred to here as DNS. We emphasize that the term DNS in this study is not meant to indicate that a fully detailed solution is being computed, but instead refers to resolving the simulation down to the Munk scale via the specified lateral eddy viscosity parameterization.
The solution algorithm has the following form
\begin{enumerate}
  \item Compute $\bar{\psi}^{n}$ using the Poisson solver for the following equation:
        \begin{equation}
        \frac{\partial^2 \bar{\psi}^{n}}{\partial x^2} + \frac{\partial^2 \bar{\psi}^{n}}{\partial y^2}
        = - \frac{1}{Ro}\left(\bar{q}^{n}-y\right).
        \label{eq:14}
        \end{equation}
  \item Compute $q^{n*}$ and $\psi^{n*}$ using $\bar{q}^{n}$ and
  	   $\bar{\psi}^{n}$, the filtering procedure in Eq.~(\ref{eq:filter}), and the
	   AD method in Eq.~\eqref{eq:6} and Eq.~\eqref{eq:7}.
  \smallskip
  \item Compute the nonlinear Jacobian terms $\overline{J(q^{n*},\psi^{n*})}$ and
  	   $J(\bar{q}^n,\bar{\psi}^n)$ using the Arakawa scheme.
  \smallskip
  \item Compute the subfilter-scale term $S^*$ using
  	   $\overline{J(q^{n*},\psi^{n*})}, J(\bar{q}^n,\bar{\psi}^n)$,
  	   and Eq.~\eqref{eq:1_ad}.
  \smallskip
   \item Compute the viscous dissipation term $\bar{D}$ using a central difference
   	    scheme and calculate the right-hand-side term $R$ in Eq.~\eqref{eq:rk}.
  \smallskip
  \item Apply the Runge-Kutta algorithm by repeating steps 1-5 for each stage of Runge-Kutta scheme.
  \smallskip
  \item Adjust the time step according to the CFL number given in Eq.~\eqref{eq:dt}.
  \smallskip
 \item Go to step 1 to begin the next time step.
\end{enumerate}

\section{Numerical Results}
\label{section_results}

The main goal of this section is to test the new AD model \eqref{eq:gef_ad}-\eqref{eq:1_ad}
in the numerical simulation of the BVE \eqref{eq:ge}.
Before testing the new model, however, we first validate the code in
Section \ref{subsection_validation}.
The new AD model is then tested in the numerical simulation of a four-gyre barotropic model
with double-gyre wind forcing in Section \ref{subsection_four_gyre}.

\subsection{Method of manufactured solutions: Code validation}
\label{subsection_validation}

To validate the code, we consider the following exact solution, which corresponds to
a steady state Taylor-Green vortex flow:
\begin{eqnarray}
q(x,y,t) &=& \left(\frac{\delta_I}{L}\right)^{2}2\,\pi^{2}\sin(\pi x)\sin(\pi y) + y ,
\label{eq:15a} \\
\omega(x,y,t) &=&2\,\pi^{2}\sin(\pi x)\sin(\pi y) ,
\label{eq:15b} \\
\psi(x,y,t) &=& \sin(\pi x)\sin(\pi y) .
\label{eq:15c}
\end{eqnarray}
The forcing function is calculated by plugging Eqs.~\eqref{eq:15a}-\eqref{eq:15c}
into the BVE \eqref{eq:ge}:
\begin{equation}
F(x,y,t)
= -\pi \, \cos(\pi x)\sin(\pi y) + \left(\frac{\delta_M}{L}\right)^{3} \, 4 \, \pi^{4}\sin(\pi x)\sin(\pi y) .
\label{eq:f1}
\end{equation}
We use  $q(x,y,0) = 0$ as the initial condition.
The spatial resolution is $64 \times 128$ and the time step $\Delta t$ is computed adaptively as given in Eq.~\eqref{eq:dt}.
We utilize the numerical algorithm in Section \ref{section_methods} to integrate
the BVE until the steady state solution is obtained.
Our numerical results closely match the exact solution.
To investigate the individual effects of the nonlinear Jacobian, dissipation, forcing and subfilter-scale terms, we compute the time series of following integral quantities:
\begin{equation}
Q_J(t)
= \frac{1}{2} \iint \left( J (\bar{q},\bar{\psi}) \right)^2 dx \, dy ,
\label{eq:41}
\end{equation}
\begin{equation}
Q_D(t)
= \frac{1}{2} \iint \bar{D}^2 dx \, dy ,
\label{eq:42}
\end{equation}
\begin{equation}
Q_F(t)
= \frac{1}{2} \iint \bar{F}^2 dx \, dy ,
\label{eq:43}
\end{equation}
\begin{equation}
Q_S(t)
= \frac{1}{2} \iint \left(S^*\right)^2 dx \, dy.
\label{eq:44}
\end{equation}
We also plot with respect to time the total energy:
\begin{equation}
E(t)
= \frac{1}{2} \iint \left(\frac{\partial \psi}{\partial y}\right)^2
+ \left(\frac{\partial \psi}{\partial x} \right)^2 dx \, dy ,
\label{eq:16}
\end{equation}
Since the exact steady state solution is given in Eqs.~\eqref{eq:15a}-\eqref{eq:15c},
we  can calculate the exact values of the above integral quantities.

\begin{figure}
\centering
\mbox{
\subfigure{\includegraphics[width=0.5\textwidth]{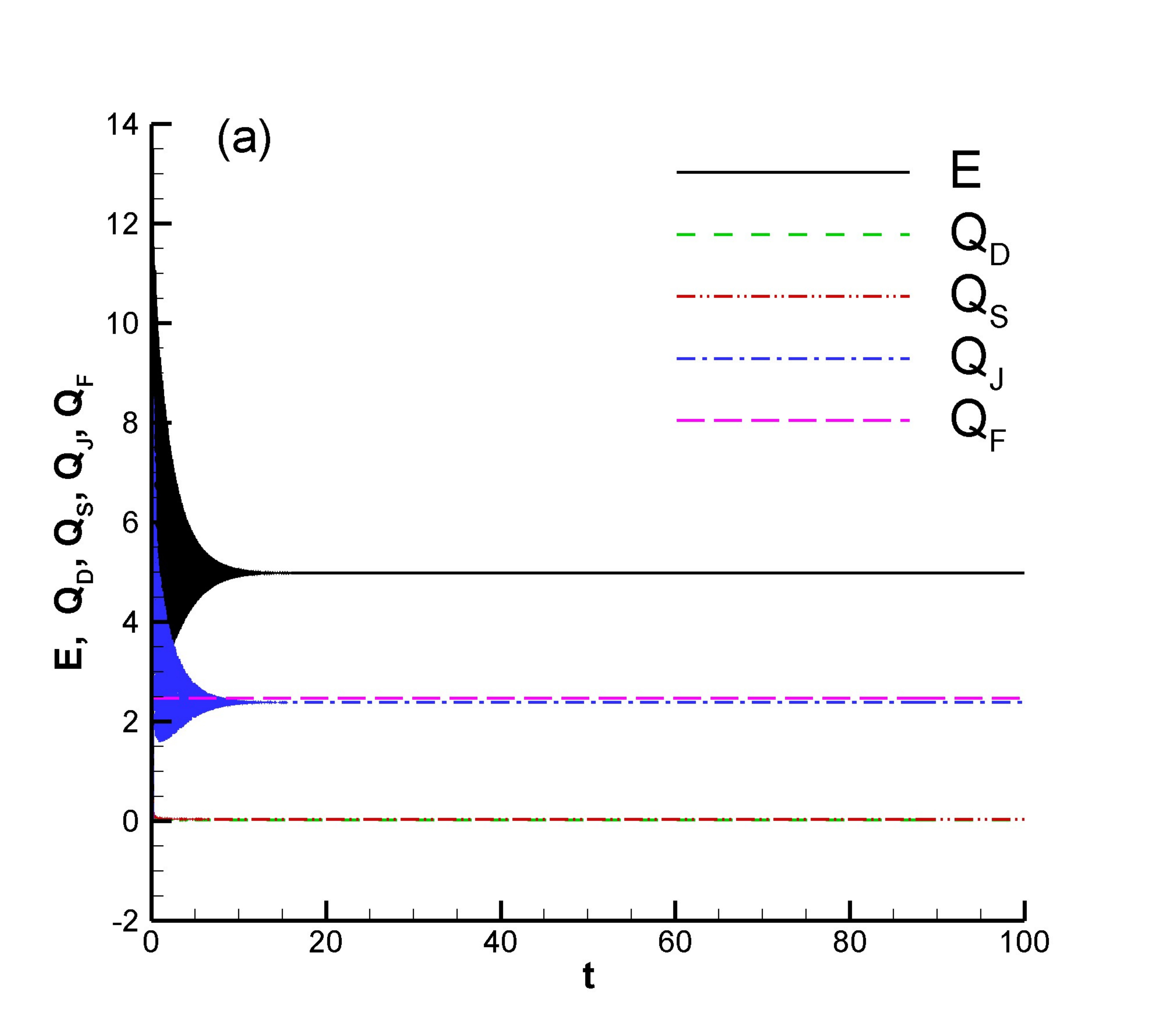}}
\subfigure{\includegraphics[width=0.5\textwidth]{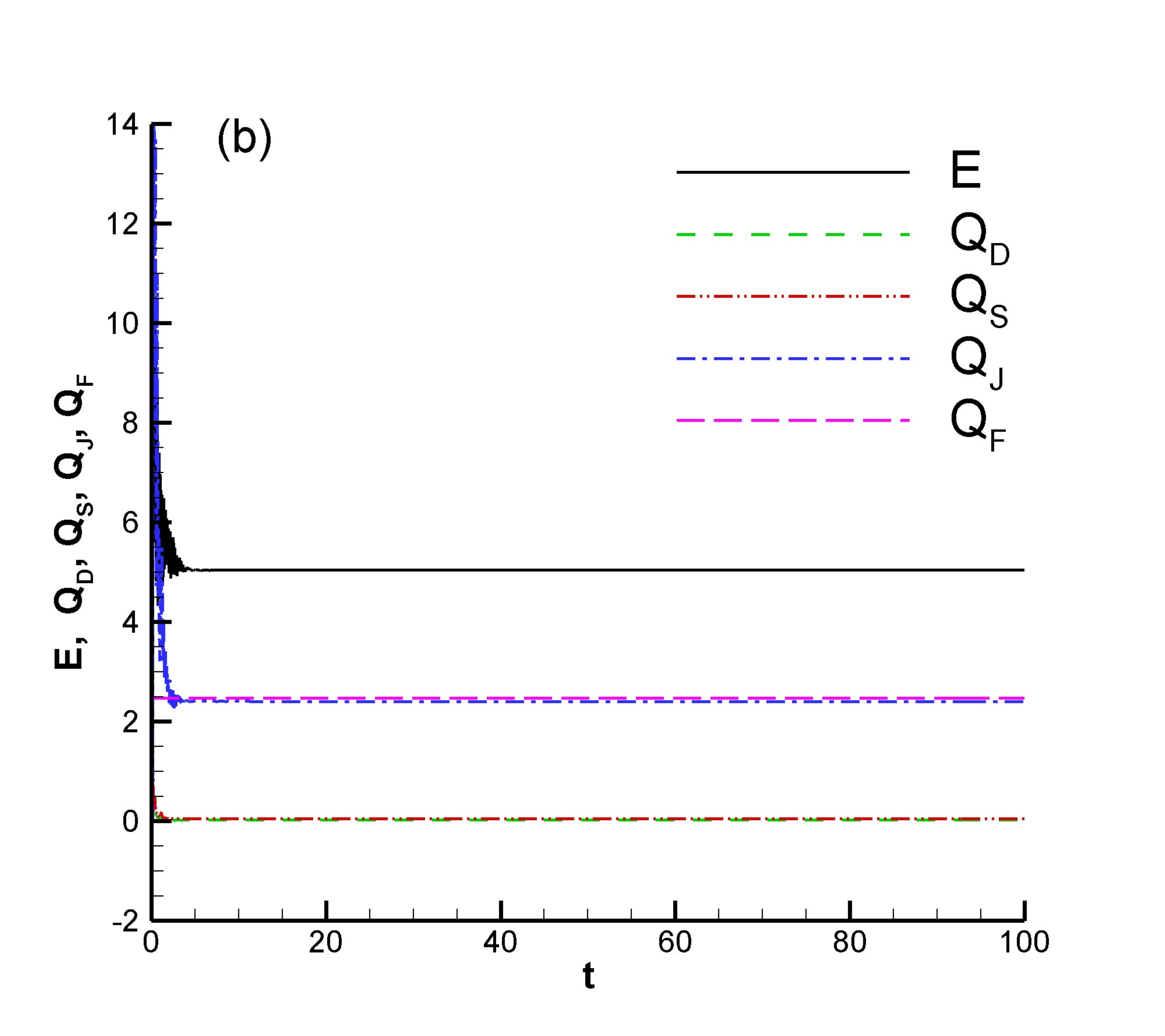}} }
\caption{
Code validation test case.Time history of the total energy, individual dissipation, Jacobian, subfilter-scale and forcing terms:
(a) $\displaystyle \delta_{I}/L = 0.04$  and $\displaystyle \delta_{M}/L = 0.02$;
and (b) $\displaystyle \delta_{I}/L = 0.06$  and $\displaystyle \delta_{M}/L = 0.02$.
Note that the numerical approximation converges to the correct value after
a short transient interval.
}
\label{fig:t}
\end{figure}

In Fig.~\eqref{fig:t} we plot the time evolution of the total energy, and contributions of the individual dissipation, subfilter-scale, nonlinear Jacobian, and forcing terms. We plot these time series for two parameter sets:
$\displaystyle \delta_{I}/L = 0.04$
and
$\displaystyle\delta_{M}/L = 0.02$ in Fig.~(\ref{fig:t}a),
and
$\displaystyle \delta_{I}/L = 0.06$
and
$\displaystyle \delta_{M}/L = 0.02$ in Fig.~(\ref{fig:t}b).
For both parameter sets, the time evolution of the above integral quantities follows the same pattern:
after a short transient interval, they converge to the correct exact values.
Thus, we conclude that the numerical algorithm produces accurate results for these
validation test cases.

\subsection{Four-gyre problem with double-gyre forcing}
\label{subsection_four_gyre}

The main goal of this section is to test the new AD model Eqs.~\eqref{eq:gef_ad}-\eqref{eq:1_ad}
in the numerical simulation of the wind-driven circulation in a shallow ocean basin, a standard
prototype of more realistic ocean dynamics.
The model employs the BVE driven by a symmetric double-gyre wind forcing, which yields a
four-gyre circulation in the time mean.
This test problem has been used in numerous studies \citep[e.g.,][]{cummins1992inertial,
greatbatch2000four,nadiga2001dispersive,holm2003modeling,fox2005reevaluating}.
This problem represents an ideal test for the new AD model.
Indeed, as showed in \cite{greatbatch2000four}, although a double gyre wind forcing is used,
the long time average yields a four gyre pattern, which is challenging to capture on coarse
spatial resolutions.
Thus, we will investigate numerically whether the new AD model can reproduce the four
gyre time average on a coarse mesh.

The mathematical model used in the four gyre problem is the BVE \eqref{eq:ge}.
The double-gyre forcing is given in Eq.~(\ref{eq:forc}).
Following \cite{holm2003modeling}, we utilize two different parameter sets,
corresponding to two physical oceanic settings: Experiment (i) with a Rhines scale of $\delta_{I}/L = 0.04$  and a Munk scale of $\delta_{M}/L = 0.02$, which corresponds
to a Reynolds number of $Re=200$ (or a boundary layer based Reynolds number of $Re_B=8$) and a Rossby
number of $Ro=0.0016$; and
Experiment (ii) with a Rhines scale of $\delta_{I}/L = 0.06$ and a Munk scale of $\delta_{M}/L = 0.02$, which corresponds
to a Reynolds number of $Re=450$ (or a boundary layer based Reynolds number of $Re_B=27$) and a Rossby
number of $Ro=0.0036$.
Since we set the Munk scale to $\delta_{M}/L = 0.02$ in our study, the uniform eddy viscosity coefficient embedded in the model can be calculated from Eq.~\eqref{eq:relation}. For example, if we set the mid-latitude ocean basin length to $L=2000$ km and the gradient of the Coriolis parameter to $\beta=1.75\times10^{-11} \ \mbox{m}^{-1}\mbox{s}^{-1}$, then our model uses $\nu=1120 \ \mbox{m}^{2}\mbox{s}^{-1}$ as its eddy viscosity parametrization.
All numerical experiments conducted here are solved for a maximum dimensionless time of $T_{max}=100$. This value corresponds to the dimensional times of $56.6$ and $25.15$ years for Experiment (i) and Experiment (ii), respectively, which are long enough to capture statistically steady states.
All computations were carried out using the gfortran compiler on a Linux cluster system (3.4 GHz/node).

To assess the new AD model, we employ the standard LES methodology:
we first run a DNS computation on a fine mesh. We then run an under-resolved numerical simulation on a much coarser mesh (denoted in
what follows as BVE$_{coarse}$), which does not employ any subfilter-scale model. We expect
BVE$_{coarse}$ to produce inaccurate results.
Finally, we employ the new AD model on the same coarse mesh utilized in BVE$_{coarse}$.
The AD model should yield results that are significantly better than those obtained with BVE$_{coarse}$
and are close to the DNS results, at a fraction of the computational cost.

\begin{figure}
\centering
\mbox{
\subfigure{\includegraphics[width=0.25\textwidth]{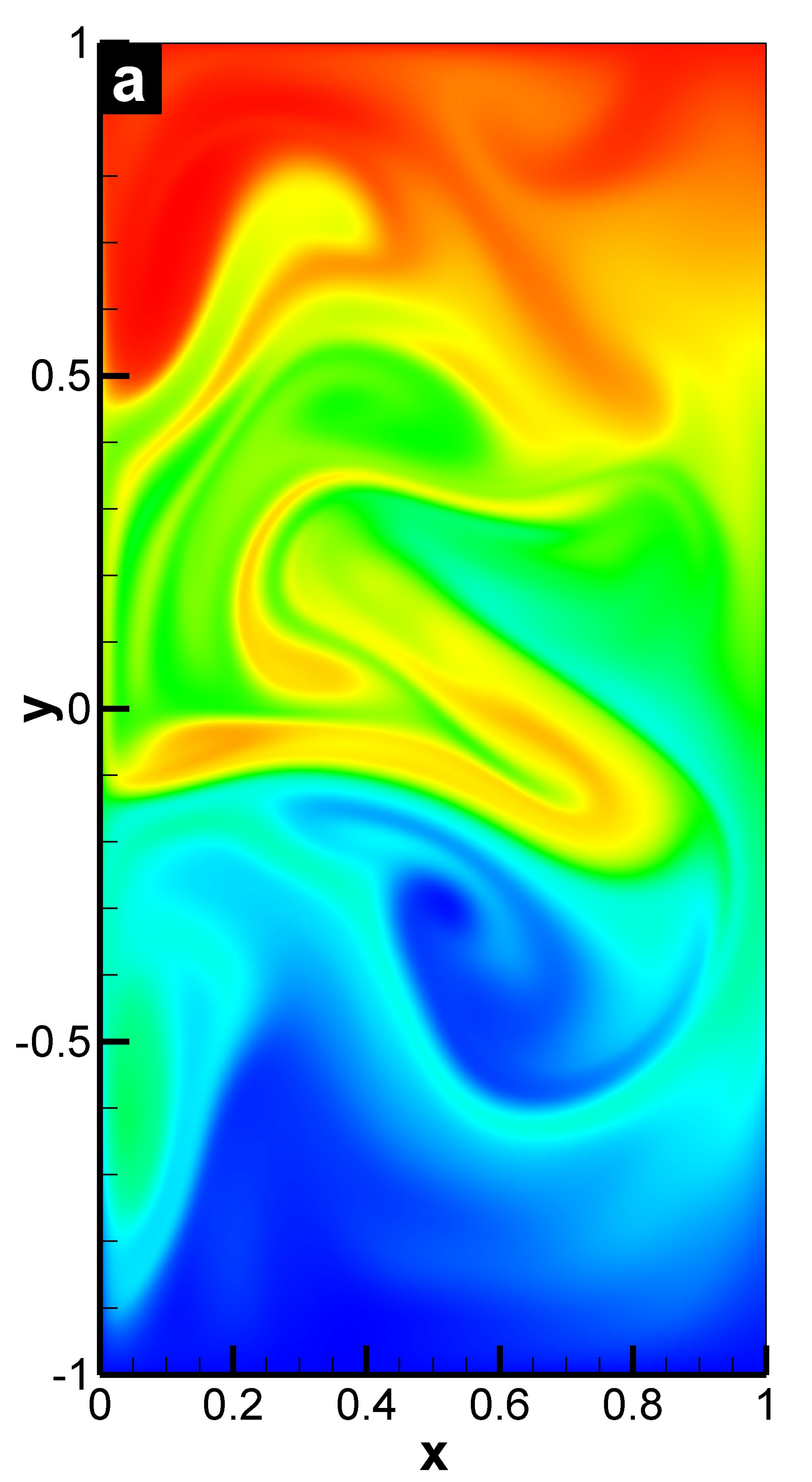}}
\subfigure{\includegraphics[width=0.25\textwidth]{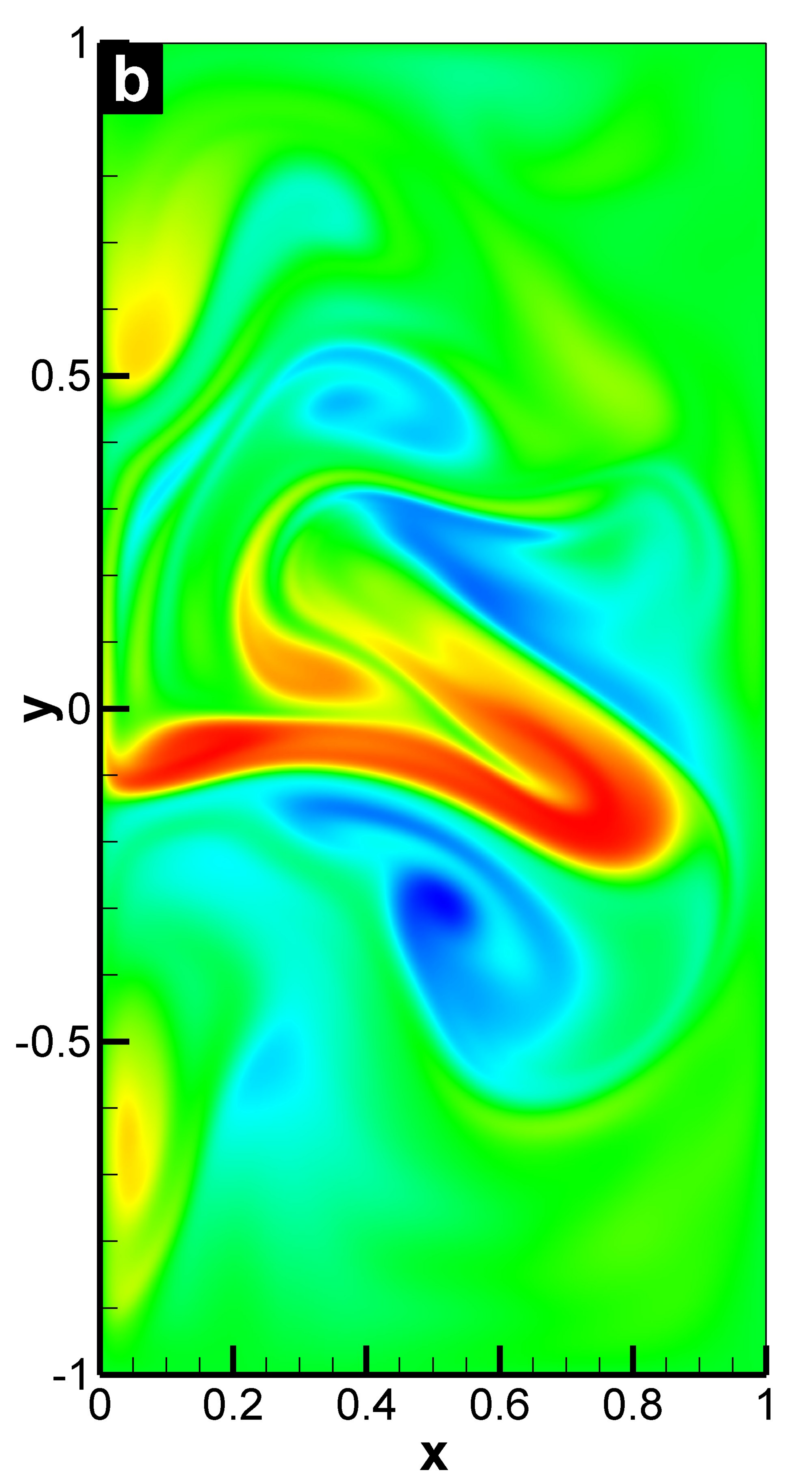}}
\subfigure{\includegraphics[width=0.25\textwidth]{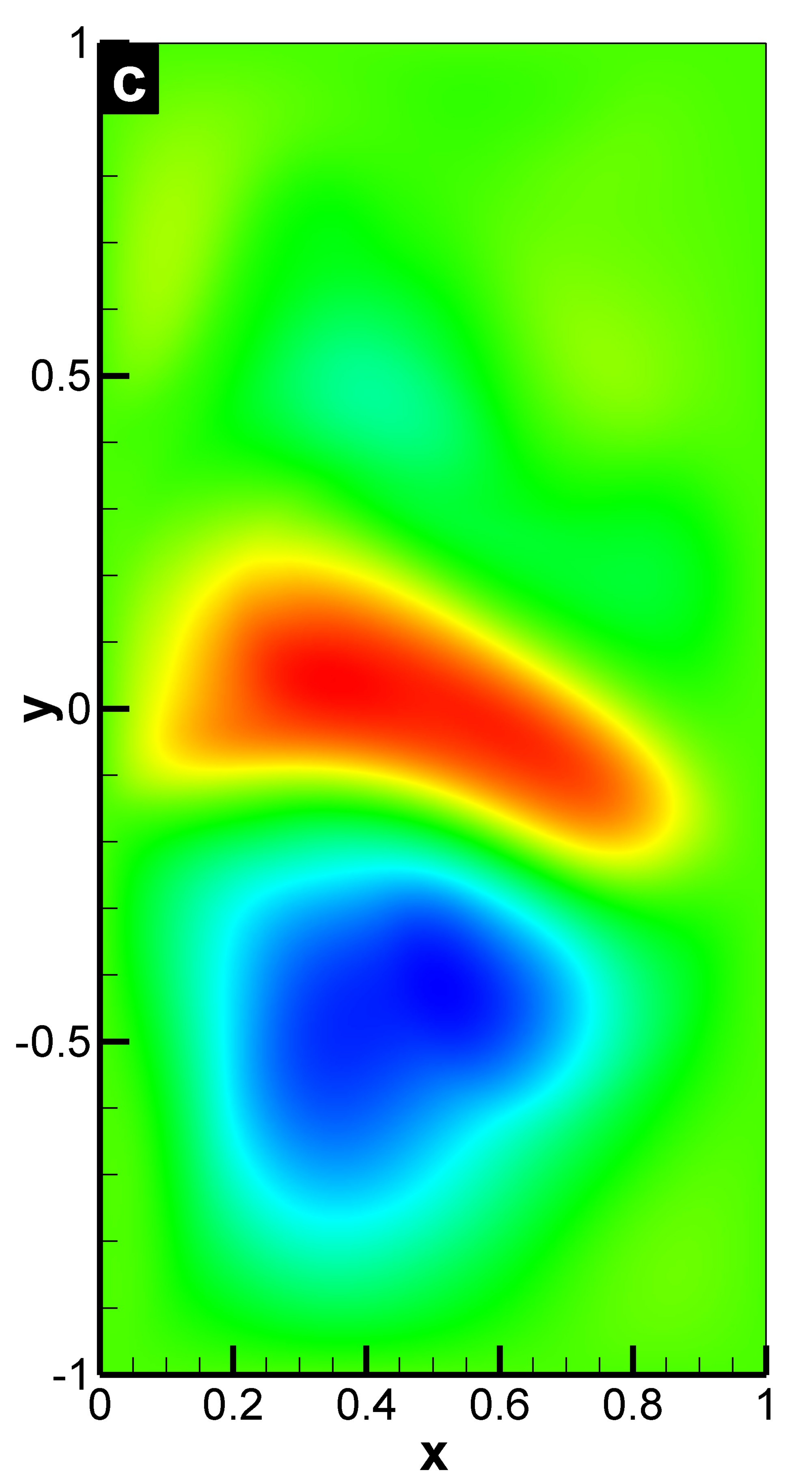}} }
\caption{
Experiment (i): \ DNS results for
$Re=200, Ro=0.0016$ and a spatial resolution of $512\times 256$.
Instantaneous field data for:
(a) potential vorticity; (b) vorticity; and (c) streamfunction.
Note that two gyres appear in the streamfunction contour plot.
}
\label{fig:ins-a}
\end{figure}

\begin{figure}
\centering
\mbox{
\subfigure{\includegraphics[width=0.25\textwidth]{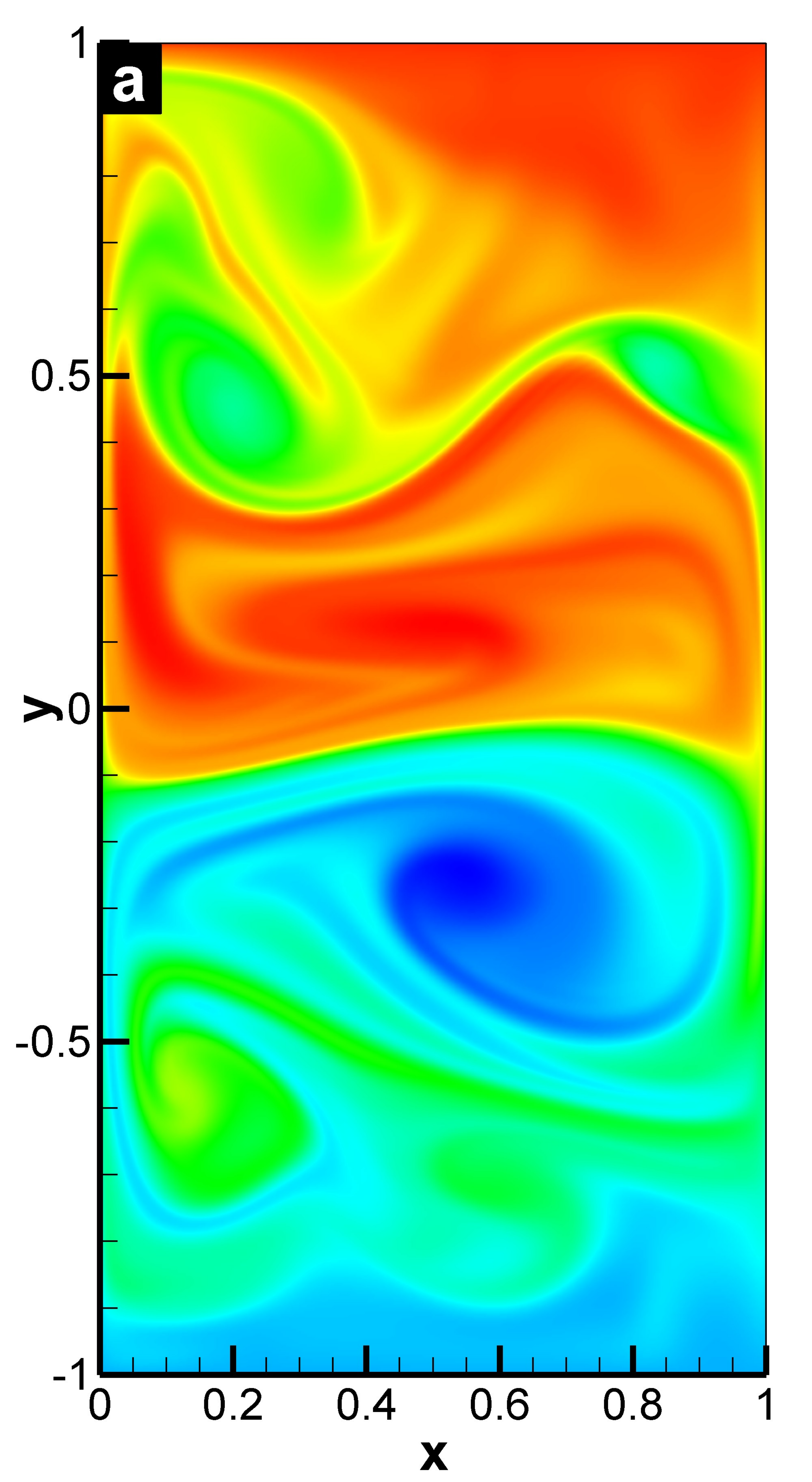}}
\subfigure{\includegraphics[width=0.25\textwidth]{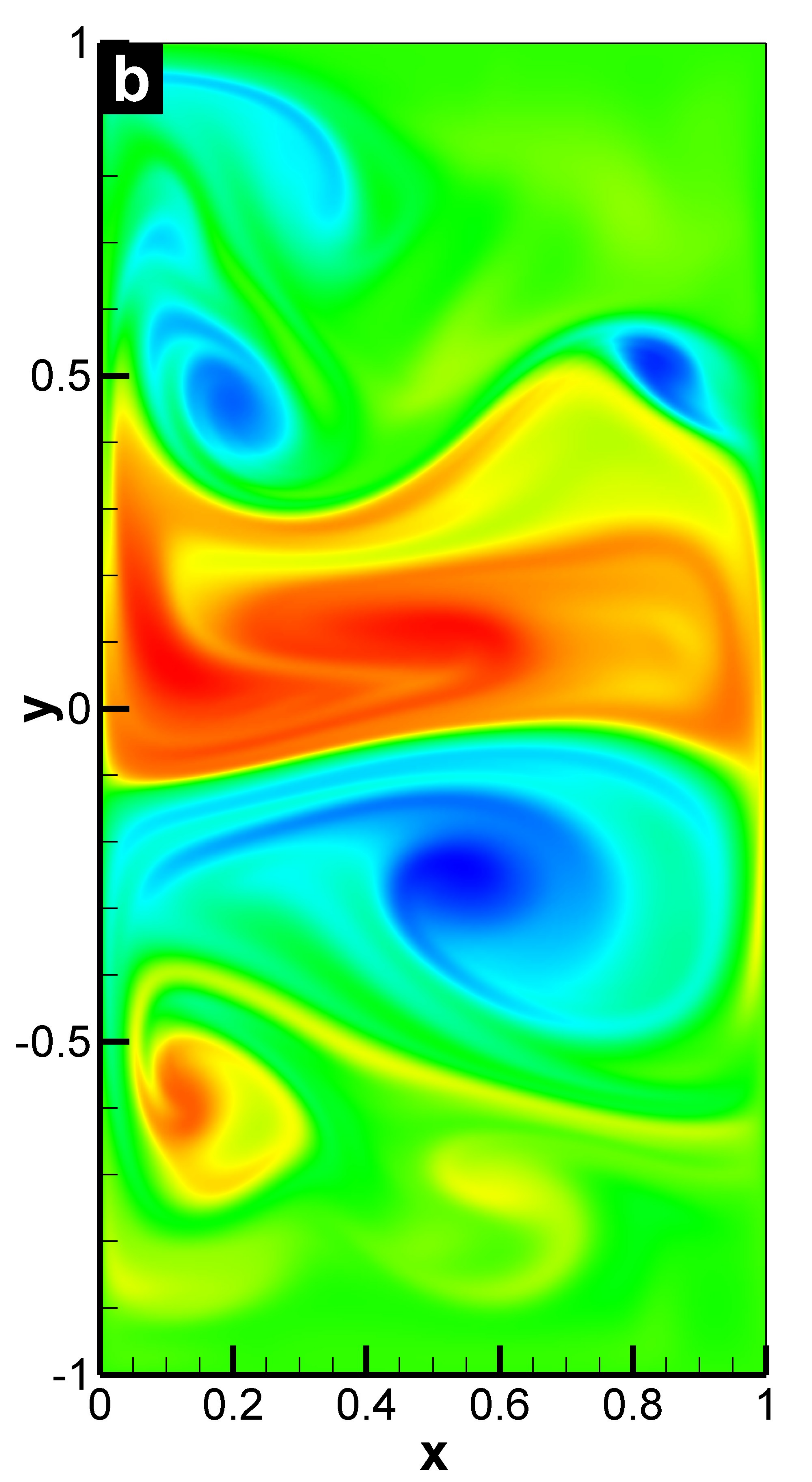}}
\subfigure{\includegraphics[width=0.25\textwidth]{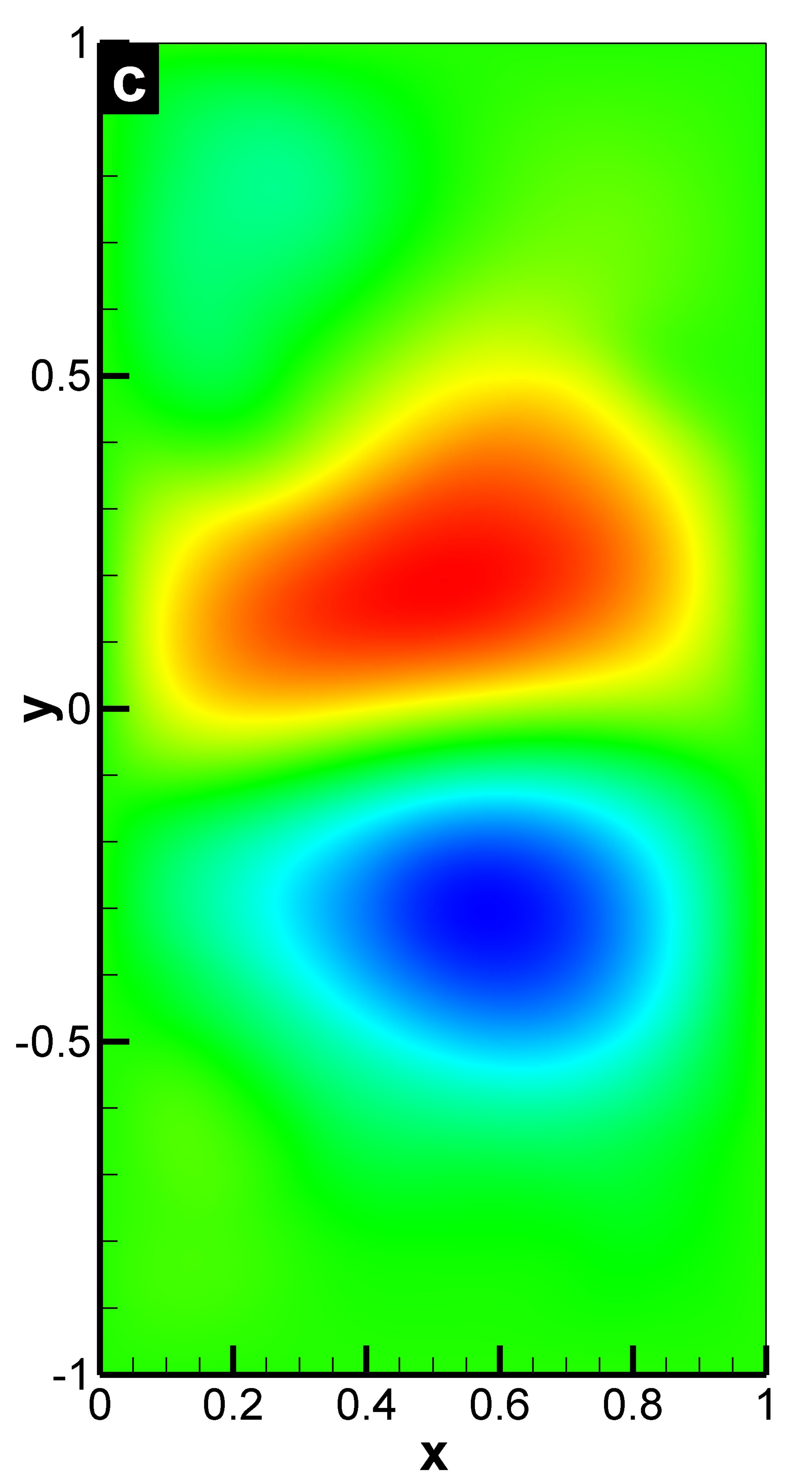}} }
\caption{
Experiment (ii): \ DNS results for
$Re=450, Ro=0.0036$ and a spatial resolution of $512\times 256$.
Instantaneous field data for:
(a) potential vorticity; (b) vorticity; and (c) streamfunction.
Note that two gyres appear in the streamfunction contour plot.
}
\label{fig:ins-b}
\end{figure}

\begin{figure}
\centering
\mbox{
\subfigure{\includegraphics[width=0.25\textwidth]{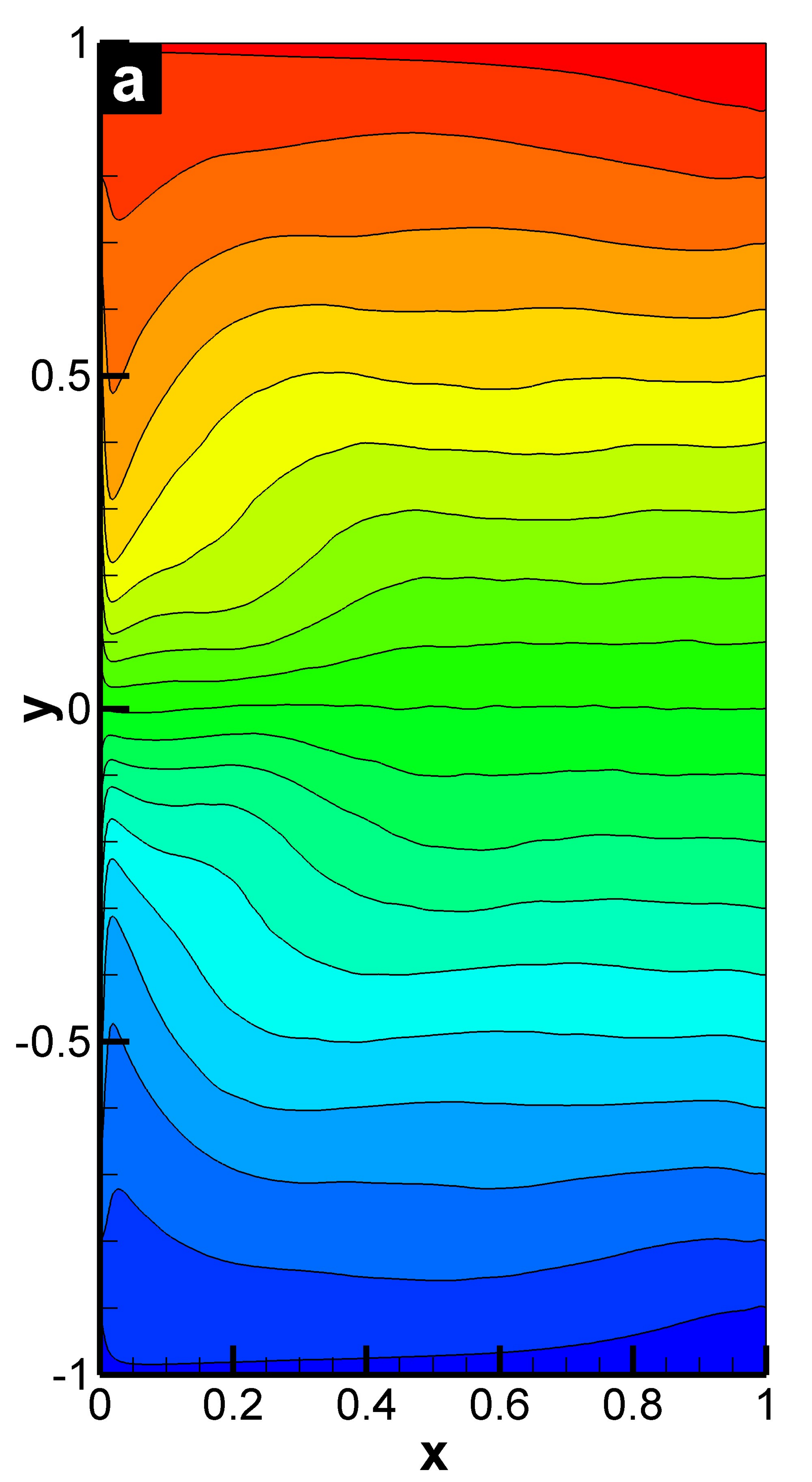}}
\subfigure{\includegraphics[width=0.25\textwidth]{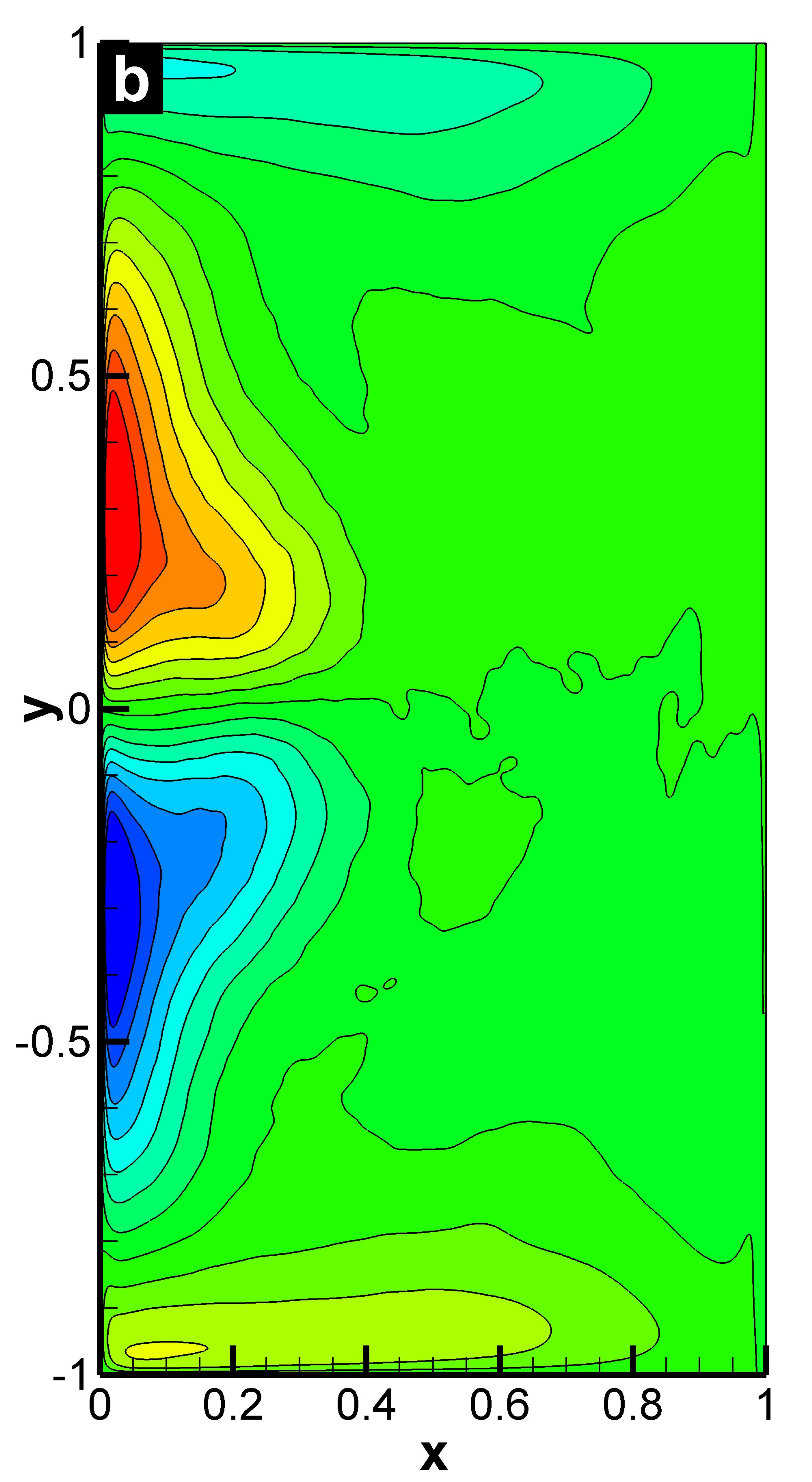}}
\subfigure{\includegraphics[width=0.25\textwidth]{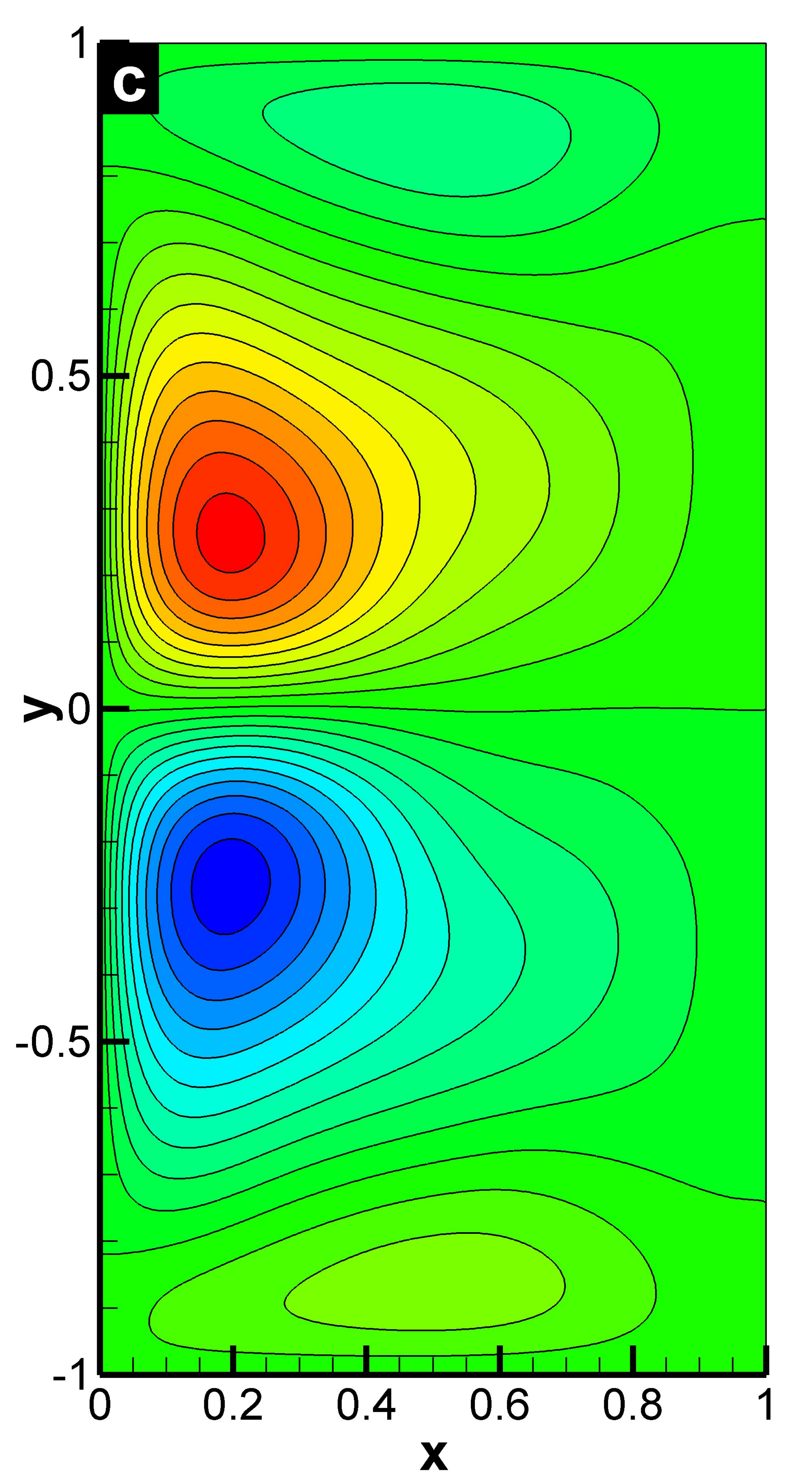}} }
\caption{
Experiment (i): \ DNS results for
$Re=200, Ro=0.0016$ and a spatial resolution of $512\times 256$.
Time-averaged field data for:
(a) potential vorticity; (b) vorticity; and (c) streamfunction.
Note that four gyres appear in the streamfunction contour plot.
}
\label{fig:dns-a}
\end{figure}

\begin{figure}
\centering
\mbox{
\subfigure{\includegraphics[width=0.25\textwidth]{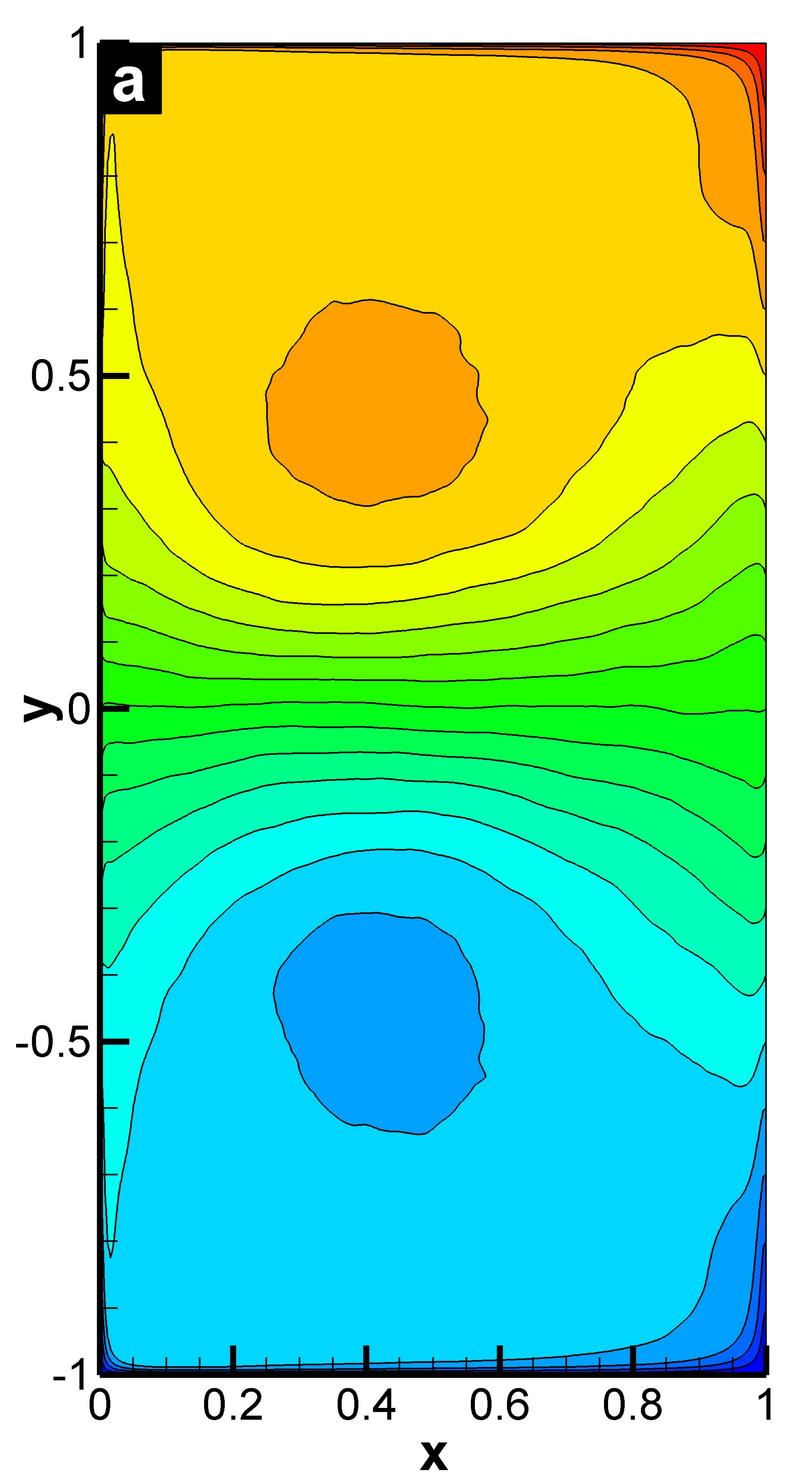}}
\subfigure{\includegraphics[width=0.25\textwidth]{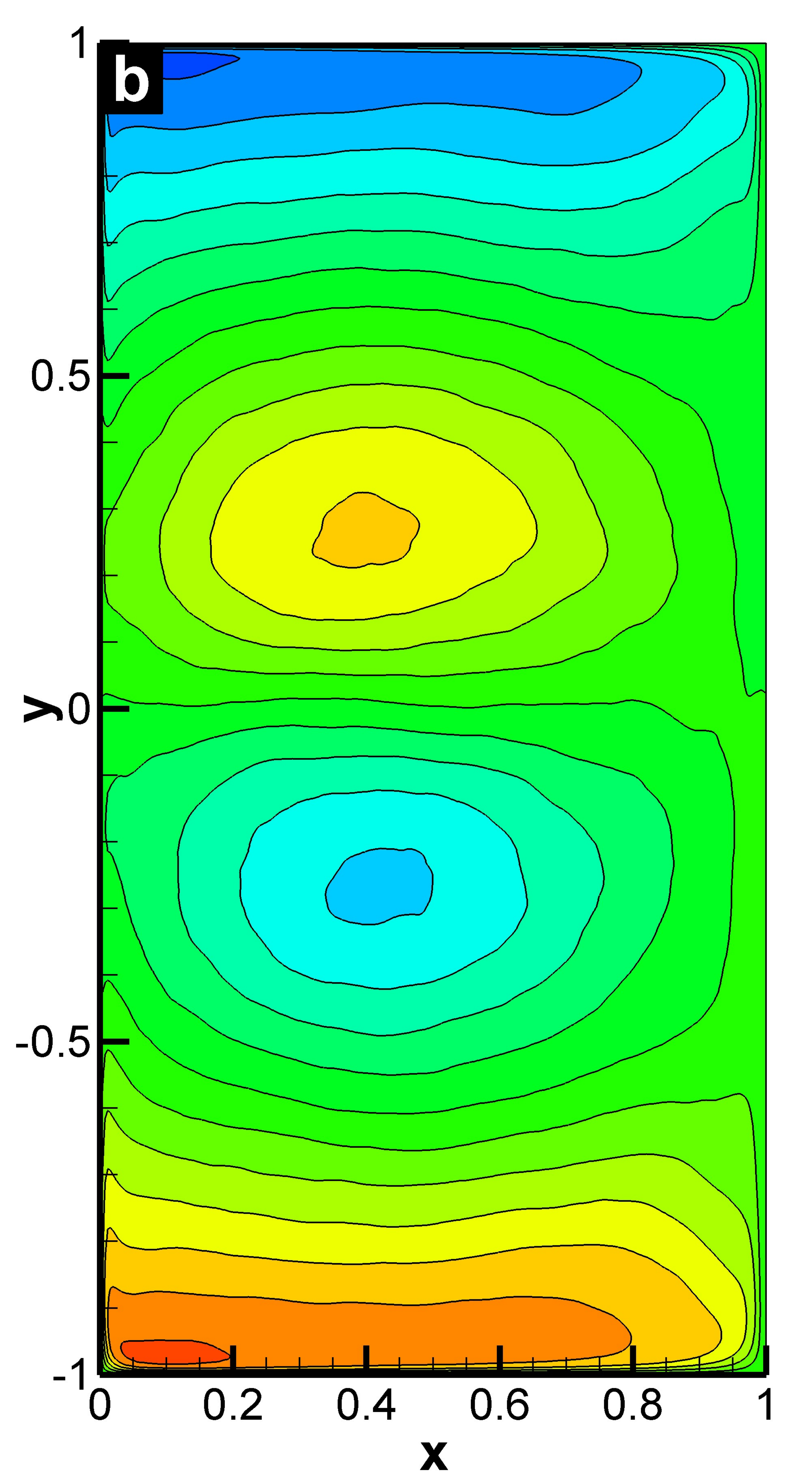}}
\subfigure{\includegraphics[width=0.25\textwidth]{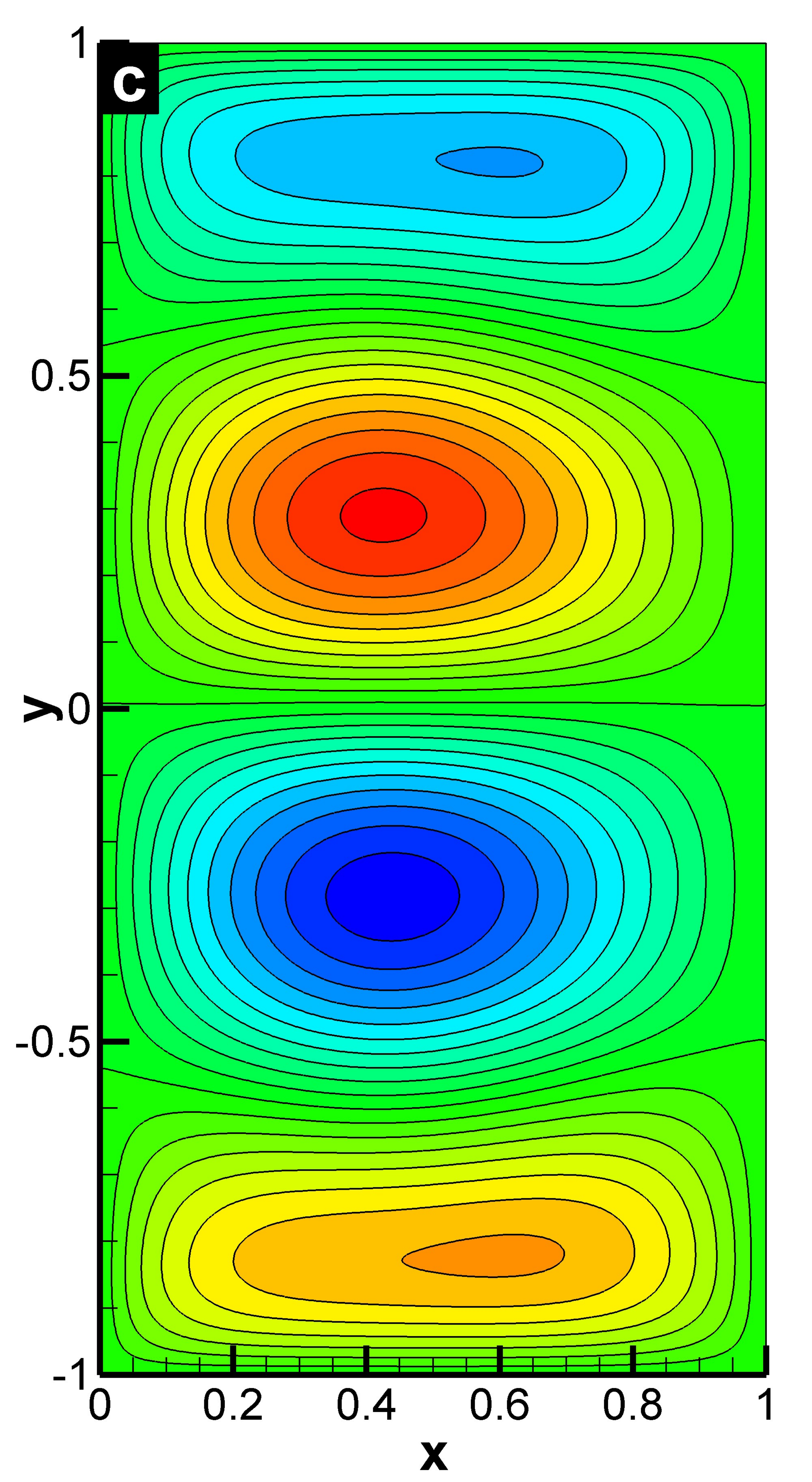}} }
\caption{
Experiment (ii): \  DNS results for
$Re=450, Ro=0.0036$ and a spatial resolution of $512\times 256$.
Time-averaged field data for:
(a) potential vorticity; (b) vorticity; and (c) streamfunction.
Note that four gyres appear in the streamfunction contour plot.
}
\label{fig:dns-b}
\end{figure}

We start by performing a DNS computation on a fine mesh ($512\times256$ spatial resolution). After a transient  period, a statistically steady state solution is obtained at a time of around $t=10$. Instantaneous contour plots at time $t=90$ for the potential vorticity, vorticity, and streamfunction are shown in Fig.~(\ref{fig:ins-a}) and Fig.~(\ref{fig:ins-b}) for Experiment (i) and Experiment (ii), respectively. Two gyres are clearly seen in the streamfunction contour plots. Next, the time average of the data is taken between time $t=20$ and $t=100$ using $8001$
snapshots.
The results are given in Fig.~(\ref{fig:dns-a}) and Fig.~(\ref{fig:dns-b}).
We emphasize that, for both parameter sets, {\em four gyres} are clearly visible in the streamfunction
plots.
This is in stark contrast with the instantaneous streamfunction contour plots in Fig.~(\ref{fig:ins-a}) and
Fig.~(\ref{fig:ins-b}), in which only two gyres are present. The computational time of each DNS run was longer than $3.5$ days.

\begin{figure}
\centering
\mbox{
\subfigure{\includegraphics[width=0.23\textwidth]{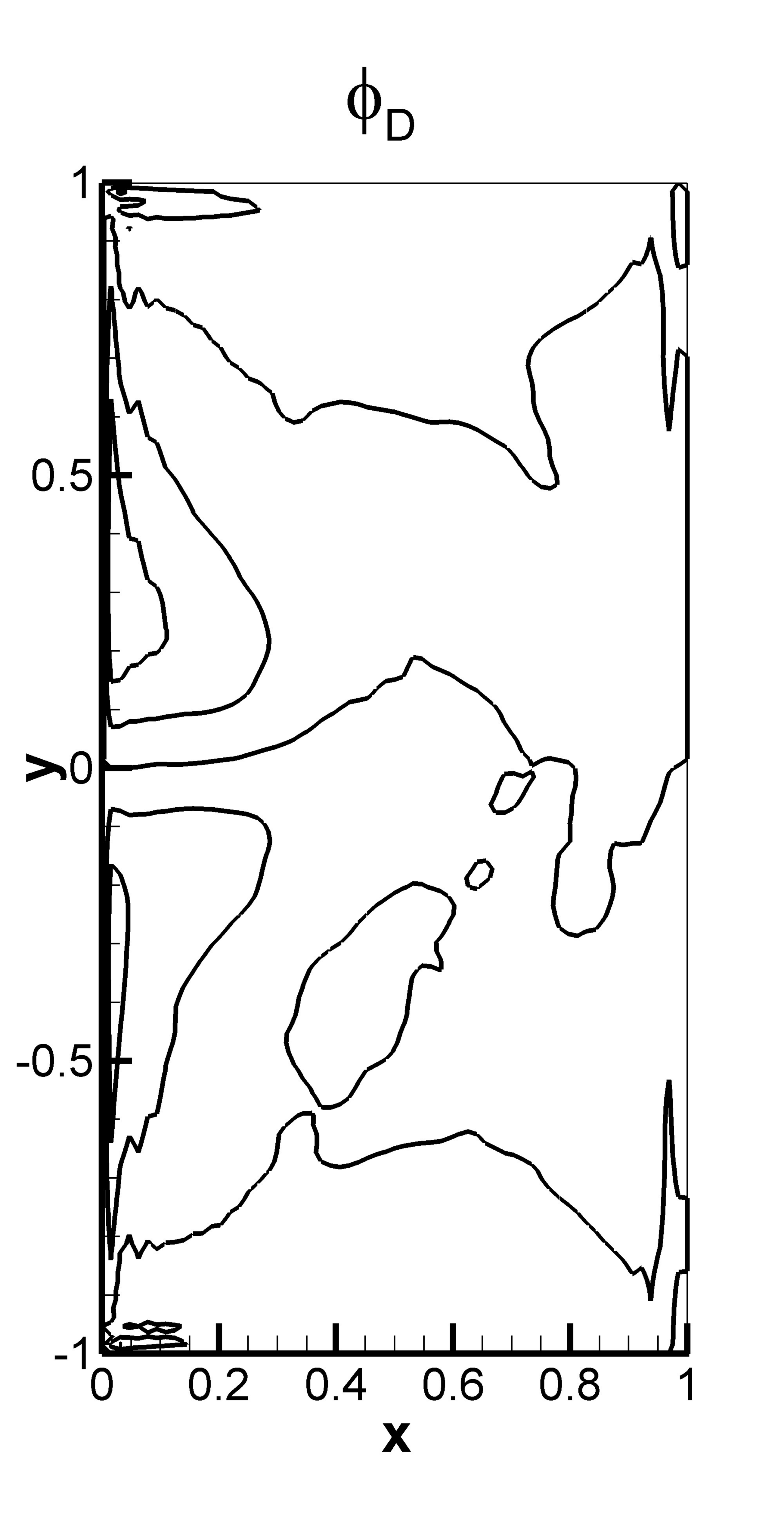}}
\subfigure{\includegraphics[width=0.23\textwidth]{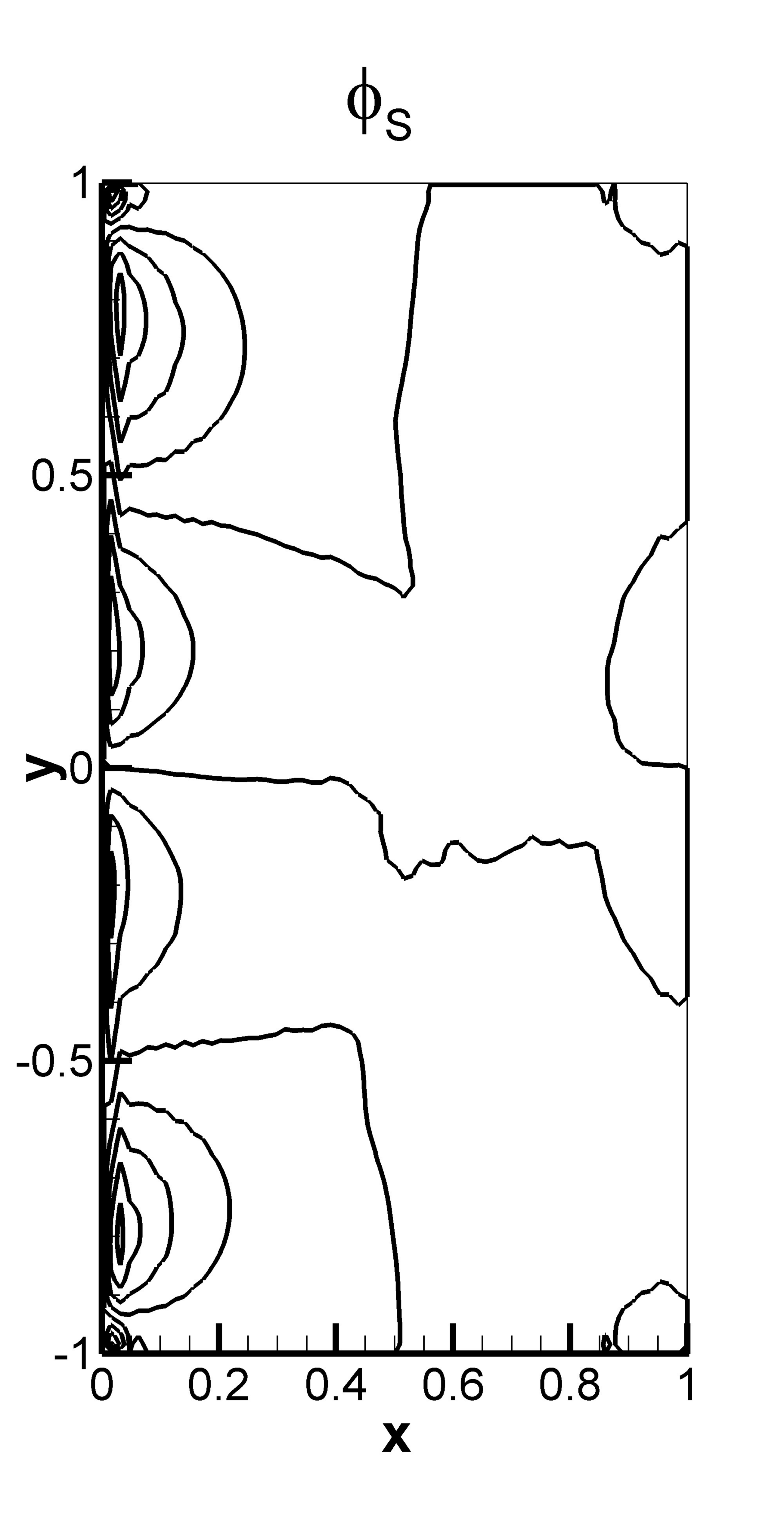}}
\subfigure{\includegraphics[width=0.23\textwidth]{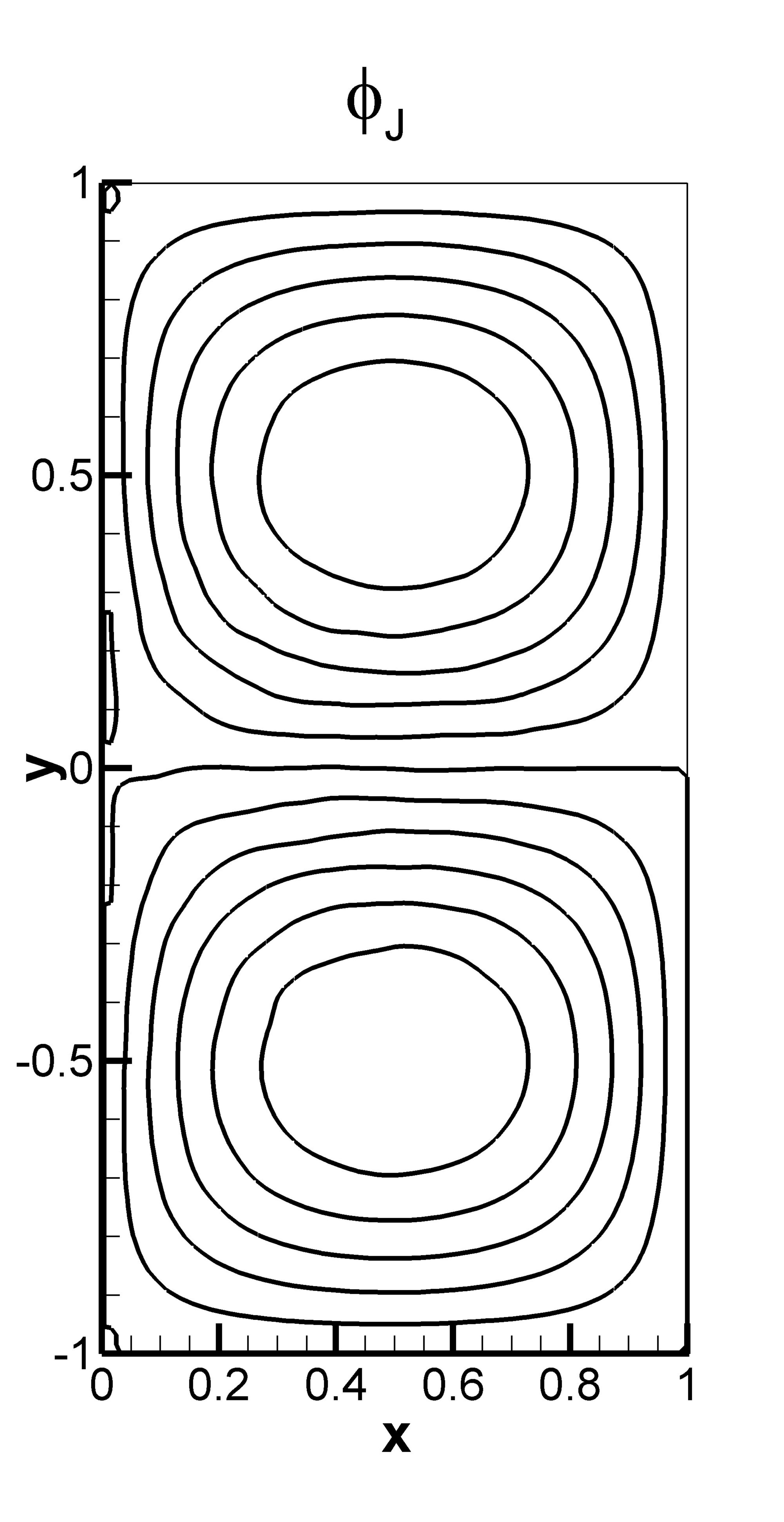}}
\subfigure{\includegraphics[width=0.23\textwidth]{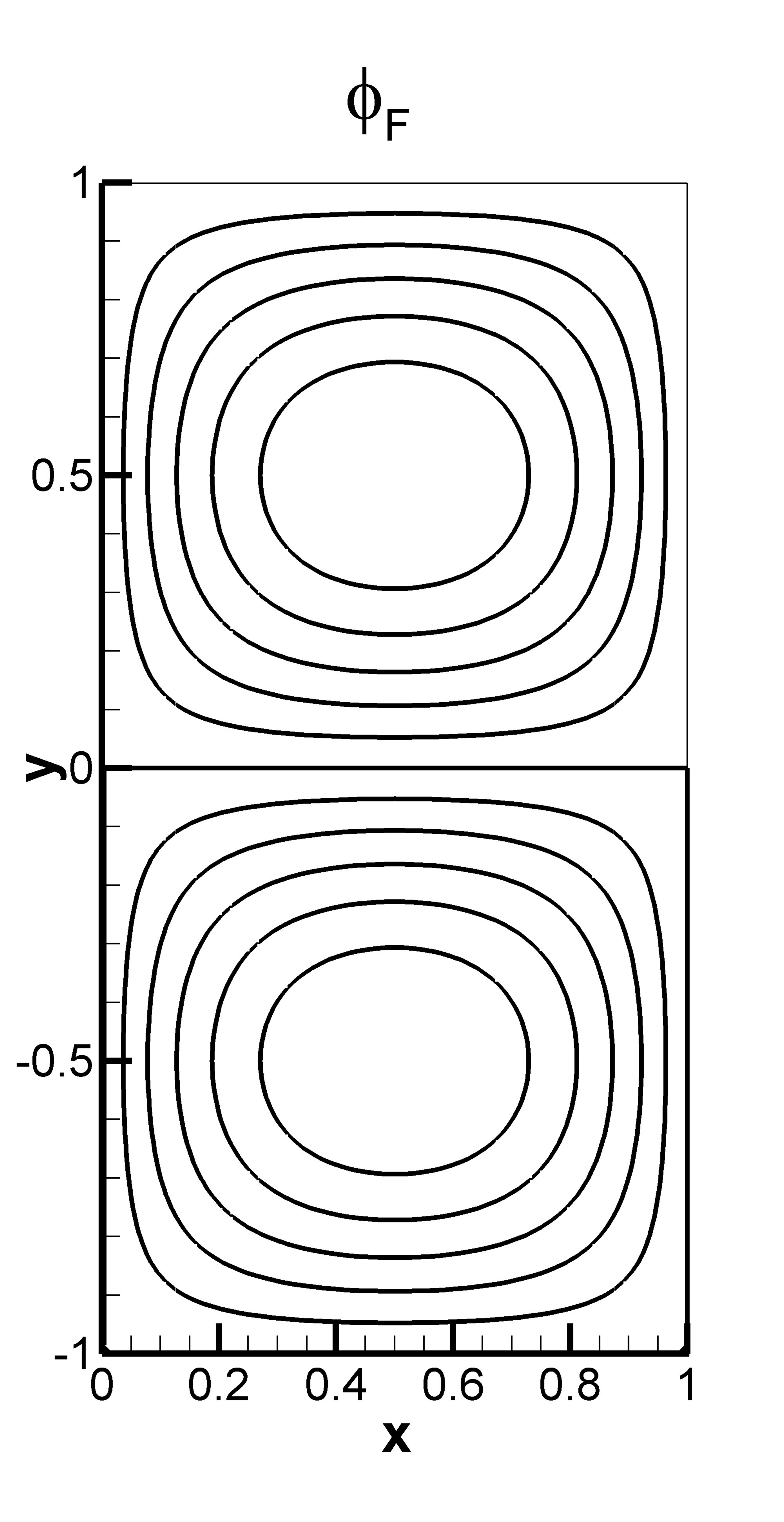}} }
\caption{
Experiment (i): \ AD results for
$Re=200, Ro=0.0016$, and a spatial resolution of $128\times 64$.
Time-averaged force function contour plots for the dissipation, subfilter-scale, Jacobian, and forcing terms. The contour intervals are between -0.05 and 0.05 for the forcing and Jacobian terms, and -0.002 and 0.002 for the dissipation and subfilter-scale terms.
}
\label{fig:ff-a}
\end{figure}

\begin{figure}
\centering
\mbox{
\subfigure{\includegraphics[width=0.23\textwidth]{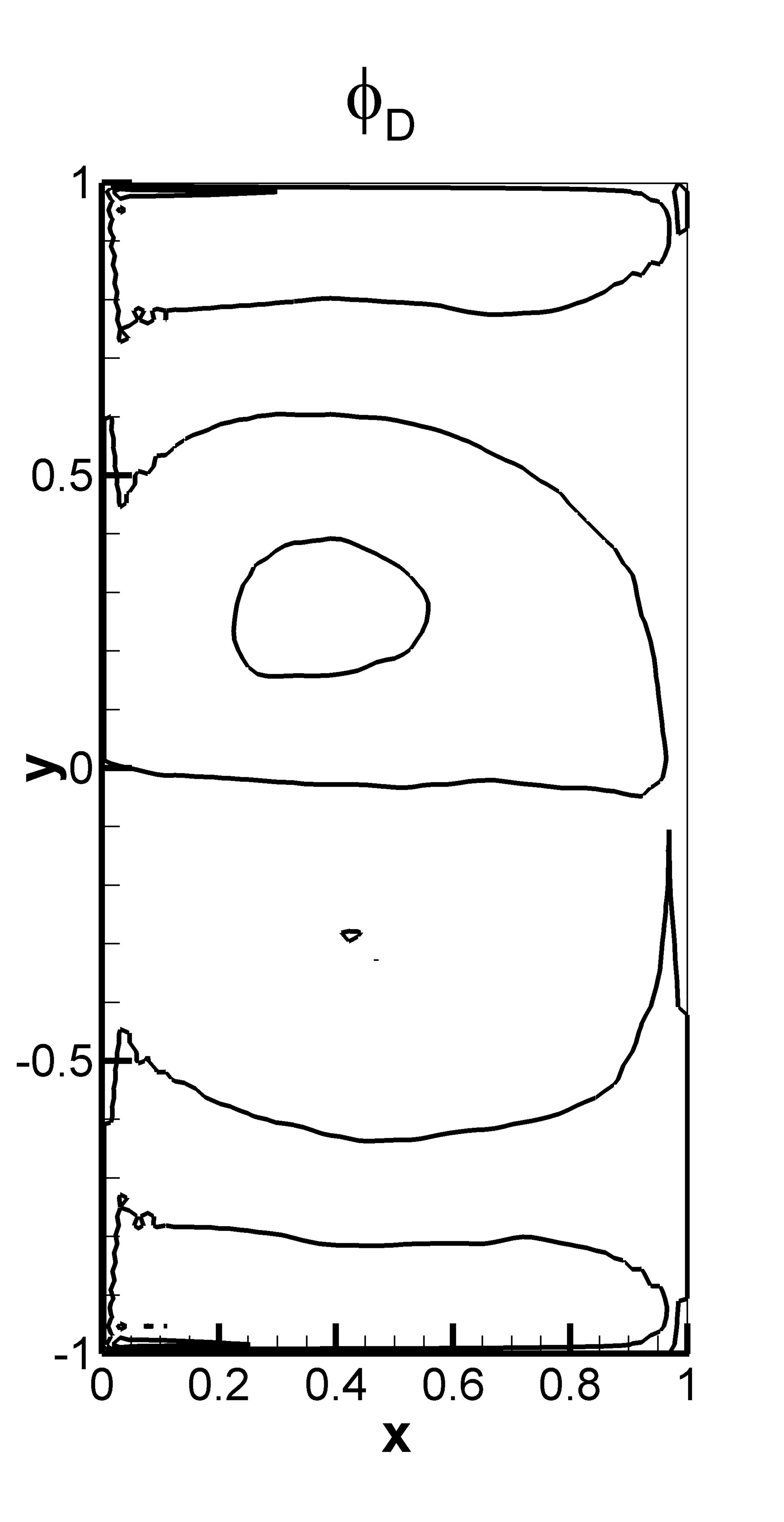}}
\subfigure{\includegraphics[width=0.23\textwidth]{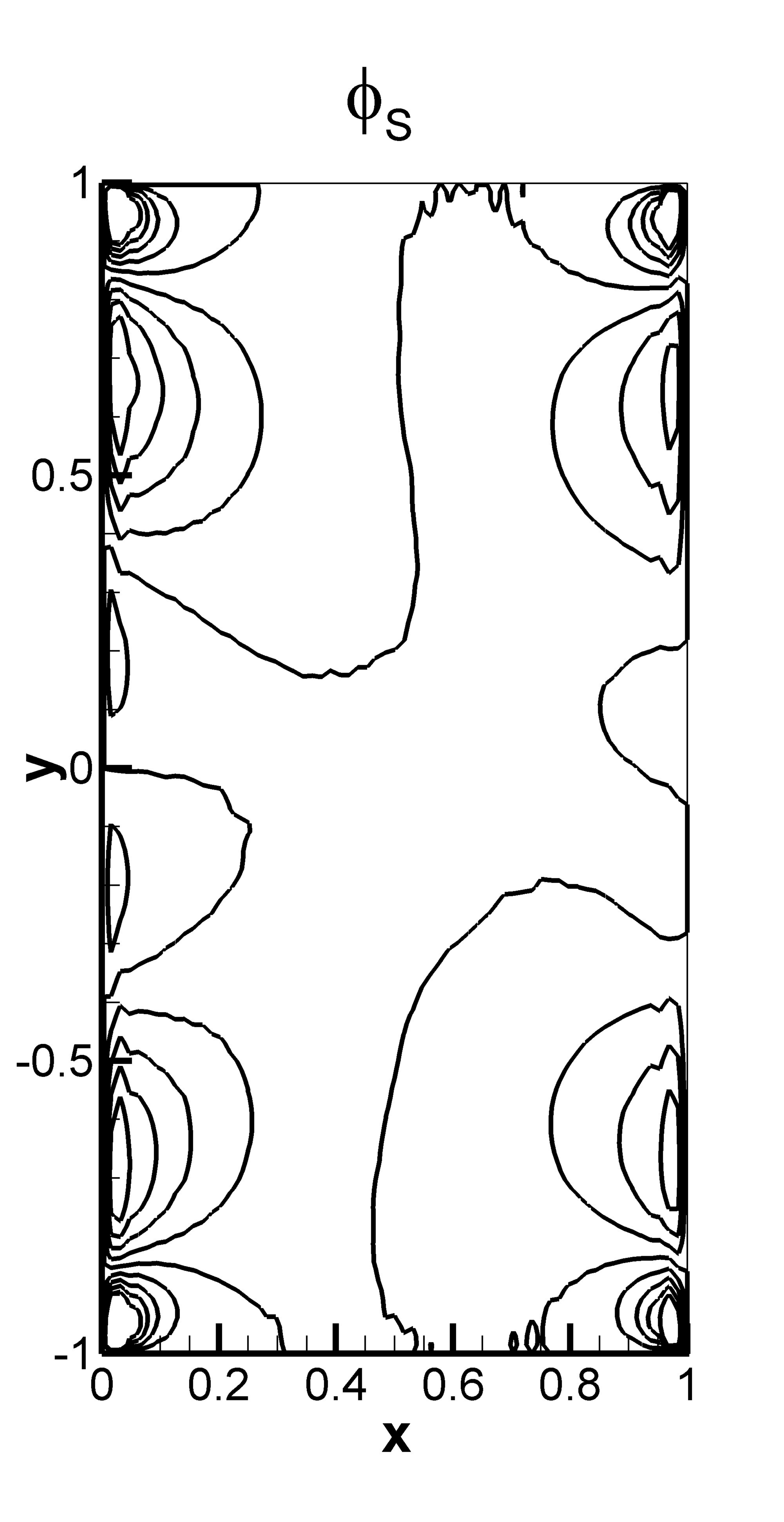}}
\subfigure{\includegraphics[width=0.23\textwidth]{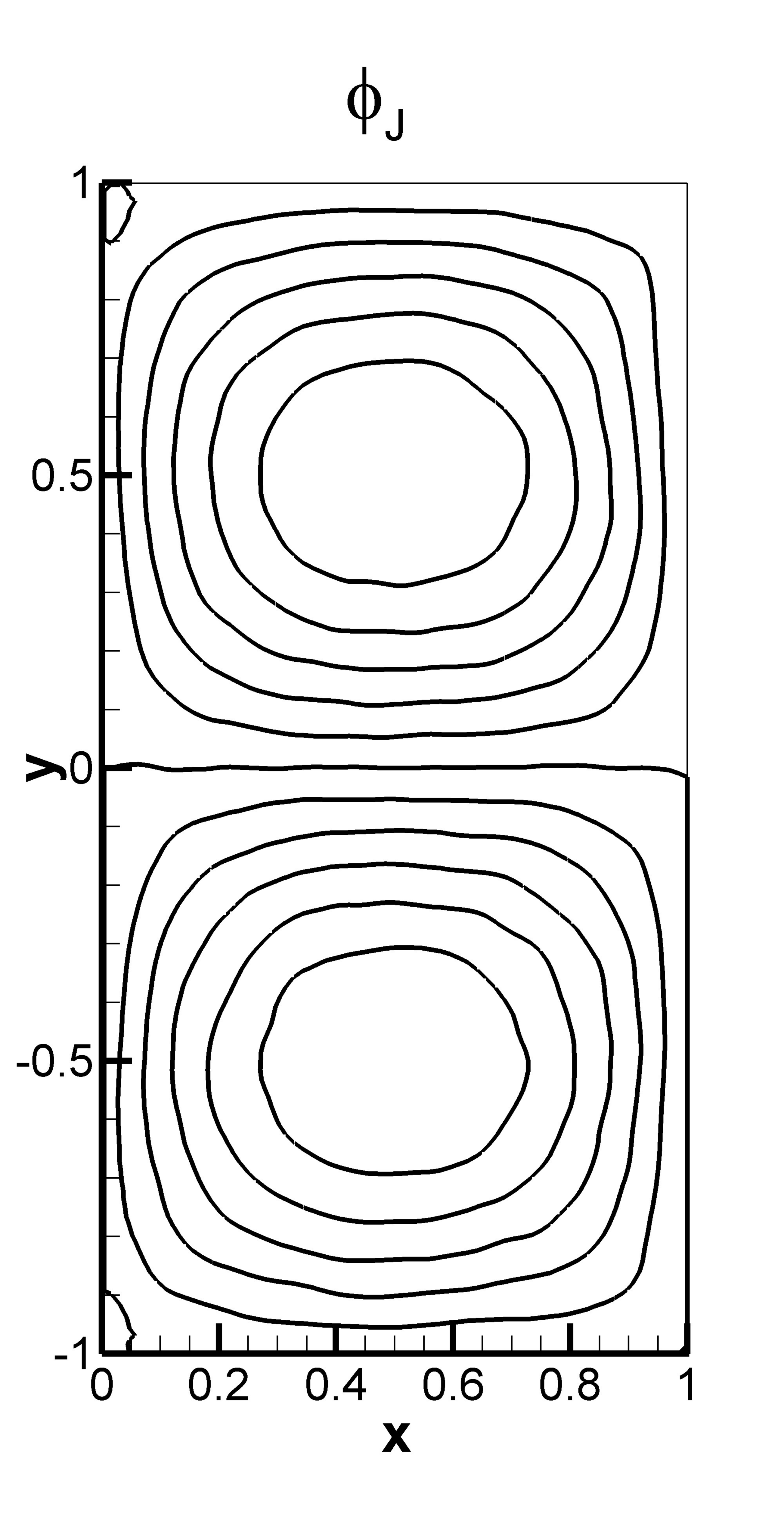}}
\subfigure{\includegraphics[width=0.23\textwidth]{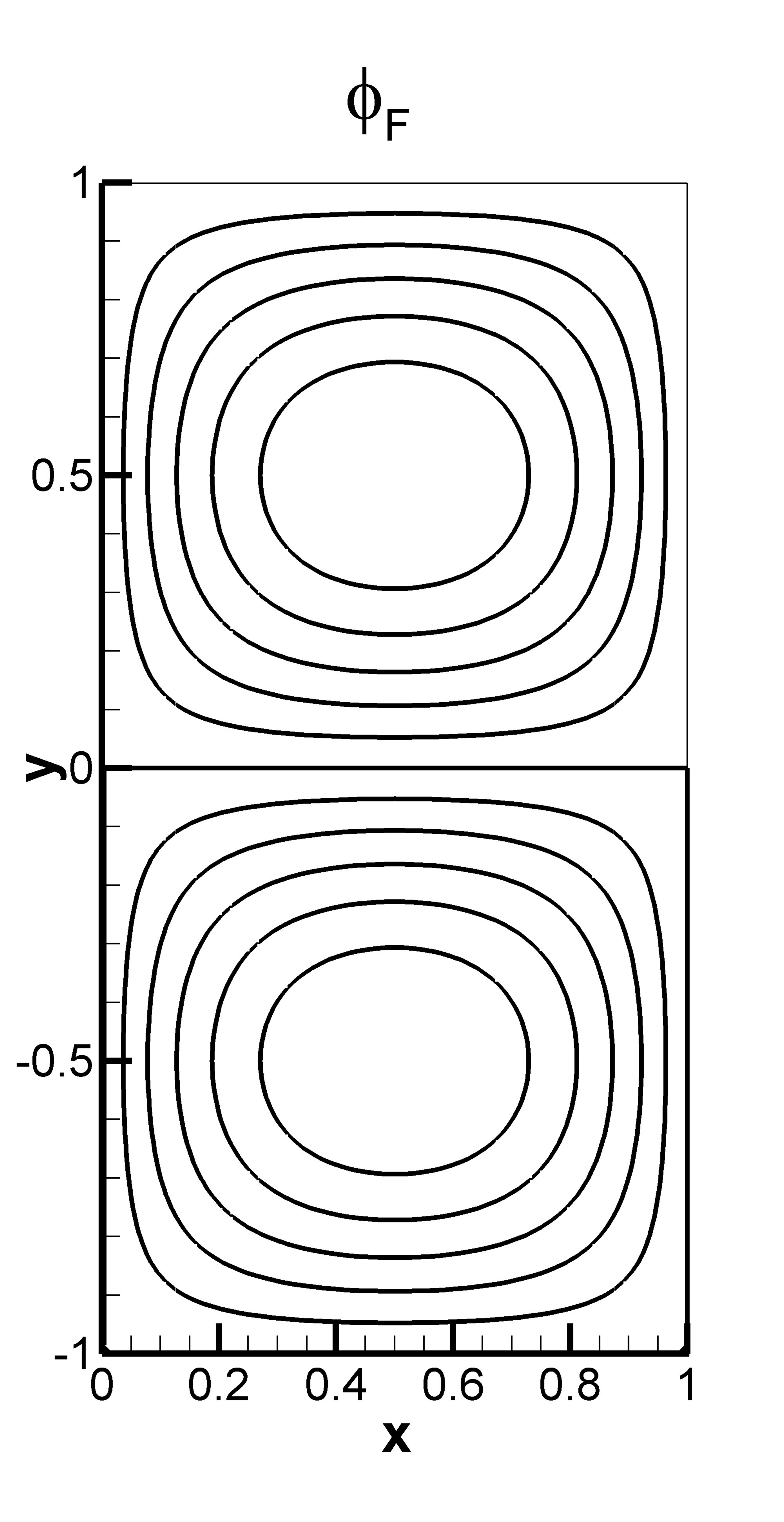}} }
\caption{
Experiment (ii): \  AD results for
$Re=450, Ro=0.0036$, and a spatial resolution of $128\times 64$.
Time-averaged force function contour plots for the dissipation, subfilter-scale, Jacobian, and forcing terms. The contour intervals are between -0.05 and 0.05 for the forcing and Jacobian terms, and -0.002 and 0.002 for the dissipation and subfilter-scale terms.
}
\label{fig:ff-b}
\end{figure}

\begin{figure}
\centering
\mbox{
\subfigure{\includegraphics[width=0.5\textwidth]{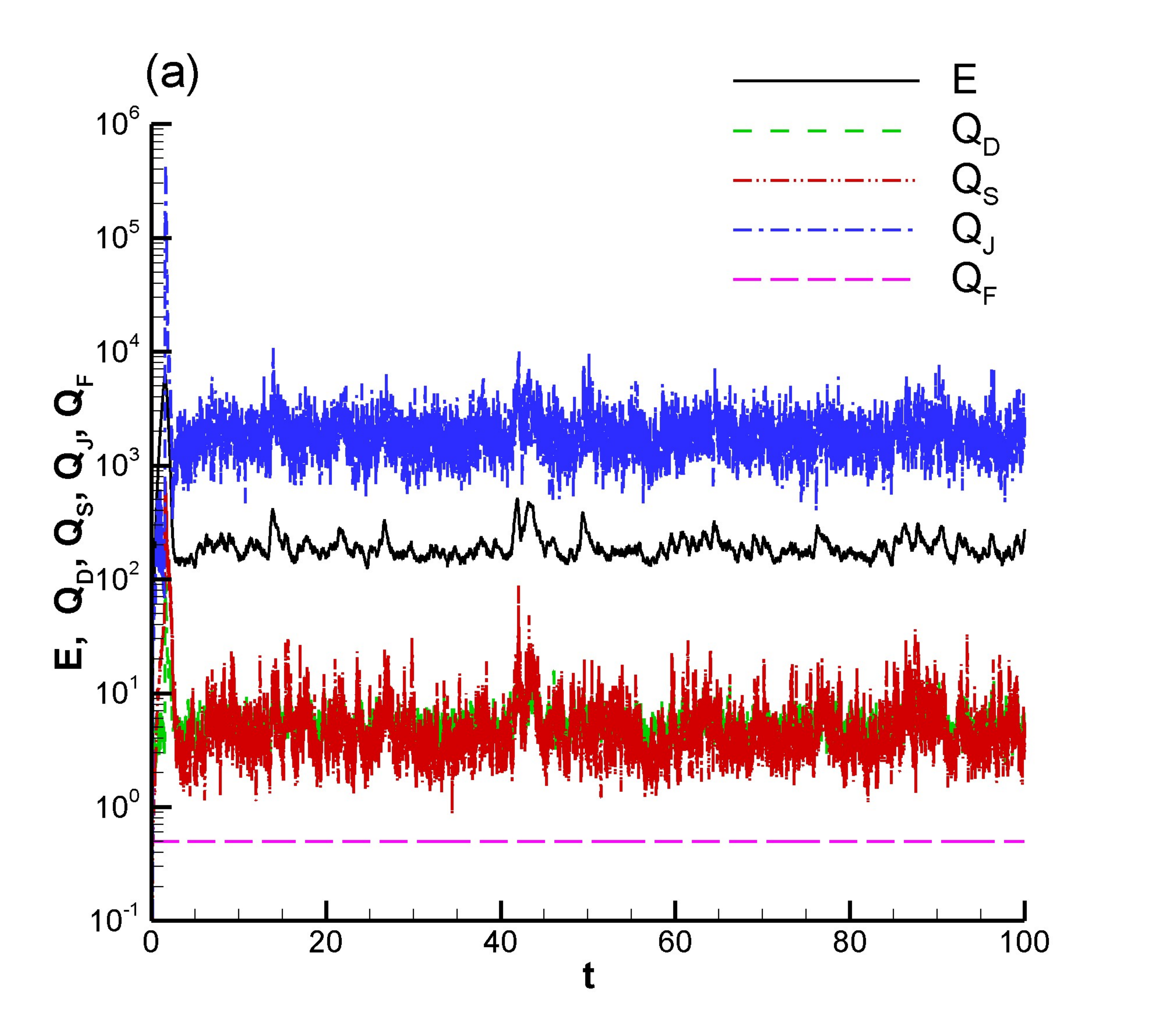}}
\subfigure{\includegraphics[width=0.5\textwidth]{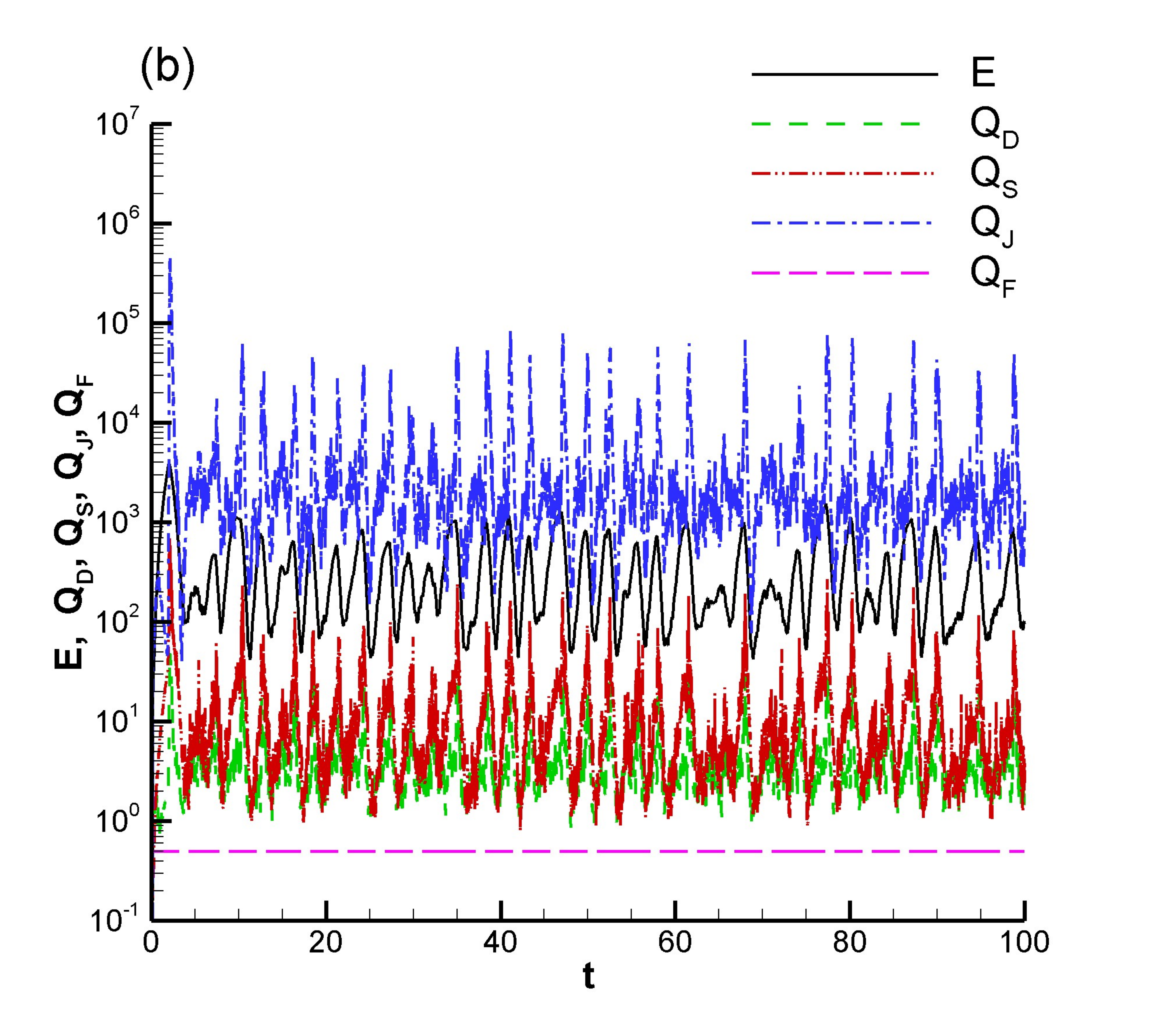}} }
\caption{
Time history of the total energy, dissipation, subfilter-scale, Jacobian, and forcing terms:
(a) Experiment (i): $\displaystyle \delta_{I}/L = 0.04$  and $\displaystyle \delta_{M}/L = 0.02$;
and (b) Experiment (ii): $\displaystyle \delta_{I}/L = 0.06$  and $\displaystyle \delta_{M}/L = 0.02$.
}
\label{fig:ad-t}
\end{figure}

In addition to the time series of each individual term given by Eqs.~\eqref{eq:41}-\eqref{eq:44}, following \cite{marshall2011momentum}, we compute the force functions in order to analyze the contribution of individual terms, including the subfilter-scale term, in the new AD model \eqref{eq:gef_ad}-\eqref{eq:1_ad}. The defining equations for the force functions corresponding to the dissipation, subfilter-scale, nonlinear Jacobian, and forcing terms are:
\begin{equation}
\nabla^2 \phi_D = \bar{D},
\label{eq:51}
\end{equation}
\begin{equation}
\nabla^2 \phi_S = S^{*},
\label{eq:52}
\end{equation}
\begin{equation}
\nabla^2 \phi_J = \bar{J},
\label{eq:53}
\end{equation}
\begin{equation}
\nabla^2 \phi_F = \bar{F}.
\label{eq:54}
\end{equation}
Solving these Poisson equations for time averaged data (between $t=20$ and $t=100$ using $8,001$
snapshots) yields the corresponding force functions. Fig.~(\ref{fig:ff-a}) and Fig.~(\ref{fig:ff-b}) illustrate these force function contour plots for Experiment (i) and Experiment (ii) for a spatial resolution of $128 \times 64$. The time series of the total energy, dissipation, subfilter-scale, Jacobian and forcing terms are plotted in Fig.~(\ref{fig:ad-t}). It can be seen that at this spatial resolution the subfilter-scale effects are on the order of dissipation term.

Next, we test the new AD model \eqref{eq:gef_ad}-\eqref{eq:1_ad} using a coarser mesh with a spatial resolution of $32 \times 16$, which is {\em $16$ times coarser} than the DNS mesh with a resolution of $512 \times 256$.
On this coarse mesh, we test both the new AD model and the under-resolved numerical
simulation BVE$_{coarse}$, which does not employ any subfilter-scale model.
To compare the models, we utilize data that is time-averaged between $t=20$ and $t=100$.
The new AD model is tested with a second-order filtering operation and $N=5$ (as in Eq.~\eqref{eq:6}
and Eq.~\eqref{eq:7}), and with the smoothing parameter $\alpha=0.25$.

\begin{figure}
\centering
\mbox{
\subfigure{\includegraphics[width=0.25\textwidth]{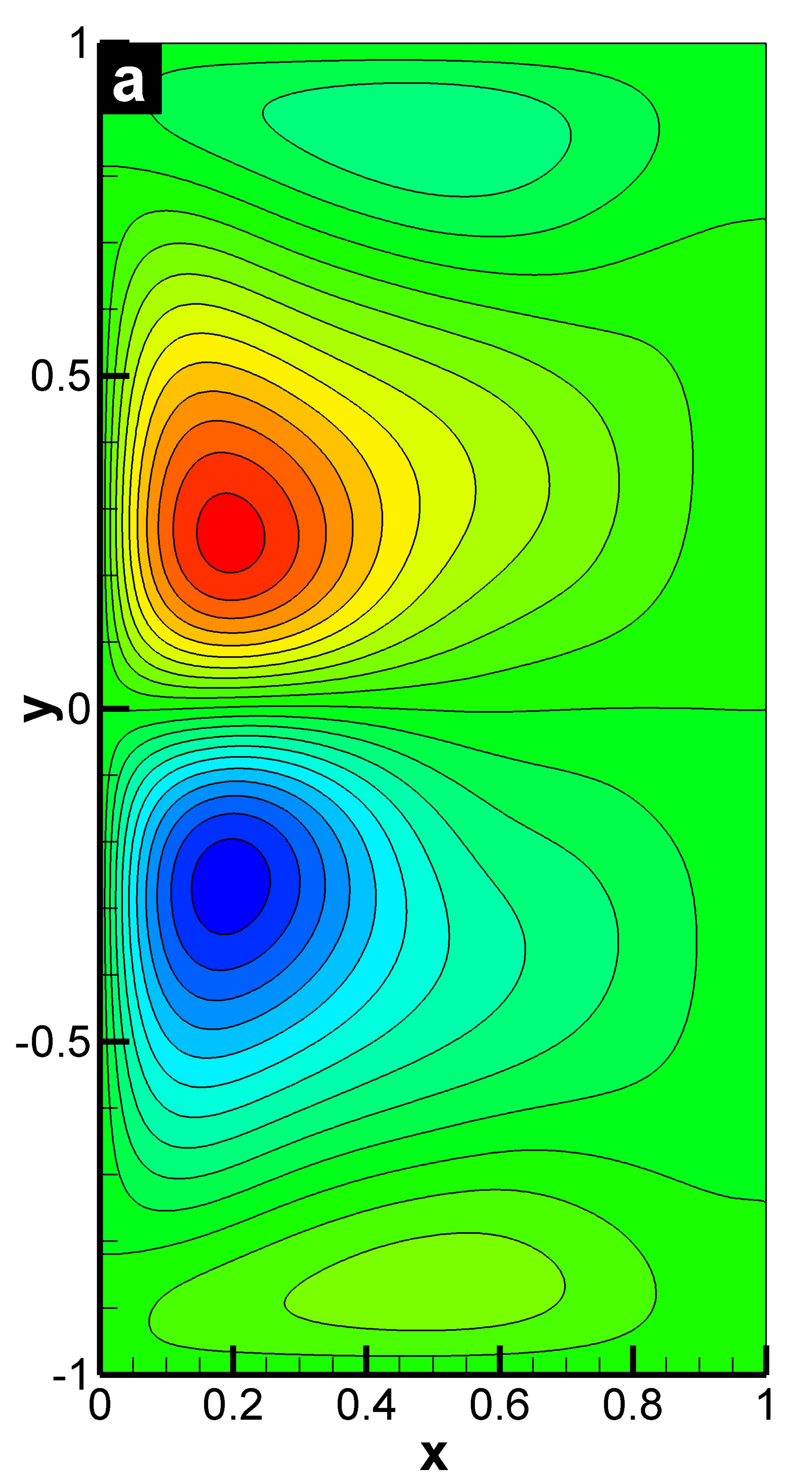}}
\subfigure{\includegraphics[width=0.25\textwidth]{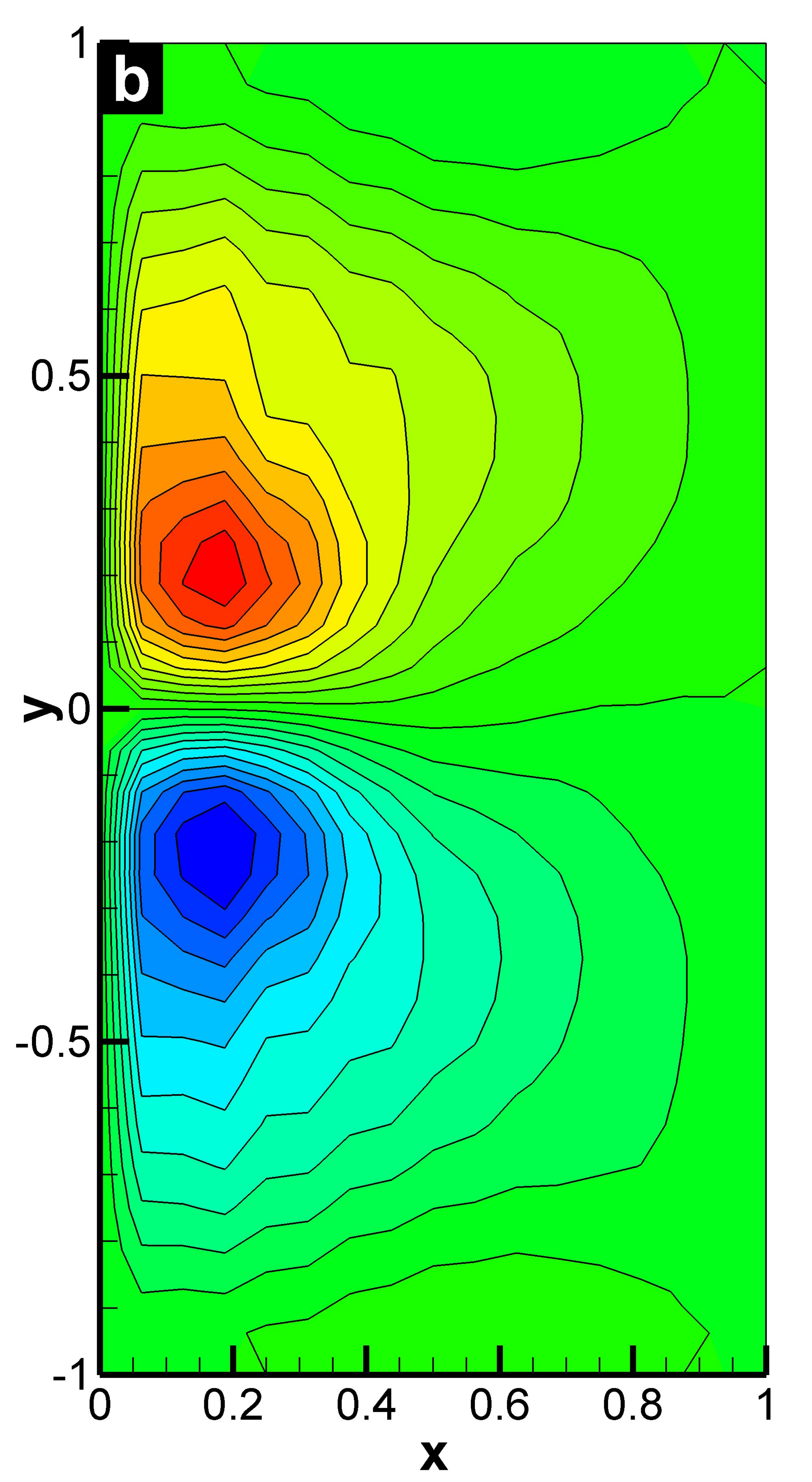}}
\subfigure{\includegraphics[width=0.25\textwidth]{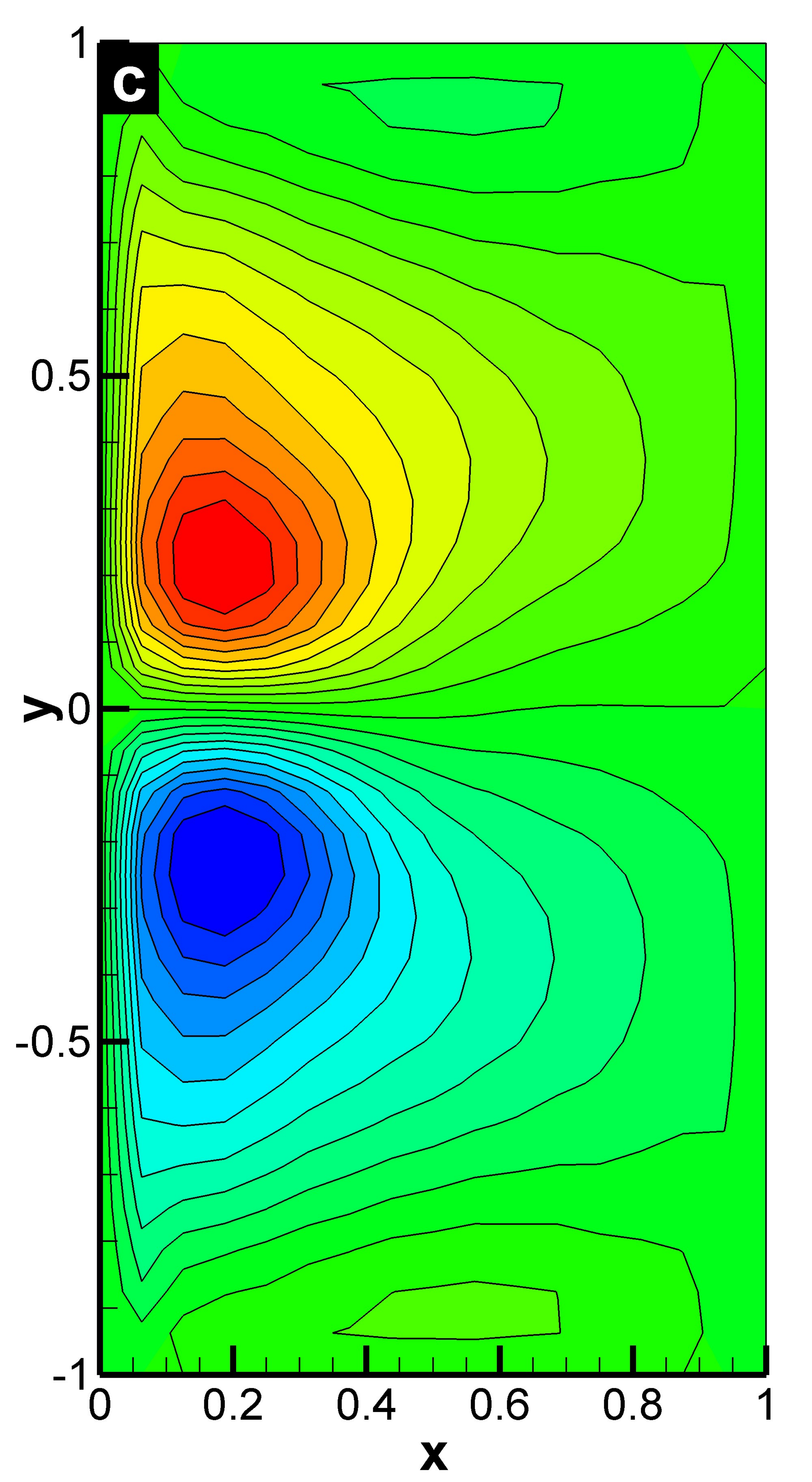}} }
\caption{
Experiment (i): \  Time-averaged streamfunction data for $Re=200$ and $Ro=0.0016$:
(a) DNS results at a resolution of $512 \times 256$;
(b) under-resolved BVE$_{coarse}$ results at a resolution of $32 \times 16$;
(c) new AD model results at a resolution of  $32 \times 16$.
Note the smoothing effect of the AD model on contour lines. Contour interval layouts are identical.
}
\label{fig:exp1-s}
\end{figure}

\begin{figure}
\centering
\mbox{
\subfigure{\includegraphics[width=0.25\textwidth]{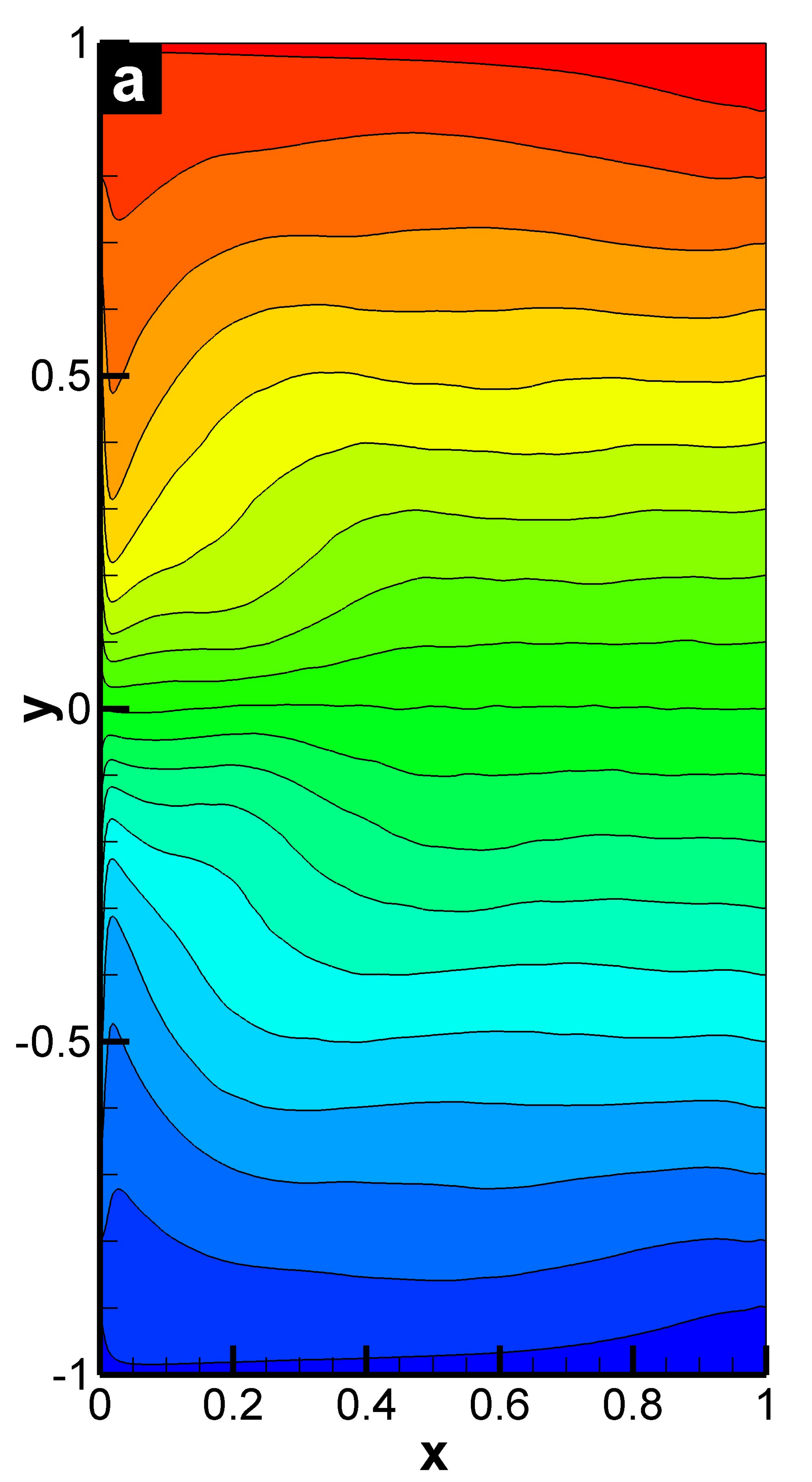}}
\subfigure{\includegraphics[width=0.25\textwidth]{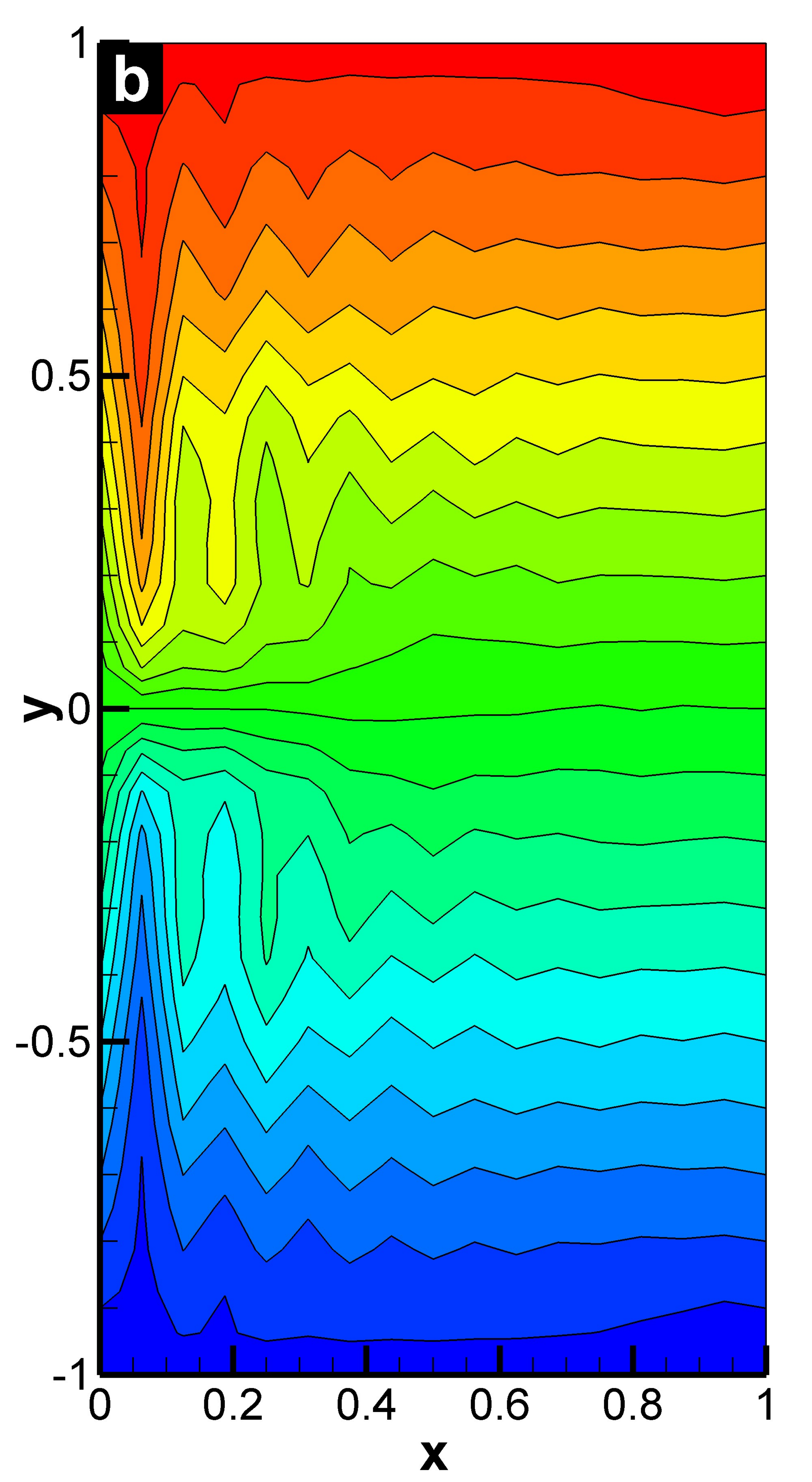}}
\subfigure{\includegraphics[width=0.25\textwidth]{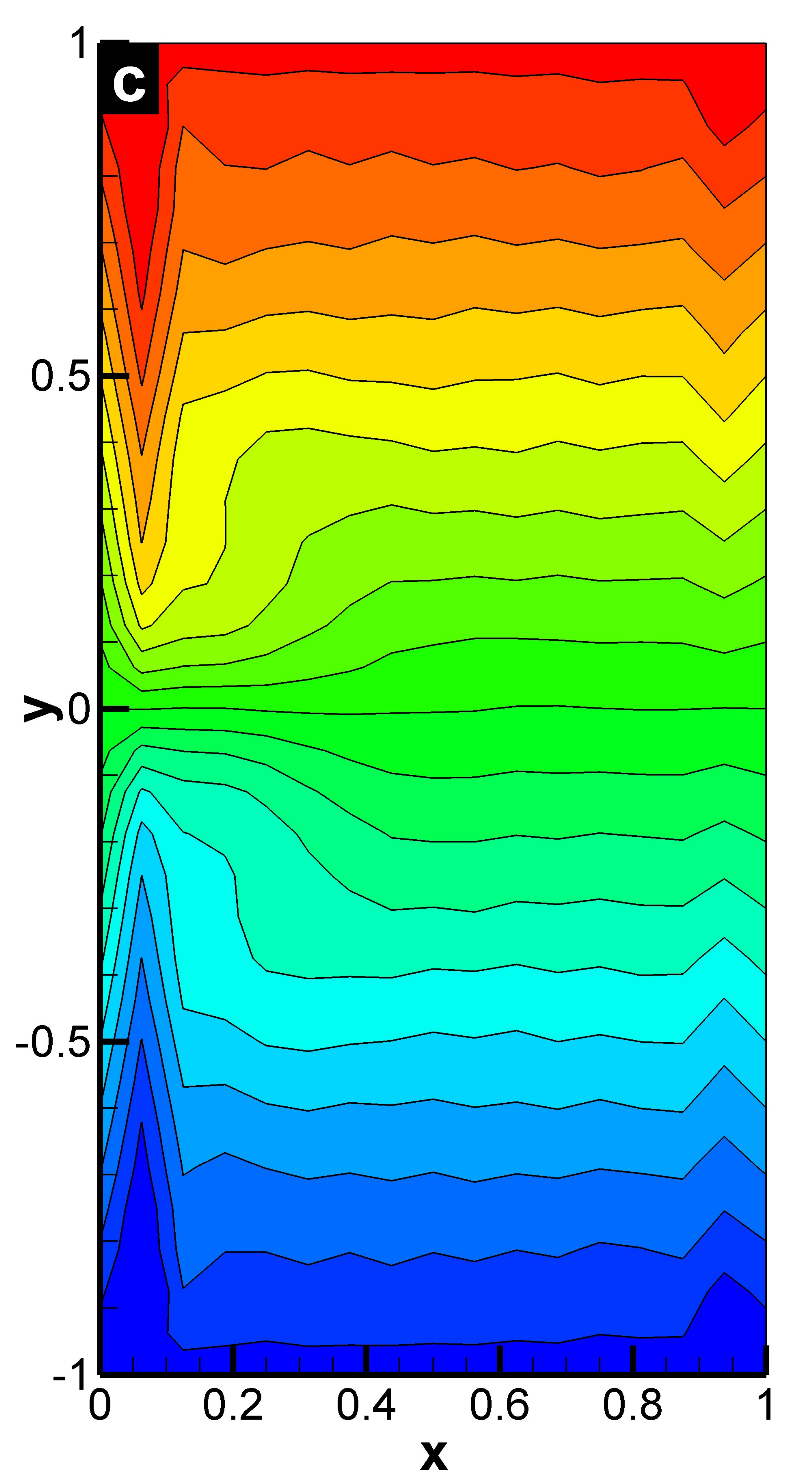}} }
\caption{
Experiment (i): \  Time-averaged potential vorticity data for $Re=200$ and $Ro=0.0016$:
(a) DNS results at a resolution of $512 \times 256$;
(b) under-resolved BVE$_{coarse}$ results at a resolution of $32 \times 16$;
(c) new AD model results at a resolution of $32 \times 16$.
Note the smoothing effect of the AD model on contour lines. Contour interval layouts are identical.
}
\label{fig:exp1-q}
\end{figure}

\begin{figure}
\centering
\mbox{
\subfigure{\includegraphics[width=0.25\textwidth]{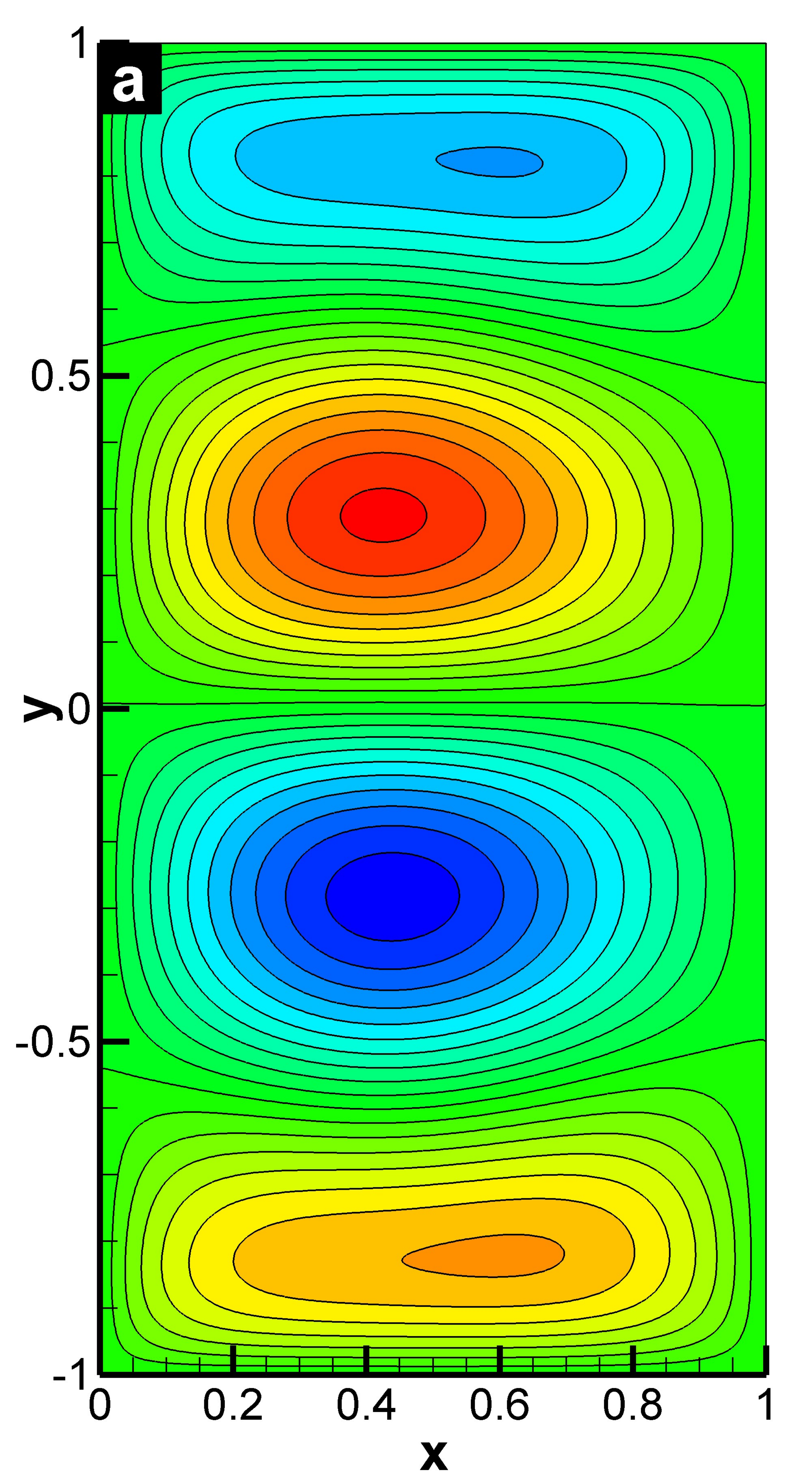}}
\subfigure{\includegraphics[width=0.25\textwidth]{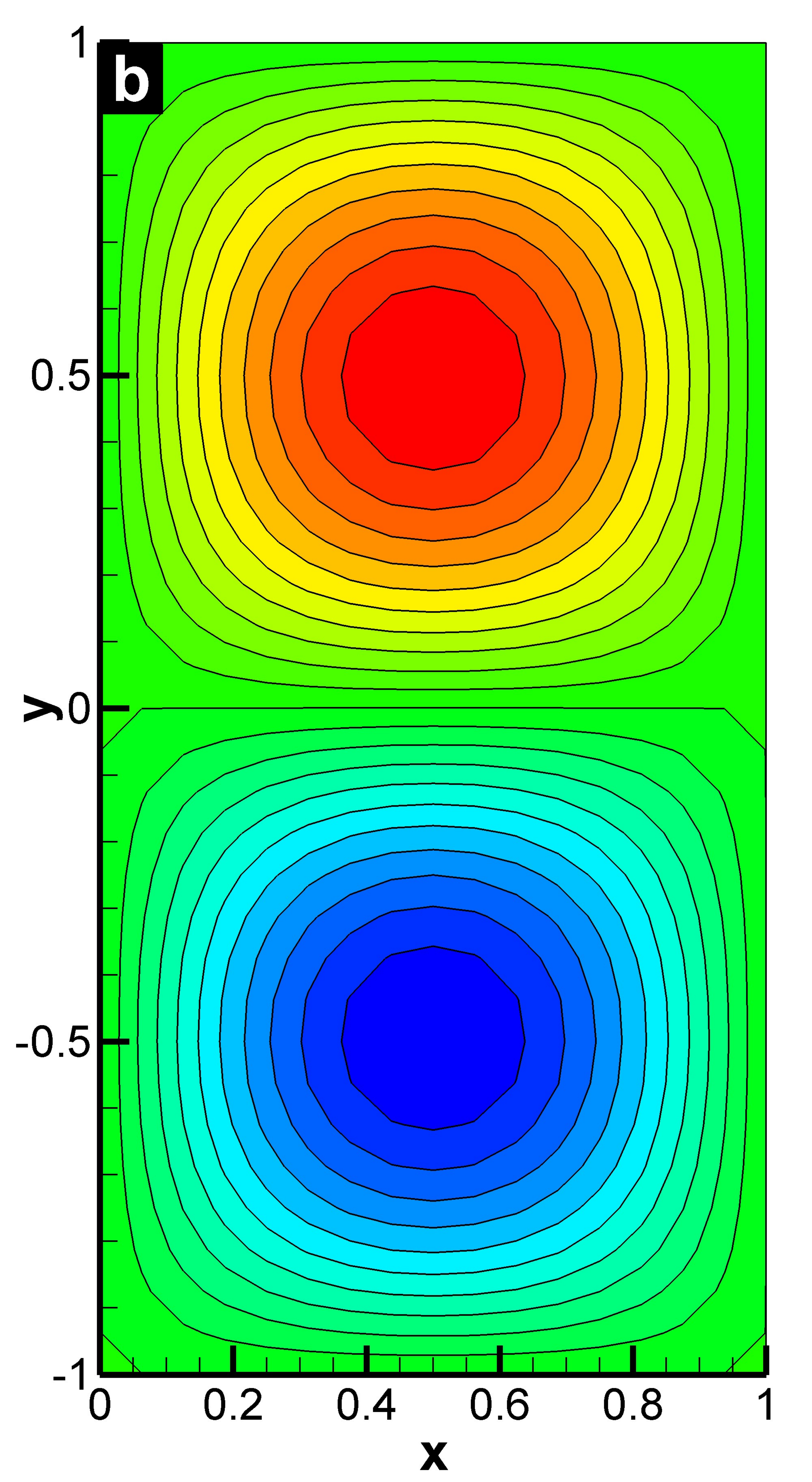}}
\subfigure{\includegraphics[width=0.25\textwidth]{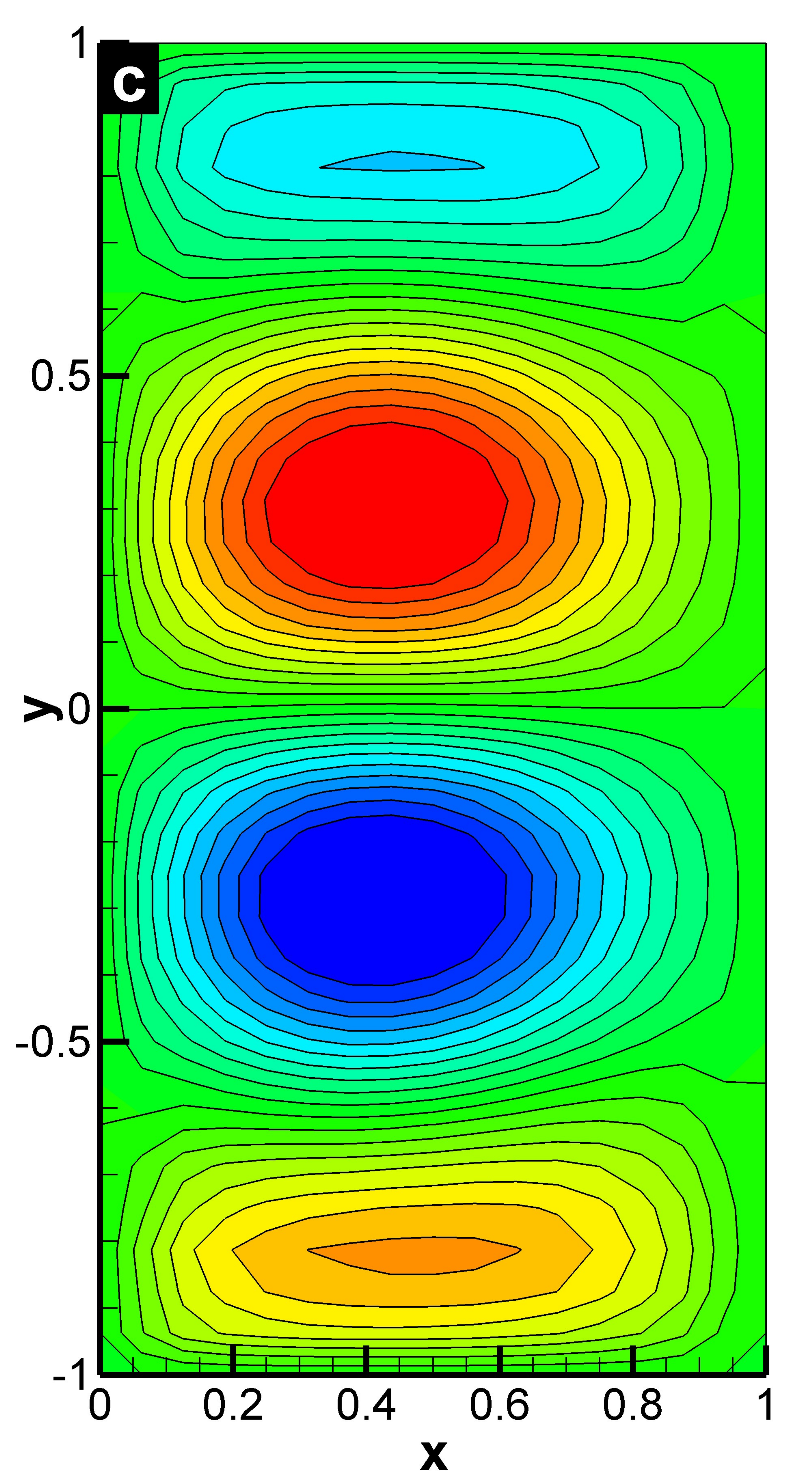}} }
\caption{
Experiment (ii): \  Time-averaged streamfunction data for $Re=450$ and $Ro=0.0036$:
(a) DNS results at a resolution of $512 \times 256$;
(b) under-resolved BVE$_{coarse}$ results at a resolution of $32 \times 16$;
(c) new AD model results at a resolution of $32 \times 16$.
Note that the BVE$_{coarse}$ results are nonphysical, whereas the
DNS and AD model results are qualitatively close. Contour interval layouts are identical only for (a) and (c).
}
\label{fig:exp2-s}
\end{figure}

\begin{figure}
\centering
\mbox{
\subfigure{\includegraphics[width=0.25\textwidth]{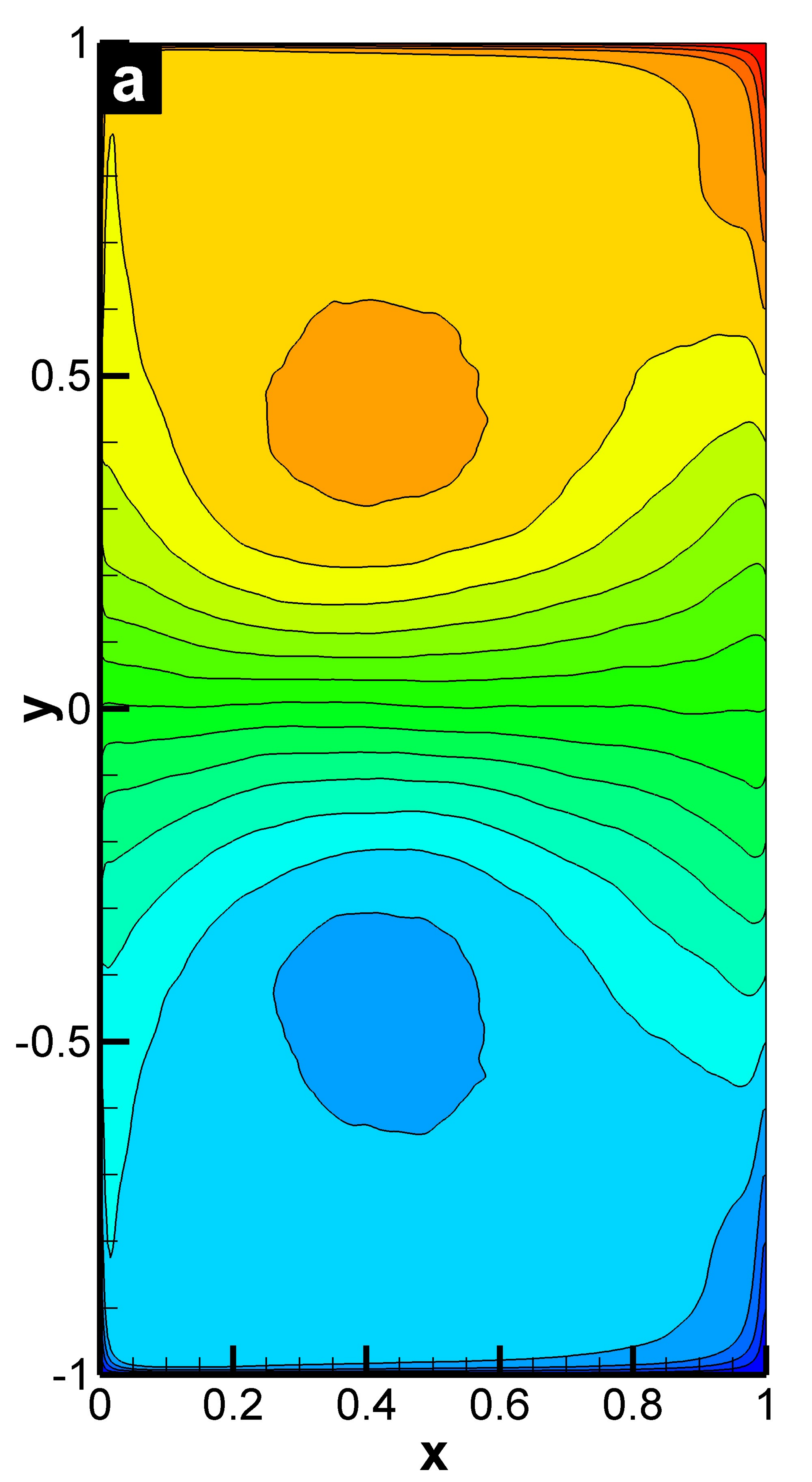}}
\subfigure{\includegraphics[width=0.25\textwidth]{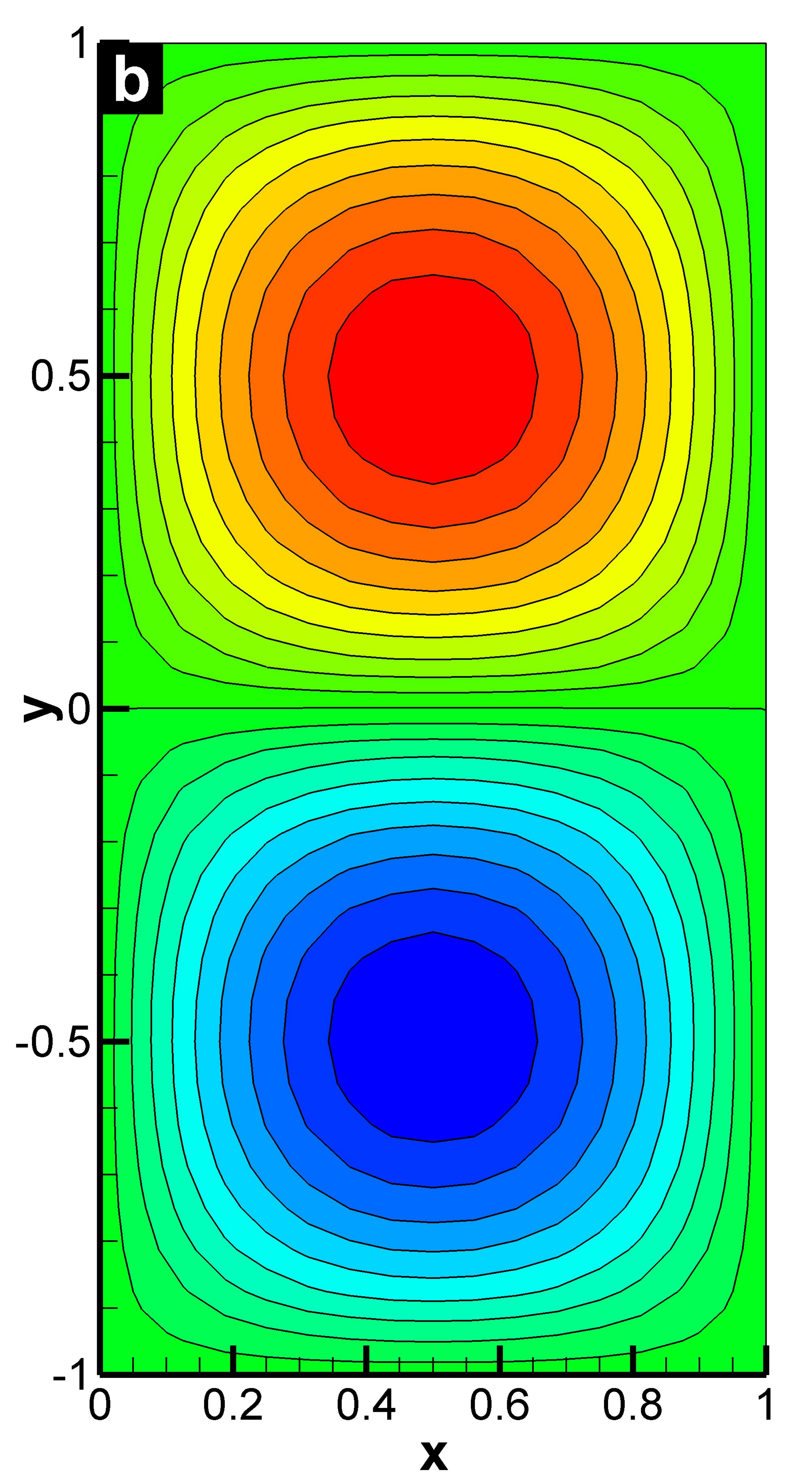}}
\subfigure{\includegraphics[width=0.25\textwidth]{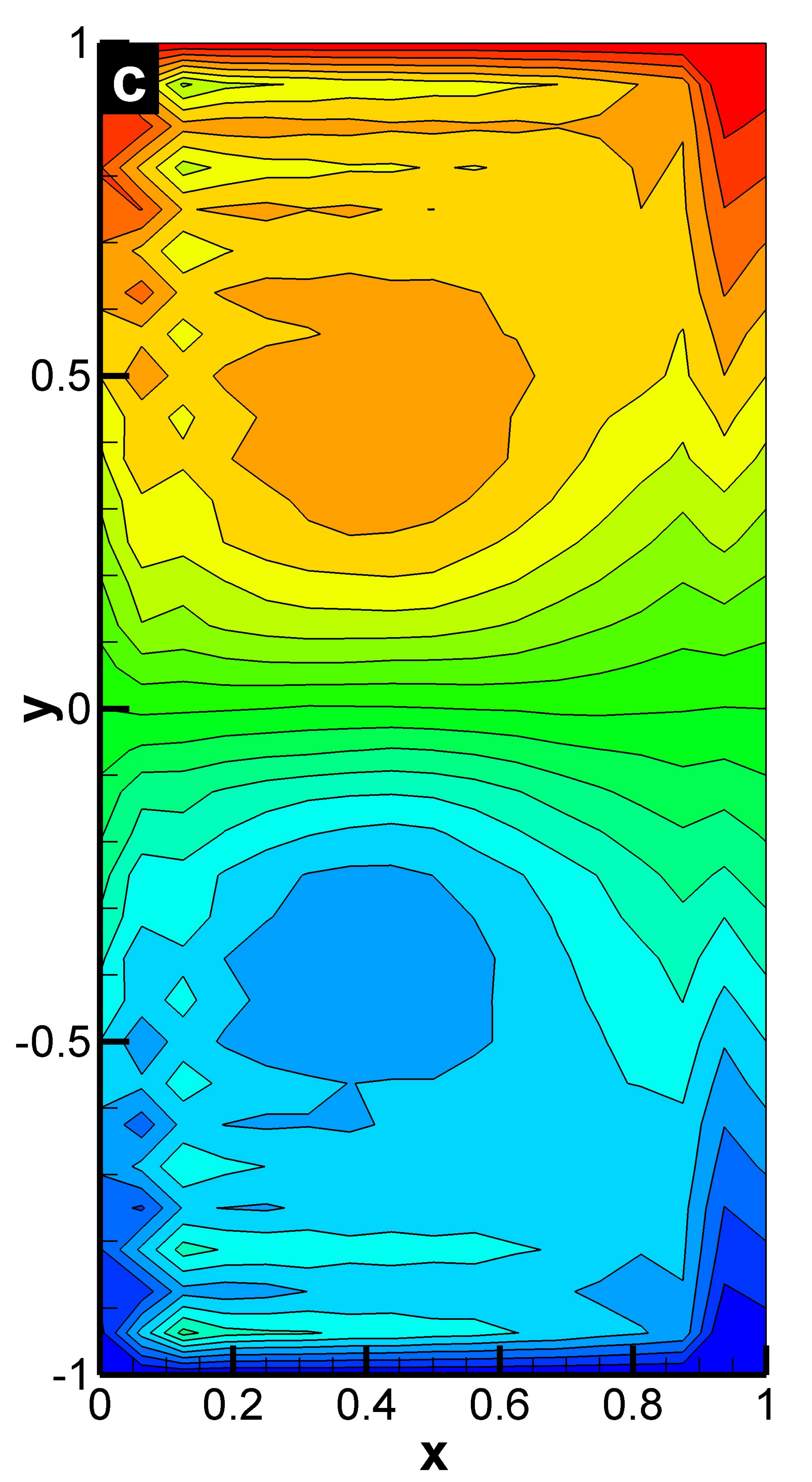}} }
\caption{
Experiment (ii): \  Time-averaged potential vorticity data for $Re=450$ and $Ro=0.0036$:
(a) DNS results at a resolution of $512 \times 256$;
(b) under-resolved BVE$_{coarse}$ results at a resolution of $32 \times 16$;
(c) new AD model results at a resolution of $32 \times 16$.
Note that the BVE$_{coarse}$ results are nonphysical, whereas the
DNS and AD model results are qualitatively similar. Contour interval layouts are identical only for (a) and (c).
}
\label{fig:exp2-q}
\end{figure}

\begin{figure}
\centering
\mbox{
\subfigure{\includegraphics[width=0.7\textwidth]{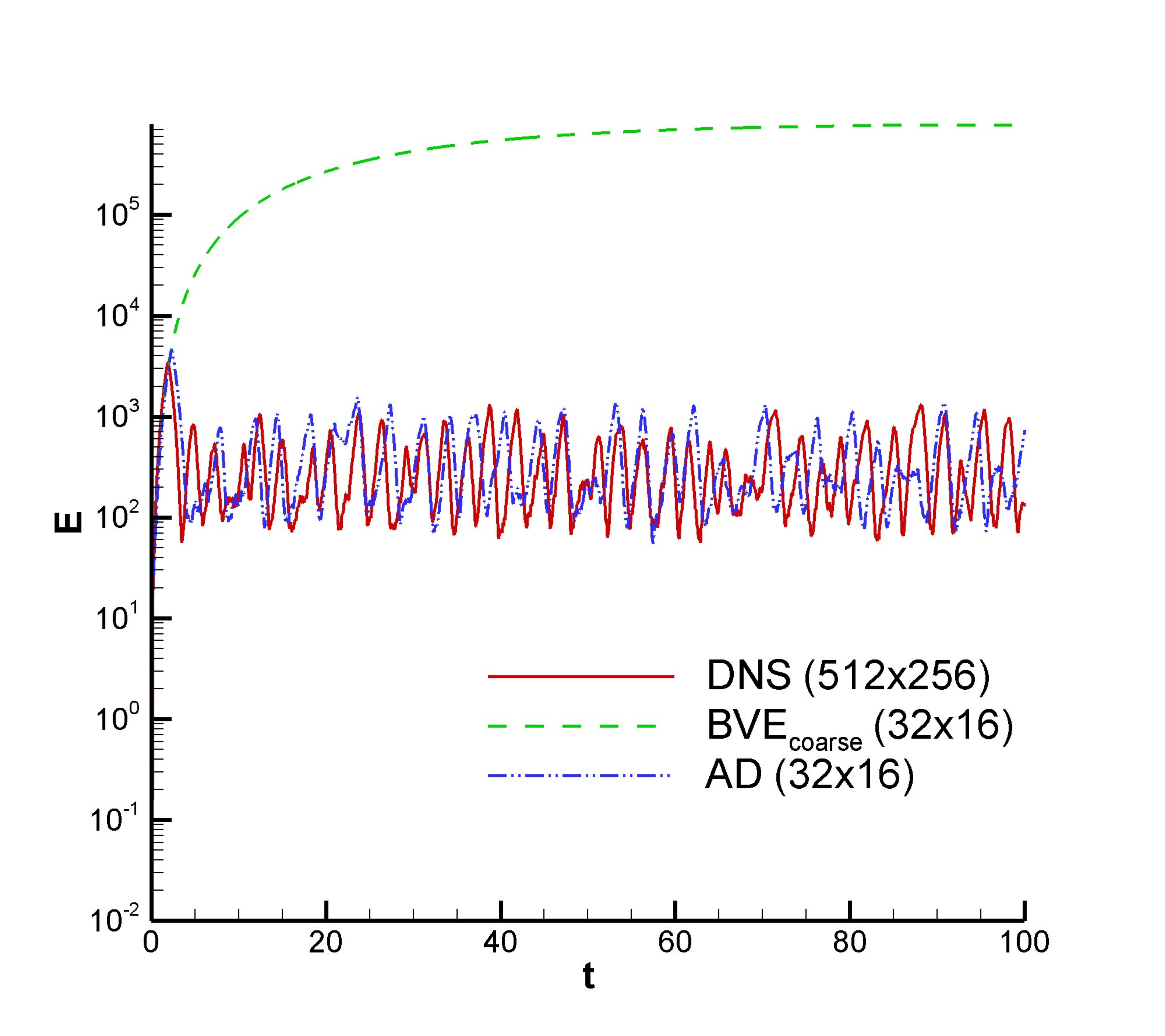}}}
\caption{
Experiment (ii): \  Time history of the total energy for $Re=450$ and $Ro=0.0036$:
DNS results at a resolution of $512 \times 256$,
under-resolved BVE$_{coarse}$ results at a resolution of $32 \times 16$,
new AD model results at a resolution of $32 \times 16$.
Note the significant improvement of the new AD model over the results from the under-resolved
BVE$_{coarse}$ run.
}
\label{fig:hist}
\end{figure}

\begin{figure}
\centering
\mbox{
\subfigure{\includegraphics[width=0.5\textwidth]{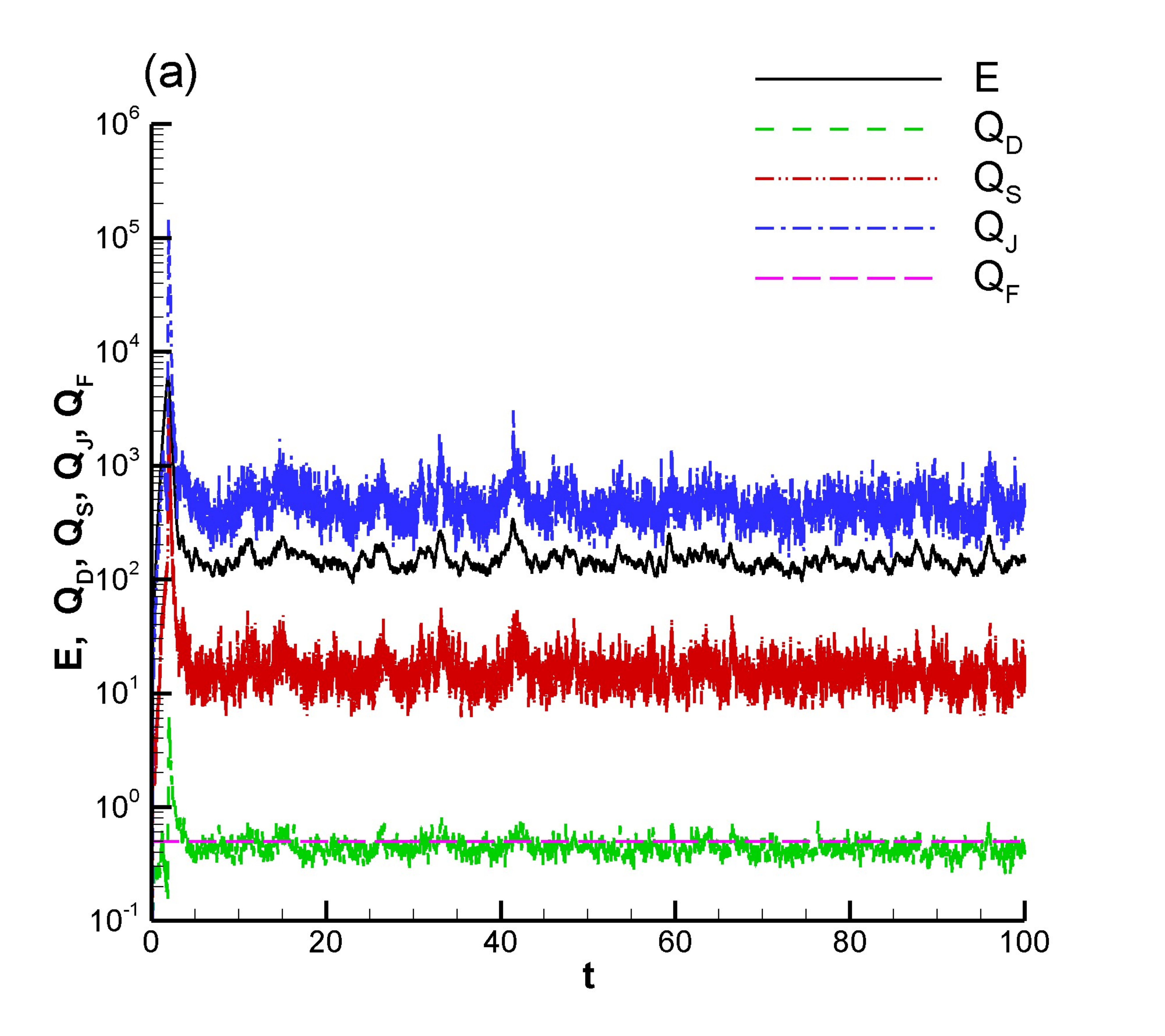}}
\subfigure{\includegraphics[width=0.5\textwidth]{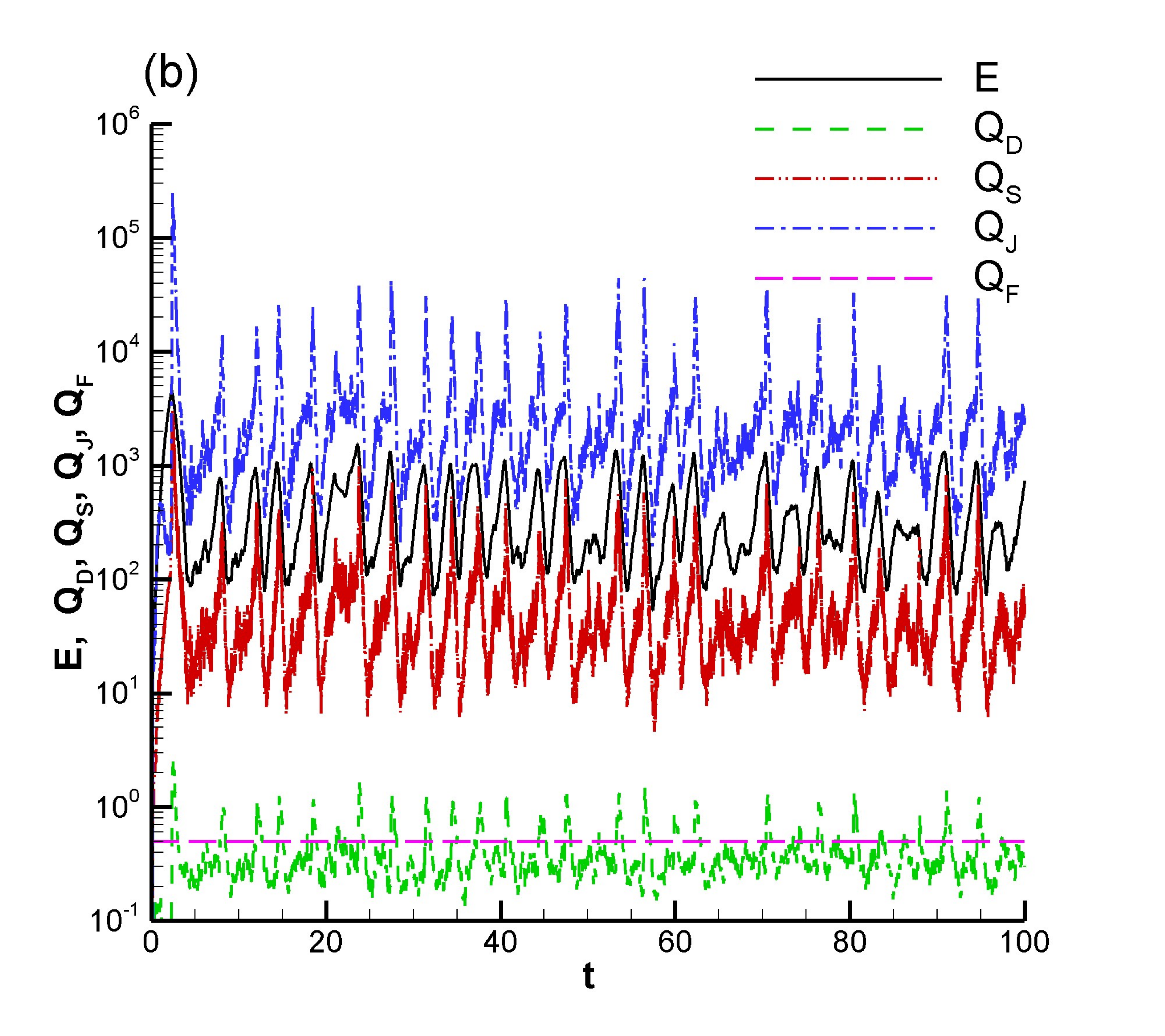}} }
\caption{
Time history of the total energy, dissipation, subfilter-scale, Jacobian, and forcing terms:
(a) Experiment (i): $\displaystyle \delta_{I}/L = 0.04$  and $\displaystyle \delta_{M}/L = 0.02$;
and (b) Experiment (ii): $\displaystyle \delta_{I}/L = 0.06$  and $\displaystyle \delta_{M}/L = 0.02$.
}
\label{fig:hist-all}
\end{figure}

For $Re=200$ and $Ro=0.0016$ we plot the time-averaged streamfunction and potential vorticity
contours in Fig.~(\ref{fig:exp1-s}) and Fig.~(\ref{fig:exp1-q}), respectively.
We note that the new AD model yields improved results by smoothing out the numerical
oscillations present in the under-resolved BVE$_{coarse}$ results.
This improvement is more clearly displayed in the potential vorticity contour plot in Fig.~(\ref{fig:exp1-q}).
For $Re=450$ and $Ro= 0.0036$ we plot the time-averaged streamfunction and potential vorticity
contours in Fig.~(\ref{fig:exp2-s}) and Fig.~(\ref{fig:exp2-q}), respectively.
The new AD model yields results that are significantly better than those corresponding to the
under-resolved BVE$_{coarse}$ run.
Indeed, in the streamfunction plot in Fig.~(\ref{fig:exp2-s}) the new AD model clearly displays the
correct four gyre pattern (as in the DNS plot), whereas the under-resolved BVE$_{coarse}$
run incorrectly yields only two gyres, which is nonphysical.
We also plot in Fig.~(\ref{fig:hist}) the time history of the total energy.
It is clear that the new AD model yields accurate results that are close to the DNS results.
The under-resolved BVE$_{coarse}$ run, however, yields totally inaccurate results. The contributions of the individual terms in Experiment (i) and Experiment (ii), obtained using the AD model for under-resolved simulations at a spatial resolution of $32 \times 16$, are also shown in Fig.~(\ref{fig:hist-all}). When we compare this to the higher resolution simulation illustrated in Fig.~(\ref{fig:ad-t}), it can be seen that the relative contribution of the subfilter-scale terms is more prominent in under-resolved simulations.

To summarize, the new AD model performs significantly better than the under-resolved
BVE$_{coarse}$ simulation in all tests.
For the higher Reynolds number case the improvement is more dramatic.
Indeed, the AD model correctly yields the four gyre pattern, just like the DNS computation,
whereas the under-resolved BVE$_{coarse}$ computation incorrectly predicts just two gyres.
We also emphasize that the new AD model (resolution of $32 \times 16$) is significantly more efficient than
the DNS (resolution of $512 \times 256$), yielding a {\em speed-up factor of more than $1,000$}.

Finally, we perform a sensitivity study with respect to the parameters used in the
new AD model:
$M$, the order of the high-order compact filter in Eq.~\eqref{eq:filter};
$N$, the order of the AD procedure in Eq.~\eqref{eq:3}; and
$\alpha$, the smoothing parameter of the second-order filter in Eq.~\eqref{eq:9}.
First, we investigate the sensitivity of the AD model with respect to $M$, the order of the high-order
compact filter in Eq.~\eqref{eq:filter}.
Since our tests show no significant differences among the different values of $M$, we are
presenting below only the results obtained with the second-order filter in Eq.~\eqref{eq:9}.

\begin{figure}
\centering
\mbox{
\subfigure{\includegraphics[width=0.23\textwidth]{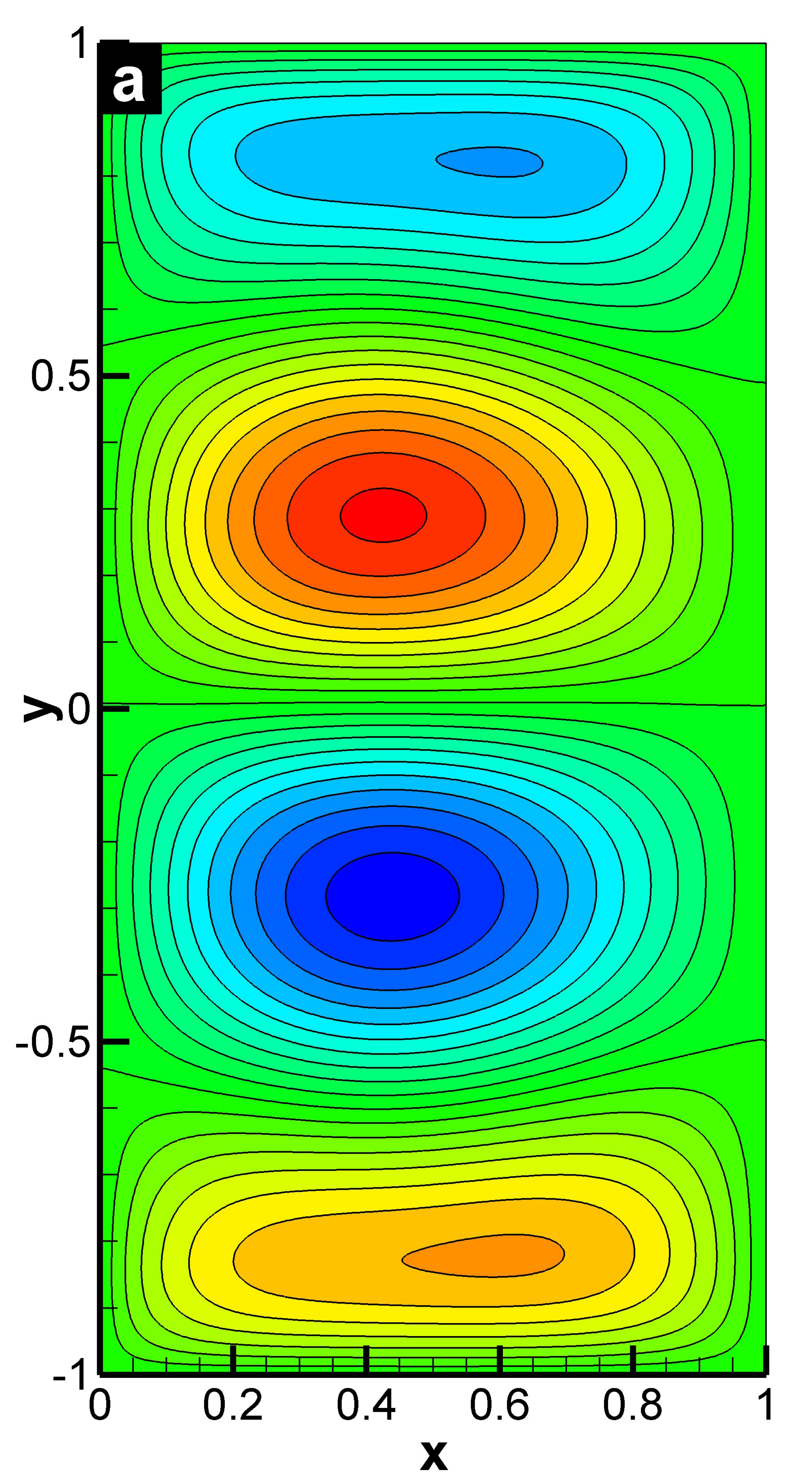}}
\subfigure{\includegraphics[width=0.23\textwidth]{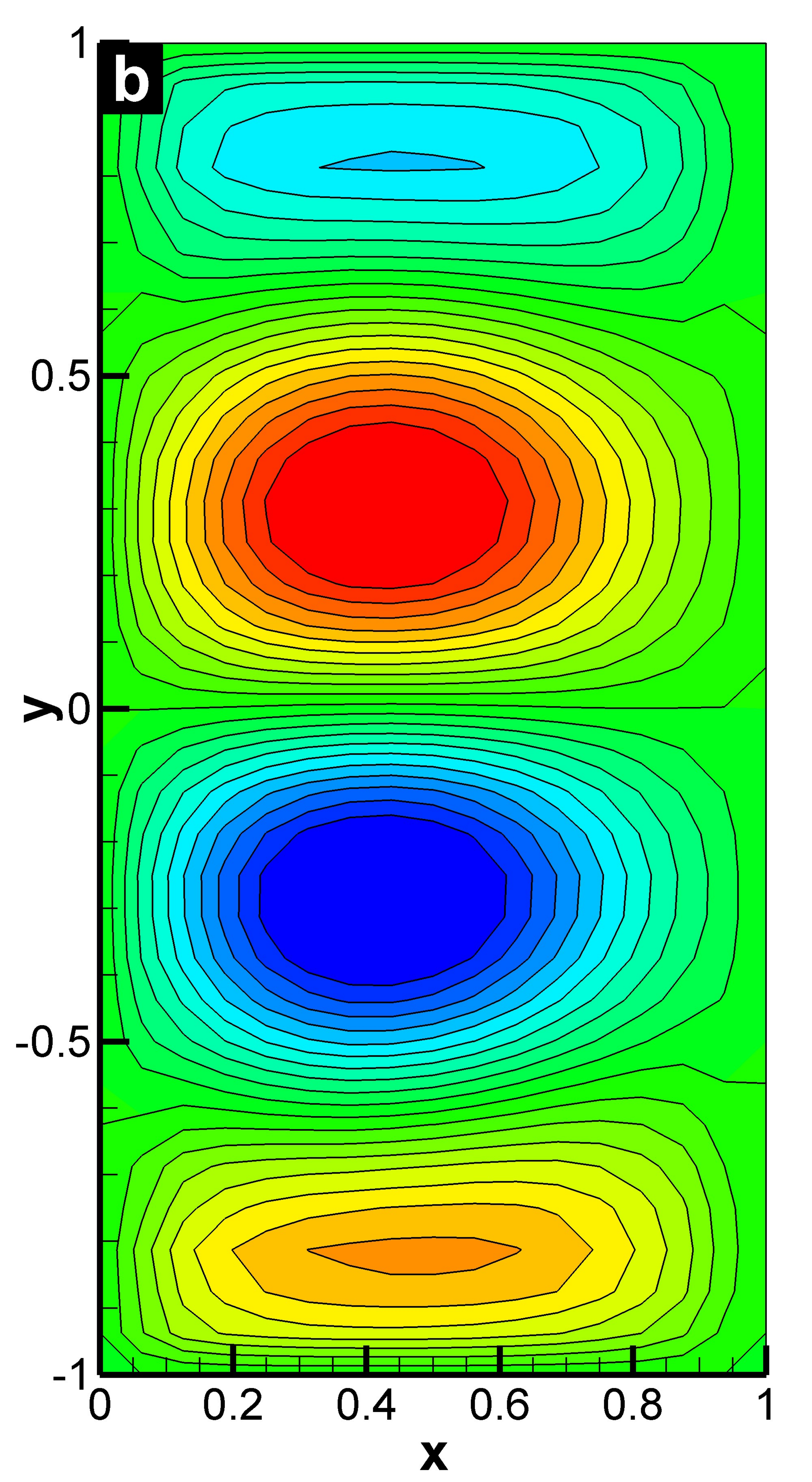}}
\subfigure{\includegraphics[width=0.23\textwidth]{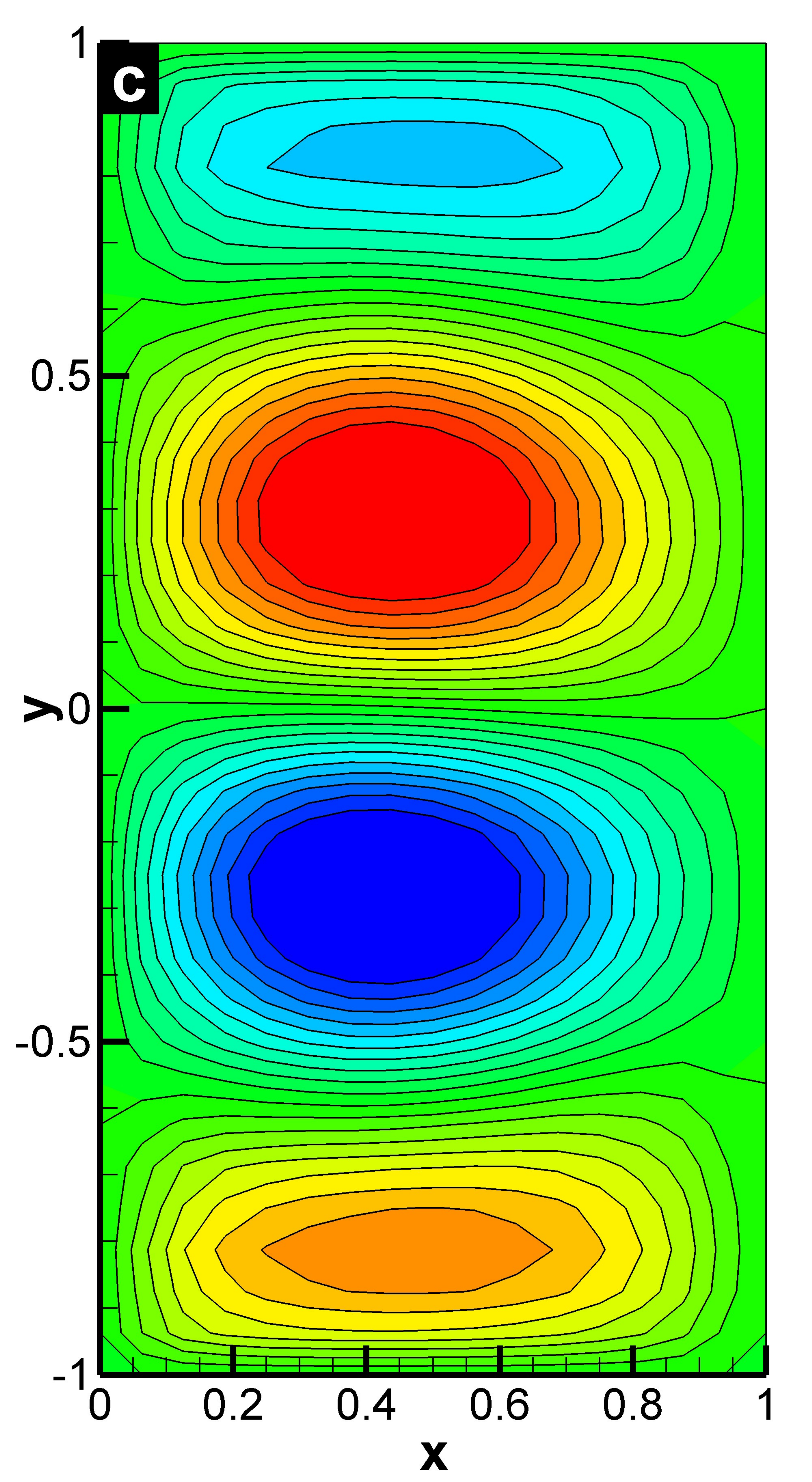}}
\subfigure{\includegraphics[width=0.23\textwidth]{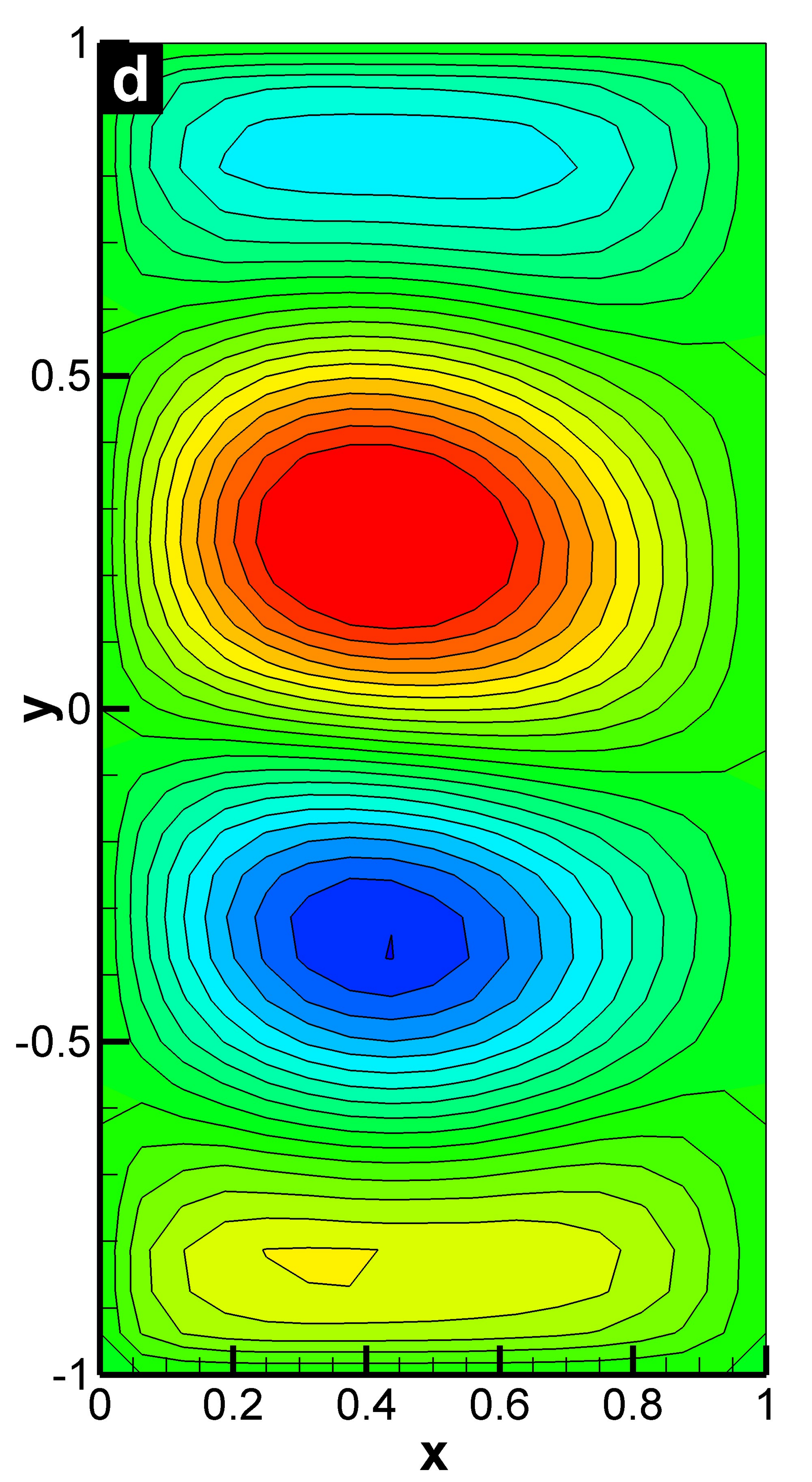}} }
\caption{
Experiment (ii): \  Time-averaged streamfunction data for $Re=450$ and $Ro=0.0036$;
Sensitivity with respect to the AD order $N$:
(a) DNS results at $512\times 256$ resolution;
(b) AD model results at a resolution of $32 \times 16$ with $N=5$,
(c) AD model results at a resolution of $32 \times 16$ with $N=3$,
(d) AD model results at a resolution of $32 \times 16$ with $N=1$.
Note the low sensitivity of the results with respect to $N$. Contour intervals are identical.
}
\label{fig:sfN}
\end{figure}

\begin{figure}
\centering
\mbox{
\subfigure{\includegraphics[width=0.7\textwidth]{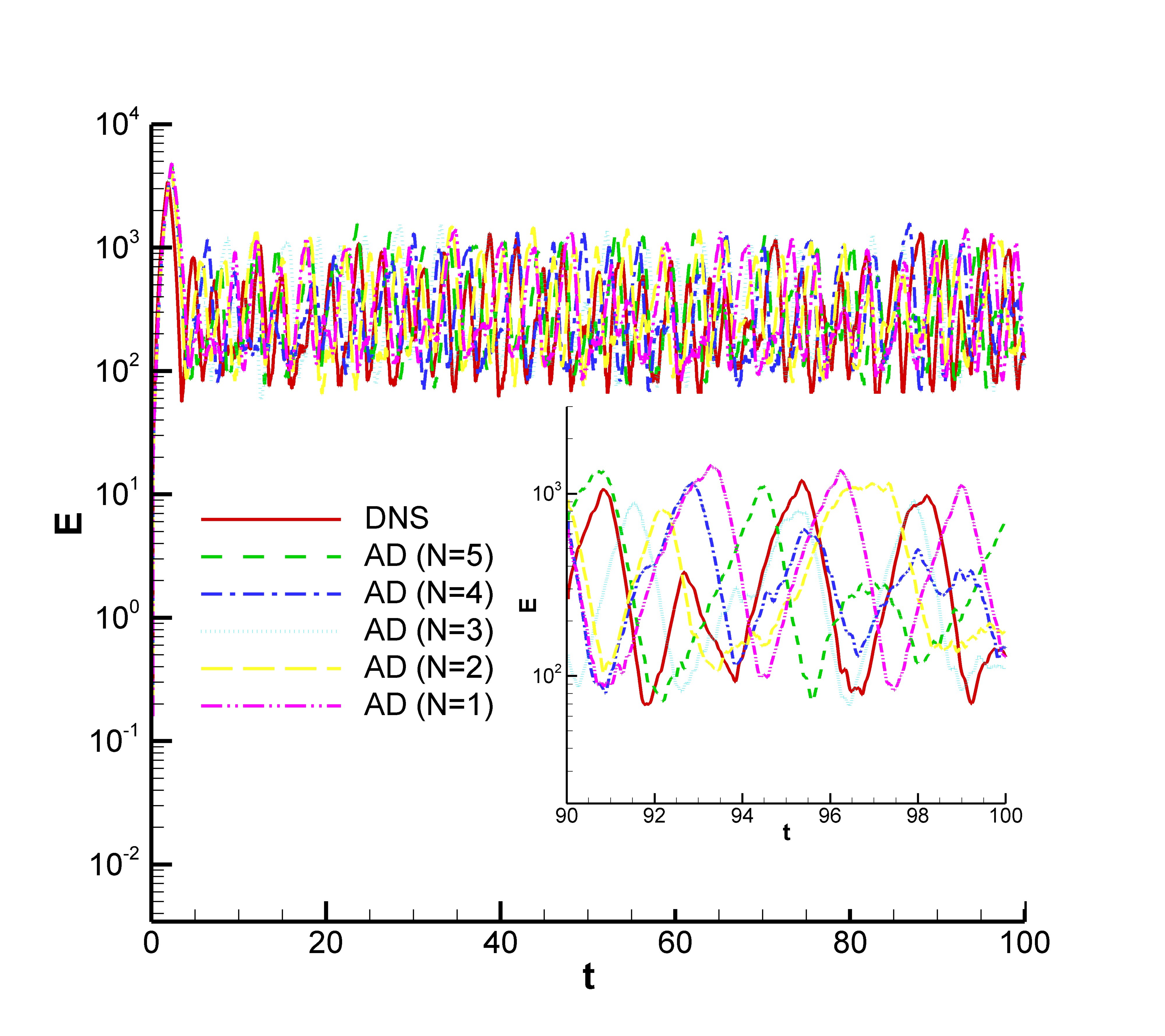}} }
\caption{
Experiment (ii): \  Time history of the total energy for $Re=450$ and $Ro=0.0036$;
Sensitivity with respect to the AD order $N$:
DNS results at a resolution of $512 \times 256$ and
new AD model results at a resolution of $32 \times 16$.
Note the non-negligible sensitivity with respect to $N$. Embedded picture shows time series between t=90 and t=100.
}
\label{fig:compN}
\end{figure}

Next, we perform a sensitivity study with respect to $N$, the order of the AD procedure in
Eq.~\eqref{eq:3}.
For a fixed smoothing parameter $\alpha=0.25$, the time-averaged streamfunction contour
plots in Fig.~(\ref{fig:sfN}) show a low sensitivity with respect to $N$.
The time history of the total energy plotted in Fig.~(\ref{fig:compN}), however, displays a non-negligible
sensitivity with respect to $N$.
Indeed, higher values of $N$ yield more accurate results.
Of course, the tradeoff is an increase in computational time due to the additional filtering
operations needed for higher values of $N$:
The CPU time is $170 s$ for $N=5$, $145 s$ for $N=3$, and $113 s$ for $N=1$.
Thus, the two numerical investigations in Fig.~(\ref{fig:sfN}) and Fig.~(\ref{fig:compN}) seem to suggest
that, in order to balance the numerical accuracy and the computational efficiency, a relative low value
for $N$ might be desirable.

\begin{figure}
\centering
\mbox{
\subfigure{\includegraphics[width=0.22\textwidth]{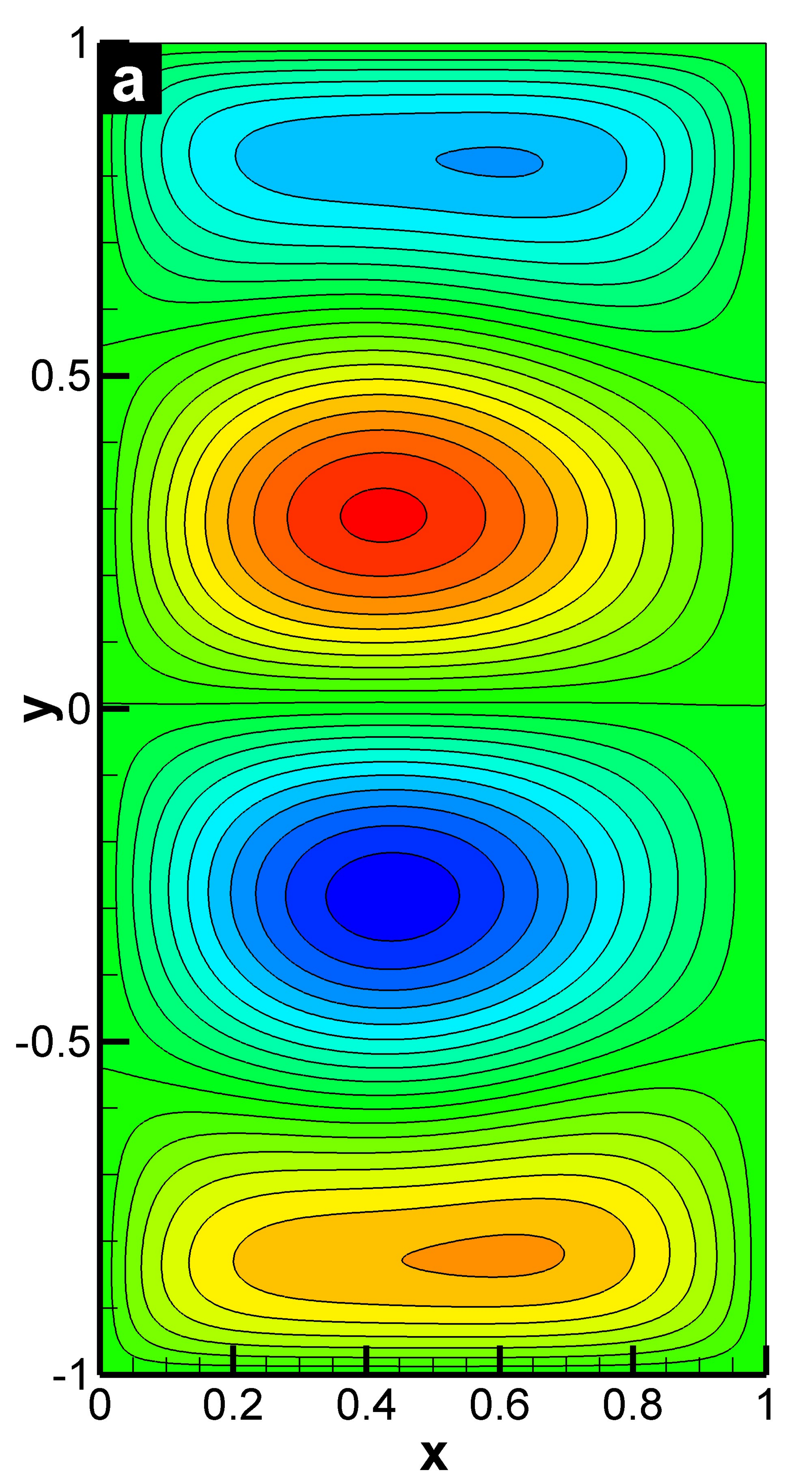}}
\subfigure{\includegraphics[width=0.22\textwidth]{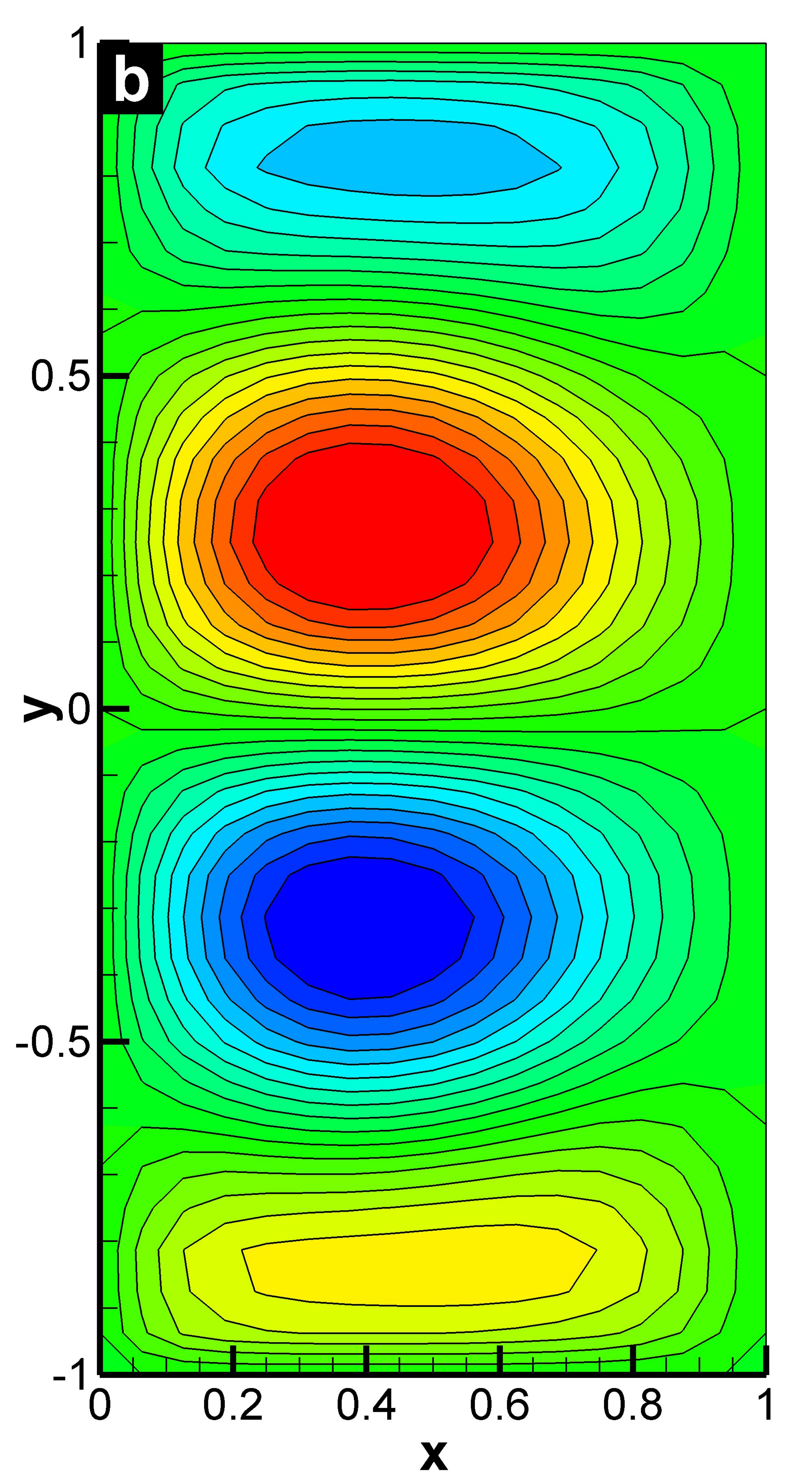}}
\subfigure{\includegraphics[width=0.22\textwidth]{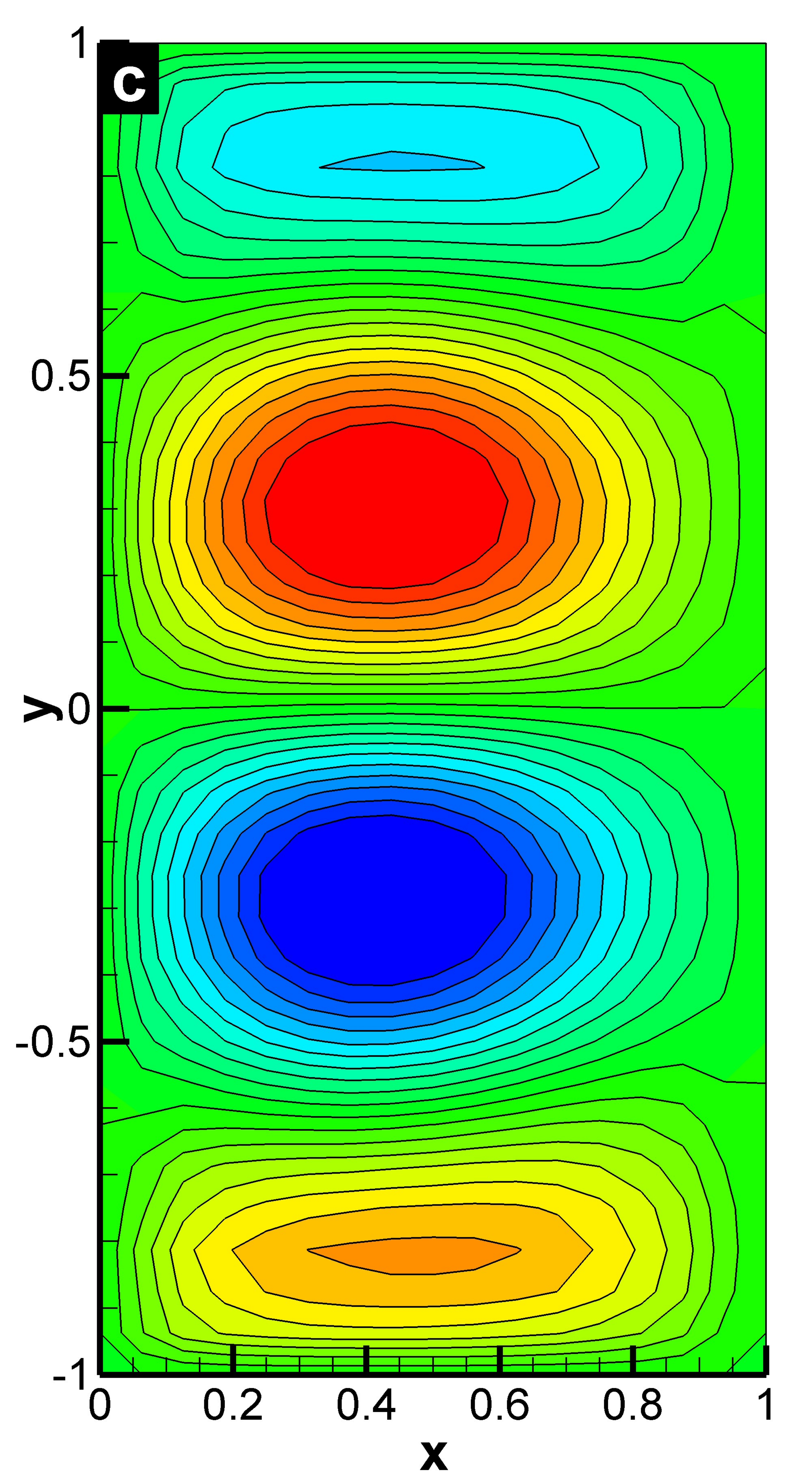}}
\subfigure{\includegraphics[width=0.22\textwidth]{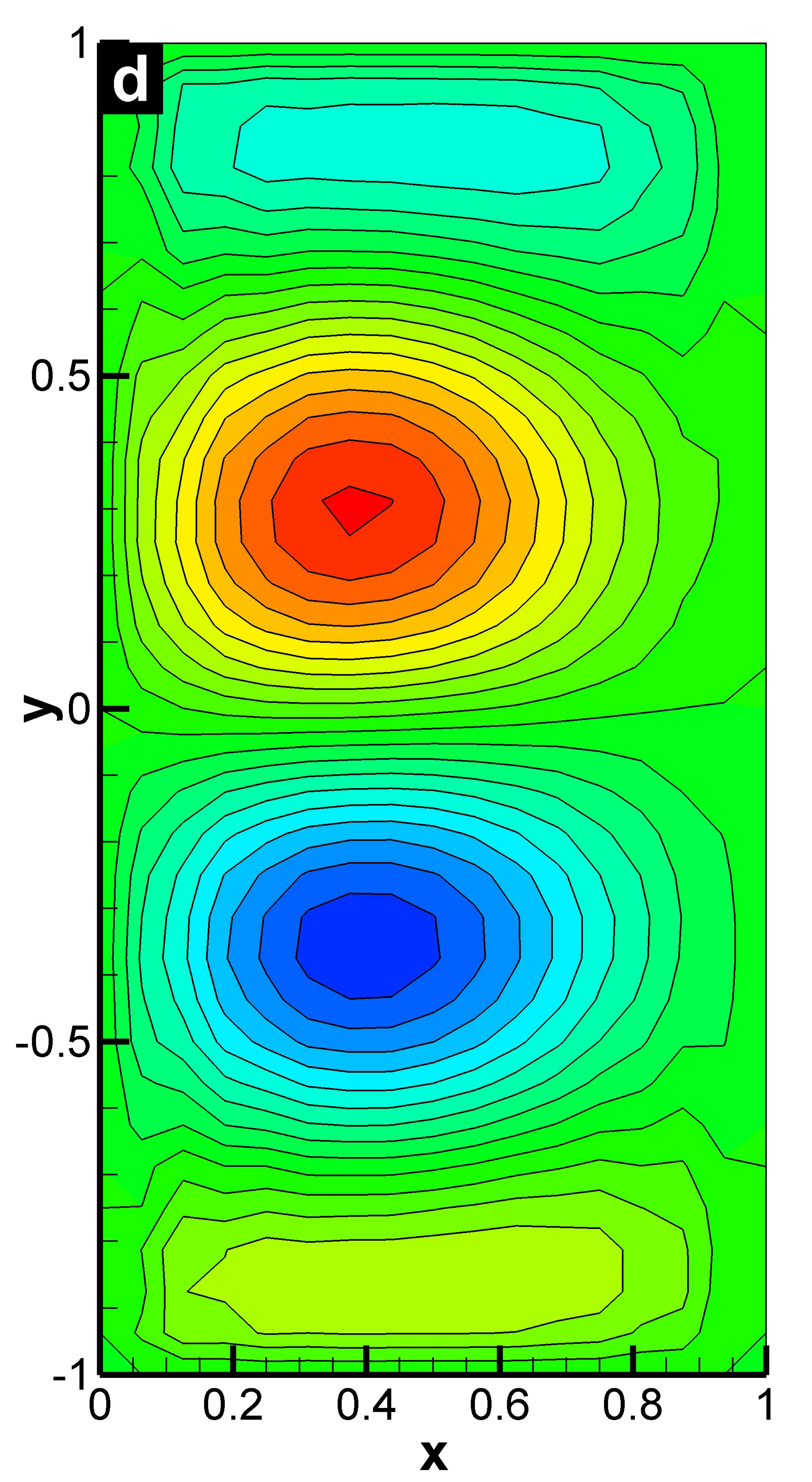}} }
\caption{
Experiment (ii): \  Time-averaged streamfunction data for $Re=450$ and $Ro=0.0036$;
Sensitivity with respect to the smoothing filtering parameter $\alpha$.
(a) DNS results at a resolution of $512\times 256$;
(b) AD model results at a resolution of $32 \times 16$ with $\alpha=0.1$,
(c) AD model results at a resolution of $32 \times 16$ with $\alpha=0.25$,
(d) AD model results at a resolution of $32 \times 16$ with $\alpha=0.45$.
Note the low sensitivity of the results with respect to $\alpha$. Contour intervals are identical.
}
\label{fig:alpha}
\end{figure}

Finally, we investigate the sensitivity of the AD model with respect to $\alpha$, the smoothing
parameter of the second-order filter in Eq.~\eqref{eq:9}.
For a fixed order of the AD procedure $N=5$, the time-averaged streamfunction contour
plots in Fig.~(\ref{fig:alpha}) show a low sensitivity with respect to $\alpha$.
Our numerical experiments, however, suggest an optimal range of $0.1 < \alpha < 0.3$, since
higher values of $\alpha$ do not eliminate numerical oscillations for coarse grids.

\section{Conclusions}
\label{section_conclusions}

This paper introduced a new approximate deconvolution (AD) model for the LES of two-dimensional turbulent geophysical
flows.
The AD model was tested in the numerical simulation of the wind-driven circulation in a
shallow ocean basin, a standard prototype of more realistic ocean dynamics.
The mathematical model employed was the barotropic vorticity equation (BVE) driven by a symmetric double-gyre wind
forcing, which yielded a four-gyre circulation in the time mean.
The AD model was tested on a mesh that was $16$ times coarser than that used
by the direct numerical simulation (DNS) run.
The AD model yielded numerical results that were in close agreement with those of the DNS.
In particular, the four gyre structure of the time-averaged streamfunction contour plots was
recovered by the AD model. Moreover, the CPU time of the AD model was 1,000 times
lower than that of DNS.
We emphasize that this combination of computational efficiency and numerical accuracy
achieved by the new AD model was due to the specific subfilter-scale model utilized.
Indeed, an under-resolved numerical simulation without any subfilter-scale model on the
same coarse mesh as that employed by the AD model produced inaccurate results.
We also performed a numerical investigation of the sensitivity of the new AD model with
respect to the parameters used and we found that the model is robust.
This first step in the numerical assessment of the new model shows that AD could represent
a viable tool in the LES of more realistic turbulent geophysical flows.

\bibliographystyle{elsarticle-harv}
\bibliography{ref}

\end{document}